\documentclass{aa}  
\usepackage{xcolor}
\usepackage{graphicx}
\usepackage{txfonts}
\usepackage{lipsum}
\usepackage{subcaption}         % necessary for continued figures, example in section 3
\usepackage{lscape}             % to rotate a single page table, example in appendix.
\usepackage{placeins}           % useful with \FloatBarrier, to keep 
\begin{document}

\title{The Domain Adaptation Problem in photometric redshift estimation: a solution applied to the HSC Survey}

   \author{M. Treyer\inst{1}
        \and R. Ait-Ouahmed\inst{1} \and S. Arnouts\inst{1} \and J. Pasquet\inst{2,3} \and E. Bertin\inst{4} \and G. Desprez\inst{5} \and V. Picouet\inst{6} \and M. Sawicki\inst{7}
        }

\institute{Aix Marseille Universit\'e, CNRS, CNES, LAM, Marseille, France 
\and AMIS, Université Montpellier Paul-Val\'ery, Montpellier, France
\and TETIS - Inrae, AgroParisTech, Cirad, CNRS, Univ. Montpellier, Montpellier, France
\and Universit\'e Paris-Saclay, Universit\'e Paris Cit\'e, CEA, CNRS, AIM 91191, Gif-sur-Yvette, France
\and Kapteyn Astronomical Institute, University of Groningen, P.O. Box 800, 9700AV Groningen, The Netherlands
\and The California Institute of Technology, 1200 E. California Blvd., Pasadena, CA 91125, USA
\and Department of Astronomy \& Physics and Institute for Computational Astrophysics, Saint Mary's University, 923 Robie Street, Halifax, Nova Scotia, B3H 3C3, Canada}

\abstract
{The multi-band HSC-CLAUDS survey comprises several sky regions with varying observing conditions, only one of which, the COSMOS Ultra Deep Field (UDF), offers extensive redshift coverage.}
{We aim to exploit a complete sample of labeled galaxies from the COSMOS UDF at $i<25$ ($z\lesssim 5$) to train a convolutional neural network (CNN) and infer more accurate photometric redshifts in the other regions than those currently available from SED-fitting methods.}
{To address the severe domain mismatch problem we observed when applying the trained CNN to regions other than the COSMOS UDF, we developed an unsupervised adversarial domain adaptation network that we grafted onto the CNN. The method is validated by three tests: the predicted redshifts are compared to the spectroscopic redshifts that are available for limited samples of mostly bright galaxies; the predicted redshift distributions of the entire galaxy population of a given field in several intervals of magnitude are compared to those of the COSMOS UDF, assumed to be representative; the redshifts predicted for a sample of galaxies selected by narrow-band filter observations sensitive to [OII] emitters at $z\sim1.47$ are compared to those of confirmed [OII] emission line galaxies.} 
{The results show successful domain adaptation: the network is able to transfer its redshift classification capability learnt from the COSMOS UDF to other regions of HSC-CLAUDS. Accuracy varies depending on magnitude and redshift, following that of the labels we used, but far exceeds that of currently available photometric redshifts. The catalogs of CNN redshifts we inferred for the XMM, DEEP2 and ELAIS fields and for the remaining COSMOS region  ($\sim 4$ million sources in total at $i<25$) are made public.}
{}

\keywords{Galaxies: distances and redshifts, Surveys, Catalogs,  Methods: data analysis, Techniques: image processing}

\maketitle

\section{Introduction}

Photometric redshifts are essential to cosmological surveys, since spectroscopy alone can no longer cover their extent. Template fitting techniques \citep[e.g.][]{Ilbert2006} have been in use for decades. They can predict redshifts in any interval, given an adequate set of spectral energy distribution (SED) templates and of photometric bands in which the galaxies are observed.
However, for a given number of bands, they are outperformed by machine learning methods, provided that the training samples cover the full range of galaxy properties whose redshifts are to be predicted, which is not often the case. These methods are fast improving. 

A multi-layer perceptron (MLP) was first used to learn the mapping between photometry and redshift from data \citep[ANNz by][]{Collister2004, Vanzella2004}. Other machine learning algorithms include Support Vector Machines \citep[SVM;][]{Wadadekar2005}, k Nearest Neighbours \citep[kNN;][]{Csabai2003, Zhang2013kNN}, Random Forest \citep{Carliles2010RF}, XGBoost \citep{Li2022XGBoost} and Catboost \citep{Li2024CatBoost}, etc.

A significant step forward was made with the use of convolutional neural networks (CNN), which can exploit the entire information content of multi-band galaxy images to refine redshift estimates \citep{Hoyle2016, DIsantoPolsterer2018}. The benefits of CNNs were clearly demonstrated by \citet{P19} (hereafter P19), who used a CNN trained with the SDSS 5-band data \citep{York2000} under a redshift bin classification loss. This proof-of-concept study illustrated how a CNN was able to make use of pixel-level information to extract features beyond colors and greatly improve redshift estimation compared to machine learning techniques. This work inspired a number of attempts to improve its results \citep[e.g.][]{Hayat2021,Dey2022,treyer2024, aitouahmed2024}, some of them proposing hybrid models combining an MLP branch tasked with processing photometric features (e.g. magnitudes, colours), and a CNN branch having access to multi-band images of the galaxies \citep{Menou2019, Henghes2021,Henghes2022,Yao2023, Zhang2024,Roster2024,Wei2025PASA}

One condition for the applicability of CNN methods in a supervised learning context is the availability of a training sample that is sufficiently large and representative of the properties of the galaxy populations for which redshifts are to be inferred. Failure to comply with this condition leads to bias.  
Unfortunately it is difficult to meet with deep observations as they span large ranges of redshift and magnitude. Spectroscopy at faint magnitude is costly, and generally limited to small regions of large deep surveys, which causes yet another issue: CNNs fail to predict accurate redshifts in survey regions that differ observationally from those hosting the training sample, even if it is large and representative of the galaxies themselves.

In machine learning, this nuisance, known as the "domain adaptation problem", occurs when the "target data" (on which the model will be used) does not come from the same distribution as the "source data" (on which the model is trained). The deep HSC-CLAUDS survey \citep{Desprez2023}, where labeled sources are essentially confined to one ultra deep field, provides us with an example of just such a problem, and an opportunity to try and remedy it.

Various deep domain adaptation (DA) approaches have been proposed to address this issue, among which:
(i) Discrepancy-based techniques, which minimize a divergence or distance between source and target feature distributions while jointly optimizing the main learning task
\citep[e.g.][]{long2015learning};
(ii) Reconstruction-based DA (autoencoder- or VAE-based adaptation), in which an encoder shared between the source and target domains learns domain-invariant latent representations by optimizing a reconstruction objective in addition to the main task
\citep[e.g.][]{ghifary2015domain};
(iii) GAN-based pixel-level DA, where generative adversarial networks \citep{goodfellow2014generative} translate source images into target-like images to reduce the appearance gap. For example, PixelDA \citep{bousmalis2017unsupervised} trains a generator to map source images into the target domain while preserving class labels;
(iv) Adversarial feature-level DA, inspired by GANs but operating in feature space. These methods train feature extractors to be simultaneously predictive for the main task and invariant to the domain through adversarial domain classification
\citep[e.g.][]{ganin2016domain, tzeng2017adversarial};
(v) Transfer learning / fine-tuning, commonly used in survey-to-survey or instrument-to-instrument adaptation. A model trained on a source dataset is partially retrained or fine-tuned on a small amount of labeled or unlabeled target data;
(vi) Semi-supervised, weakly-supervised, or universal / open-set DA, which addresses scenarios where the target domain may contain new, missing, or unknown classes, or where only limited or weak labels are available.

These DA methods, or combinations thereof, have been used in various astrophysical endeavors, mostly for classification tasks. Indeed, domain shifts are ubiquitous in astronomy: between simulated and observed data, between different instruments or surveys, between clean and noisy images, between different calibration systems, etc. Discrepancy-based methods have been used, e.g., to improve the robustness of galaxy morphology classification under degraded image quality \citep{Ciprijanovic2022}, to extract domain-independent cosmological information from different hydrodynamical cosmological simulations \citep{Roncoli2023arXiv}, and to align simulated and noisy images to estimate Einstein radii \citep[in a regression task,][]{Swierc2024}. They have also been used in combination with adversarial DA between simulations and SDSS observations to classify galaxy mergers \citep{Ciprijanovic2021}.
Instance reweighting/density-ratio DA (a statistical discrepancy-based method) has been employed between simulated and observed SEDs to improve the recovery of star-formation histories \citep{Gilda2024}. Reconstruction-Based DA (VAE) was used for cross-survey galaxy morphology classification \citep{Xu2023}. 
Adversarial feature-level DA has been applied between photo-ionisation models and integral-field observations to classify ionised nebulae \citep{Belfiore2025}, between HST and JWST images to transfer morphological classification \citep{Huertas-Company2024}, and between simulated lensing data and HSC data to find strong gravitational lenses 
\citep{Alexander2023}. Semi-supervised open-set DA has been used for cross-survey galaxy morphology classification enabling anomaly detection (e.g., mergers or strong lenses in the target domains) \citep{Ciprijanovic2023}. Transfer learning and fine-tuning approaches have been applied between the DESI Legacy Imaging Surveys to adapt Galaxy Zoo labels to the other DESI surveys \citep{Ye2025}, and to identify blended sources using networks pre-trained on natural images \citep{Farrens2022}.

Here, we aim to implement domain adaptation to estimate CNN photometric redshifts across the entire HSC-CLAUDS survey from a unique labeled field. Discrepancy-based methods require a choice of distance metric that would effectively create domain invariant representations and they are computationally expensive. Reconstruction-based methods may loose the redshift information in the reconstruction process as it is a subtle signal in the correlations between bands. The same goes for GAN-based methods, as it may be challenging to translate images from one field to another while preserving the redshift information. Transfer learning with fine tuning proved ineffective at faint magnitude given the shortage of labels in the target fields. Thus we chose to develop an adversarial approach, which would allow us to directly optimize a representation space that does not distinguish between fields and can be used for redshift estimation. 

The paper is organized as follows: the HSC-CLAUDS data and the training set are described in Section \ref{sec:data} and \ref{sec:refsample}, respectively; the baseline redshift-estimating CNN and its domain adaptation issue are introduced in Section \ref{sec:baseline}; our adversarial solution and its application to the different HSC regions are presented in Section \ref{sec:adversarial} and Section \ref{sec:otherfields}, respectively. We highlight shortcomings in Section \ref{sec:qa} and conclude in Section \ref{sec:conclusion}. Samples of images and results are displayed in Appendix \ref{sec:images_pdf}.

\section{The HSC data}
\label{sec:data}

\subsection{The photometric data}
\label{subsec:phot}

\begin{figure*}
\includegraphics[width=\textwidth]{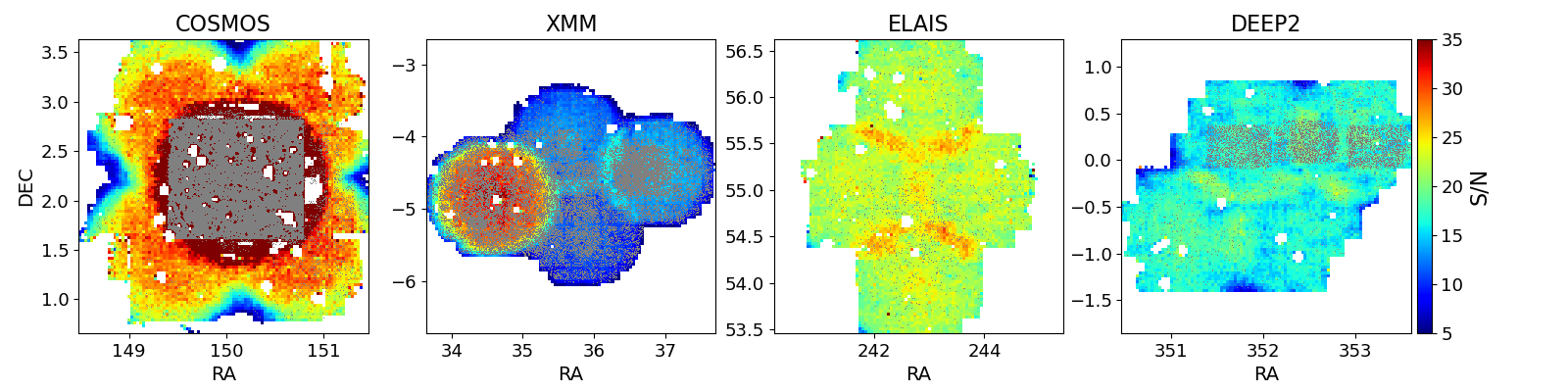}
\includegraphics[width=\textwidth]{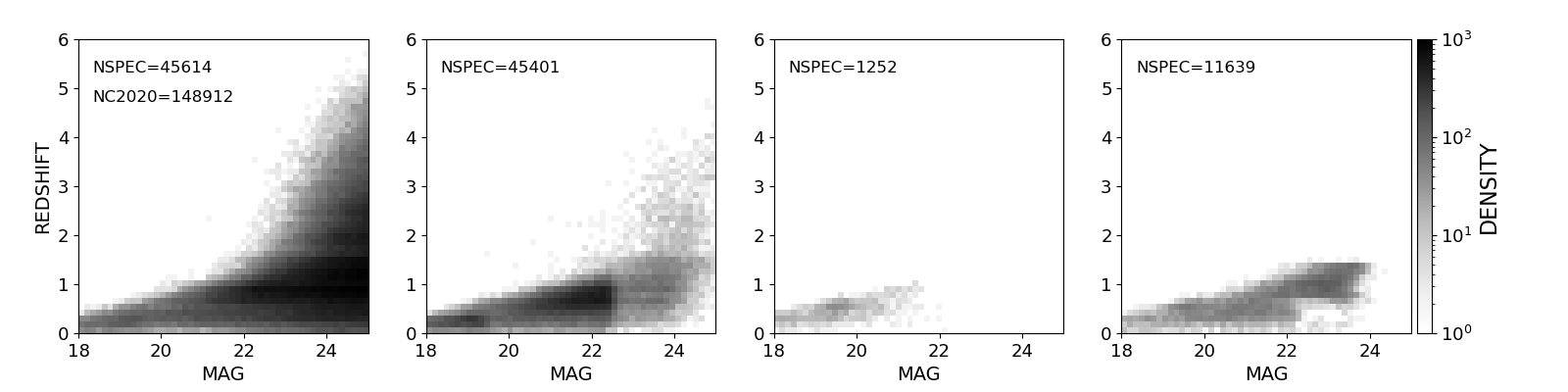}
\caption{{\bf Top:} the four HSC regions color-coded by the $i$-band S/N of sources at $24<i<25$. Only sources imaged in all 6 bands are shown. Several masks are applied to rid the catalogs of stellar sources. The dark red circles in COSMOS and XMM are the 2 UDFs. The gray dots mark the position of sources with known redshifts at $i<25$. {\bf Bottom:} their distributions in $i$-band magnitude (MAG) - redshift space. NSPEC is the number of galaxies with spectroscopic redshifts, NC2020 is the number of galaxies with redshifts from the COSMOS2020 photometric redshift catalog (introduced in Section \ref{subsec:spec}).}
\label{fig:hscfields}
\end{figure*}

The Hyper Suprime-Cam Strategic Survey Program on the Subaru telescope (HSC-SSP, \citealt{Aihara2018a}) represents the current state of the art in deep, wide-area imaging surveys. The HSC Deep component is an imaging survey in the $grizy$ filters covering 26 deg$^2$ to a limiting magnitude of $i_{AB}\sim$26. 
This is an unprecedented combination of area and depth that will remain unmatched for many years after LSST scientific operations begin. It consists of four regions: COSMOS, XMM-LSS, ELAIS-N1 and DEEP2-3, each mapped by several pointings. We refer to these regions as simply COSMOS, XMM, ELAIS et DEEP2. An Ultra Deep layer, 1 magnitude deeper, completes the survey in two small sub-regions (covering $\sim$3.5 deg$^2$) of the COSMOS and XMM fields. We refer to these as the COSMOS and XMM ULTRA DEEP fields (UDF), and to the none UDF parts as COSMOS DEEP and XMM DEEP. We use the public DR2 \citep[][]{Aihara2019} for the Deep and Ultra Deep layers of the survey. These have median depths $g=26.5$, $y=24.5$, and $g=27$, $y=25.5$, respectively. 

Follow-up observations in the $u$-band and in the slightly redder $u^{\star}$-band from the deep CLAUDS survey using the CFHT MegaCam imager \citep[][]{Sawicki2019} were added to this dataset. COSMOS was observed with both filters, while XMM was only covered by the filter $u^{\star}$, and ELAIS and DEEP2 by the filter $u$. CLAUDS covers 18 deg$^{2}$ of the four fields to a median depth $u=27$, and 1.6 deg$^2$ of the two UDFs to $u=27.4$. 

Deep near-infrared (NIR) observations in 3 bands ($JHK_s$) are also partially available on the COSMOS field \citep[][]{McCracken2012} from the UltraVISTA\footnote{\url{https://ultravista.org}} survey, and on XMM \citep[][]{Jarvis2013} from the VIDEO\footnote{\url{http://www.eso.org/sci/observing/phase3/data_releases.html}} survey. We did not use them in the present analysis due to their limited coverage, although we tested that adding them noticeably improved the CNN performance in the COSMOS field (training is described in Section \ref{subsection:training}), particularly at faint magnitude and high redshift. 

\begin{figure}
\includegraphics[width=\linewidth]{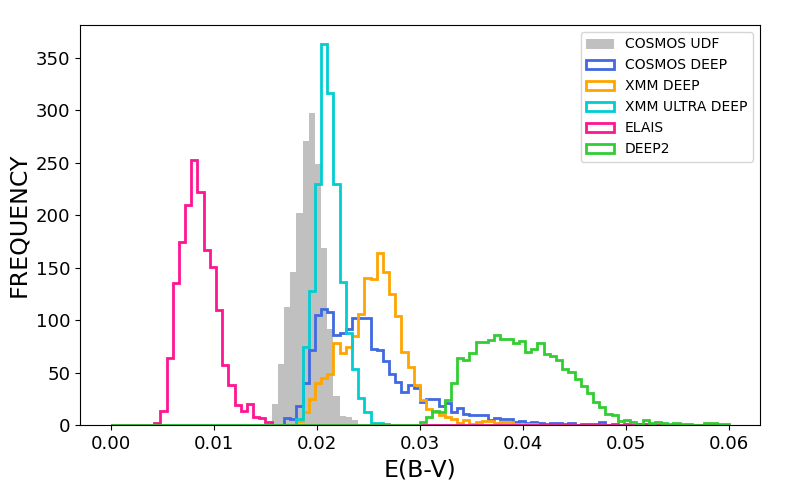}
\caption{The normalized E(B-V) distributions in the different HSC sub-regions.}
\label{fig:EBV_DIS}
\end{figure}

The top panels of Fig.~\ref{fig:hscfields} show the four HSC regions color-coded by the $i$-band signal-to-noise ratio (S/N) of sources at $24<i<25$. The two UDFs are well identified as dark red circles (highest S/N) within COSMOS and XMM. There are also variations in seeing and observing conditions in the other filters. The spatial distributions of sources with known redshifts are overlaid. Their distributions in magnitude-redshift space are displayed in the bottom panels and discussed in the next section. The E(B-V) distributions in the various sub-regions are shown in Fig.~\ref{fig:EBV_DIS}. 

For each source at $i<25$, we created a datacube of $6\times64\times 64 $ pixels, that includes one CLAUDS image ($u^{\star}$ or $u$) and five HSC images ($grizy$). All images were projected onto the same HSC reference pixel grid, using \textsc{SWarp} \citep{2002ASPC..281..228B}, with a scale of 0.168~arcsec/pixel. 
Sources whose images presented obvious defects were excluded. Most stellar and dubious sources were also removed with the following mask: ${\tt CLASS\_STAR\_HSC\_I}$ < 0.9 \& ${\tt COMPACT}==0$ \& ${\tt OBJ\_TYPE}==0$ \& ${\tt MASK}==0$. We chose to limit our analysis to $i=25$ because the accuracy of our redshift estimation is too severely degraded beyond this magnitude, as will be shown in following sections.

SED-fitting redshifts were computed for all sources in all four regions, based on $ugrizy$ magnitudes or on $ugrizy$ + $JHK_s$ magnitudes when available in the COSMOS and XMM fields. The purpose of this work is to supersede them with estimates of significantly higher quality. A full description of the HSC Deep survey and its ancillary data can be found in \citet[][]{Desprez2023} and \citet[][]{Picouet2023}.

\subsection{The labeled data}
\label{subsec:spec}
 
\begin{figure}
\centering
\includegraphics[width=\linewidth]{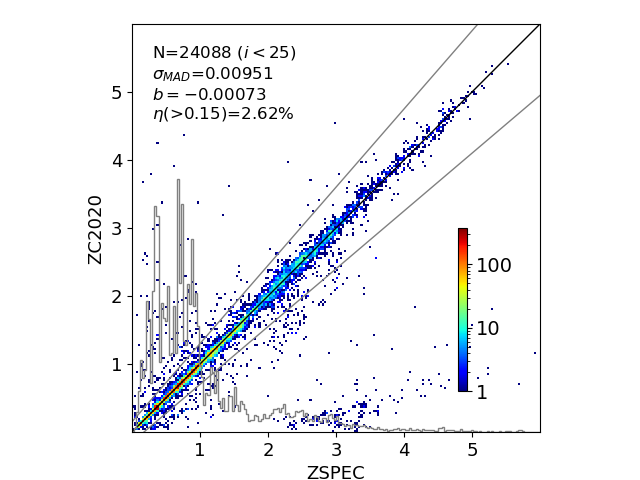}
\caption{C2020 photometric redshifts (ZC2020) compared to spectroscopic redshifts (ZSPEC) when available in the COSMOS UDF at $i<25$. Their distribution is overlaid in gray. The number of sources, N, and the C2020 performance metrics (Section \ref{subsection:training}) are quoted in the upper-left corner.}
\label{fig:zc2020_zspec}
\end{figure}

The spectroscopic redshifts available in the HSC fields come from a compilation of spectroscopic surveys summarized in Table~\ref{table:zspec}. The COSMOS field is completed with redshifts from the \citet{Khostovan2025} compilation. 
A total of $\sim$104,000 secure spectroscopic redshifts are available at $i<25$, unevenly distributed among the 4 regions and in magnitude-redshift space (84\% are in COSMOS and XMM), as can be seen in the lower panels of Fig.~\ref{fig:hscfields}.  
In the COSMOS UDF, the lack of completeness in magnitude-redshift space is compensated for by the COSMOS2020 photometric redshift catalog  \citep{Weaver2022}. These redshifts were computed from 30 photometric bands, ranging from UV to IR, via SED-fitting techniques in the UDF. The authors estimated four photometric redshifts based on two different multiband photometric catalogs (using two distinct flux extraction software packages) and two different SED-fitting codes. To construct a reliable sample, we computed the mean and standard deviation of these four redshifts, $\bar z$ and $\sigma_z$, and retained those with $\sigma_z\le 0.1(1+\bar z)$. Figure \ref{fig:zc2020_zspec} illustrates the quality of this selection (hereafter the C2020 sample) for sources with measured spectroscopic redshifts\footnote{While we use $z$ as the usual notation for redshift in the text, with specifiers when necessary, e.g. $z_{\rm SPEC}$, in all figures we prefer to use capital letters for greater visibility: thus ZSPEC stands for spectroscopic redshift ($z_{\rm SPEC}$), ZC2020 for C2020 redshift ($z_{\rm C2020}$), ZCNN for CNN predicted redshift ($z_{\rm CNN}$), and ZPHOT for $ugrizy$ or $ugrizy$+NIR SED-fitting photometric redshift ($z_{\rm PHOT}$). MAG stands for $i$-band magnitude.}. The global deviation, $\sigma_{MAD}$, and median bias, $b$, (see Section \ref{subsection:training}) quoted in the figure, are remarkably low. The fraction of catastrophic outliers, $\eta$, could be improved by restricting the sample to the most secure spectroscopic redshifts but we chose to keep them all as our labels are imperfect in any case. Indeed, the spectroscopic sample is essentially confined to $i<22.5$ and $z<2$ and the C2020 metrics degrade with magnitude and redshift.

\begin{figure}
\centering
\includegraphics[width=\linewidth]{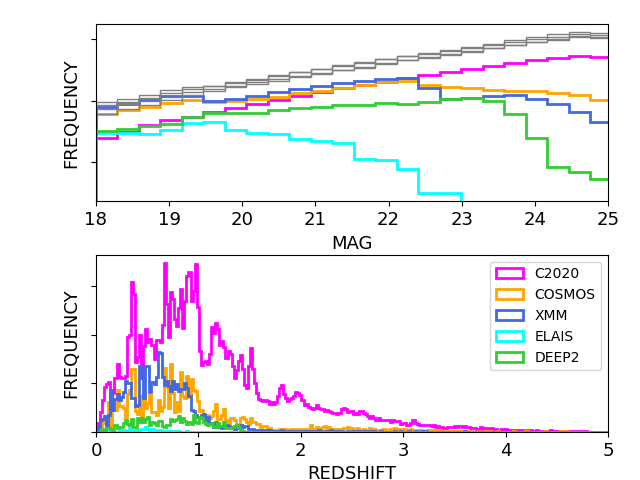}
\caption{{\bf Top:} the $i$-band magnitude distributions of the photometric (gray), C2020 (pink) and spectroscopic samples per region in log scale; {\bf Bottom:} the C2020 and spectroscopic redshift distributions per region at $i<25$.}
\label{fig:hscfields_magz}
\end{figure}

The $i$-band magnitude and redshift distributions of the spectroscopic samples per region and of the C2020 sample are displayed in Fig. \ref{fig:hscfields_magz}. The C2020 sample is complete to $i=25$, as shown in the top panel where its magnitude distribution (in pink) can be compared to those of the complete photometric samples in the 4 regions (gray lines). 

Finally, deep observations in the narrow band filter NB921, picking out strong [OII] emitters at $z\sim 1.47$ \citep[]{Hayashi2020}, complete the spectroscopic data. Although they may include a fraction of other galaxy types at other redshifts, they provide an additional test population.

\begin{table}
\centering
\begin{tabular}{|l|c|c|r|}
%%%%%%%%%%%%%%%%%%%%%%%%%%%%%%%%%
\hline  
\multicolumn{4}{|c|}{Spectroscopy } \\
\hline 
Survey               & Res.       &  z-range           & Selection  \\%[0.1cm] 
\hline 
SDSS DR12$^{(1)}$    &  2000      &  $z \le0.4$        & $r\le 17.8$     \\
SDSS-BOSS$^{(2)}$    &  2000      &  $0.3\le z \le0.7$ & LRGs            \\  
GAMA$^{(3)}$         &  1300      &  $z \le0.7$        & $r\le 19.8$     \\  
WIGGLEZ$^{(4)}$      &  1300      &  $z \le1.2$        & $NUV\le 22.8$   \\  
zCOSMOS$^{(5)}$      &  650       &  $z \le 1.2-5$     & $r\le 22.5-25$  \\  
VANDELS$^{(6)}$      &  650       &  $1\le z \le6$     & $H\le 25$       \\  
UDSz$^{(7)}$         &  650       &  $z\le4$           & $K\le 23$       \\  
DEEP2$^{8)}$         &  6000      &  $0.7\le z\le1.5$  & $r\le 24$       \\  
VVDS$^{(9)}$         &  230       &  $z\le 1.2-6$      & $i\le 22.5-24$  \\  
VIPERS$^{(10)}$      &  230       &  $0.4\le z\le1.5$  & $i\le 22.5$     \\  
VUDS$^{(11)}$        &  230       &  $2\le z\le6$      & $K\le 23$       \\  
CLAMATO$^{(12)}$     &  1100      &  $2\le z\le3.5$    & LBGs            \\  
C3R2$^{(13)}$        &  1100      &  $z\le 4$          & SOM             \\
COSMOS$^{(14)}$      &  multiple  &  $z\le 4$          & multiple        \\%[0.2cm] 
3DHST$^{(15)}$       &  130       &  $z\le4$           & $H\le24$        \\
PRIMUS$^{(16)}$      &  40        &  $z\le0.9$         & $i\le22.5$      \\
COSMOS20$^{(17)}$    &  photo-z   &  $z\le6$           & $i\le26.5$      \\%[0.1cm] 
\hline 
%%%%%%%%%%%%%%%%%%%%%%%%
\end{tabular}
\caption{
  Summary of the spectroscopic surveys with their typical spectral resolution, redshift range and main target selection criteria. }
\label{table:zspec}
\end{table}

\section{The training set}
\label{sec:refsample}

Short of representative spectroscopy, the C2020 photometric redshift sample is currently the best training set at hand. To make the most of our resources, we replaced these redshifts by spectroscopic ones where available, essentially at low redshift and bright magnitudes ($\sim$ 19\% of the C2020 sample at $i<25$), and added $\sim 2700$ spectroscopic redshifts of bright COSMOS UDF galaxies without C2020 measurements. All the redshifts in this sample of $\sim 179,600$ galaxies, whether spectroscopic or C2020, were used as one-hot labels (assigned to a unique class of our classification scheme, described in Section \ref{sec:baseline}). As the COSMOS field was mapped with both the $u$-band and $u^\star$-band filters, whereas the other fields were mapped with only one of them, we created 2 sets of training data cubes: one containing $ugrizy$ images and one containing $u^\star grizy$ images ($\sim 179,600\times 6\times64\times 64 $ pixels each). 

\begin{figure}
\centering
\includegraphics[width=\linewidth]{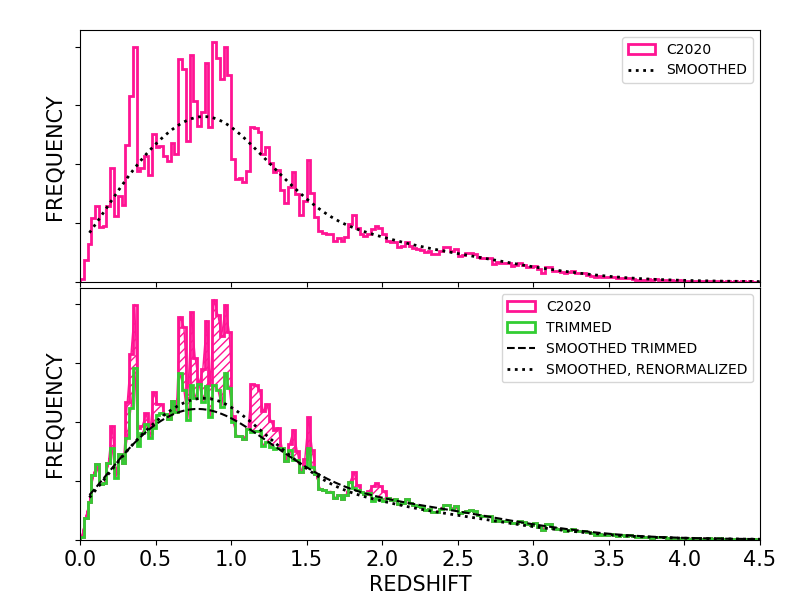}
\caption {{\bf Top panel:} simple smoothing of the training redshift distribution using SciPy's KDE function. {\bf Bottom panel:} galaxies from the highest density regions in each of the large spikes protruding above the smoothed distribution, are removed from the initial sample, resulting in the green histogram, which is then smoothed using KDE (dashed line). The pink hashed spikes show the number and redshift location of the discarded galaxies ($\sim 14\%$ in total at $i<25$).}
\label{fig:smoothing}
\end{figure}

As a narrow pencil beam survey, the COSMOS UDF intercepts several structures resulting in large redshift spikes clearly visible in the bottom panel of Fig. \ref{fig:hscfields_magz} (the pink histogram). These redshift over-densities tend to bias the photometric redshift estimates of trained algorithms. They are also likely to encompass dense galaxy clusters containing an excess of massive red galaxies relative to the general population, potentially inducing an additional bias by increasing the probability of classifying such galaxies in other fields at these particular redshifts. We therefore chose to selectively smooth the data as follows. We first smoothed the original training redshift distribution in several bins of magnitude using the kernel density estimation (KDE) function from the SciPy library \citep{2020SciPy-NMeth}. This is shown as the dotted line in the upper panel of Fig.~\ref{fig:smoothing} for the full sample at $i<25$. In the redshift intervals protruding above the smoothed distribution, we randomly removed galaxies from the highest-density regions identified in RA-Dec space, and then smoothed the trimmed distribution using KDE. The result is illustrated in the lower panel of Fig.~\ref{fig:smoothing}. The pink hashed spikes show the number and redshift location of the galaxies in spatial over-densities that were excluded from the original distribution ($\sim 14\%$ in total at $i<25$). The dashed line shows the final smoothed distribution, while the dotted line shows the same smoothed distribution as in the upper panel, renormalized to match the number of galaxies in the final one: $\sim 155,000$. The difference is mild, as is the difference between the redshifts resulting from the first and second smoothed training distributions. However it is systematically in favor of the second step when appraisable. In the following, the training set refers to this trimmed and smoothed dataset of $\sim$155,000 galaxies from the COSMOS UDF, with C2020 redshifts for the most part, or spectroscopic redshifts. We expect this training set to be representative of the redshift distribution of galaxies at $i<25$ in all fields. 
As the process involves random draws from the full sample, it varies slightly at the start of each training epoch.

\section{The baseline model}
\label{sec:baseline}

\subsection{Baseline network}

\begin{figure}
\centering
\includegraphics[width=\linewidth]{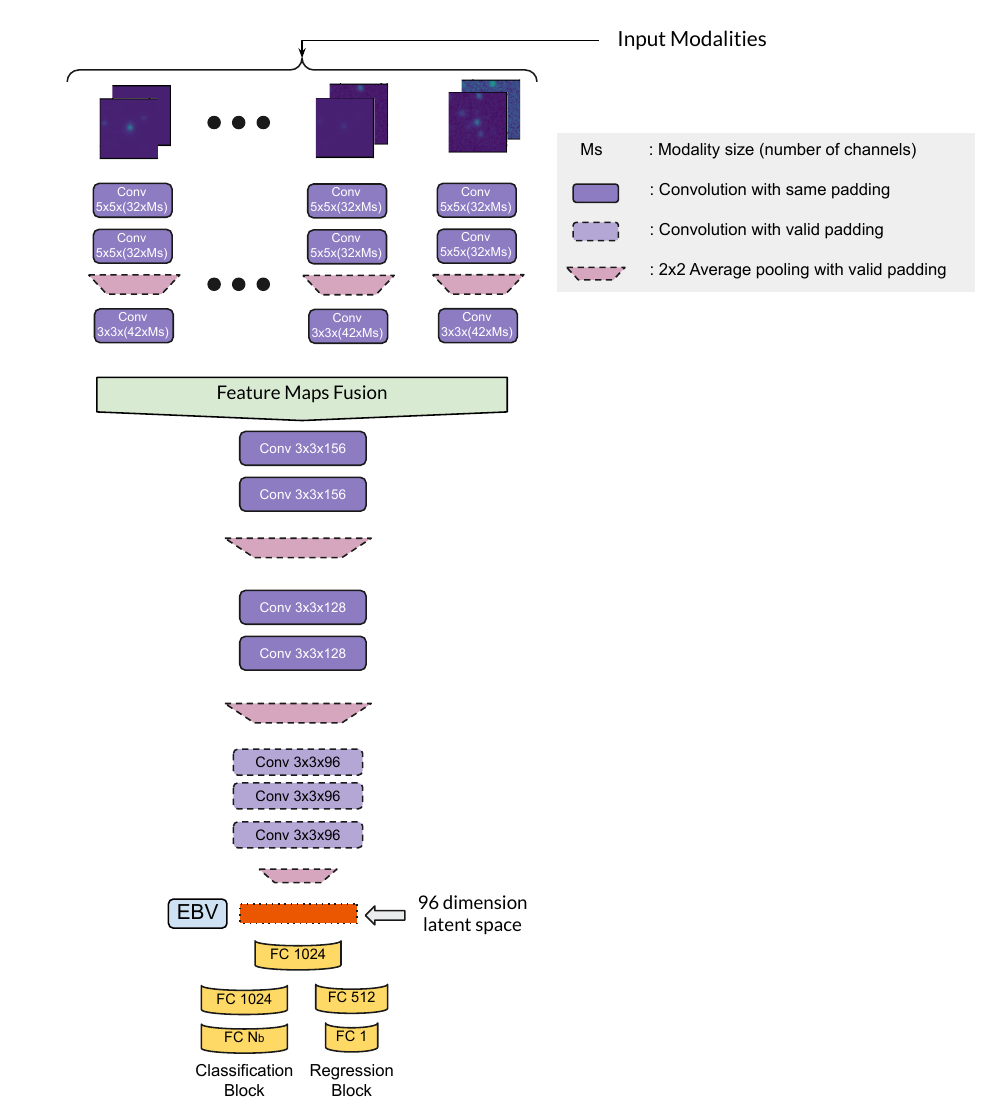}
\caption {The baseline model \citep{aitouahmed2024}: several modalities first process subsets of the input images consisting of two adjacent bands (5 in our case). Their respective outputs are then merged and fed into a common series of convolutions.}
\label{fig:10conv_archi}
\end{figure}

Our baseline model, shown in Fig. \ref{fig:10conv_archi}, is the multi-modal CNN introduced by \citet{aitouahmed2024}, which improved on P19. It features 5 multi-modalities upstream, each processing pairs of input images from two adjacent photometric bands. Their respective features maps are merged and fed into the main network. These multi-modalities added to the main CNN were shown to improve redshift estimation. The network is trained as a classifier into 360 contiguous redshift bins between $z=0$ and 6. We consider the normalized output of the classifier to be a PDF and use its median value as optimal redshift point estimate \citep{treyer2024}.

\subsection{Training}
\label{subsection:training}

\begin{figure}
\includegraphics[width=\linewidth]{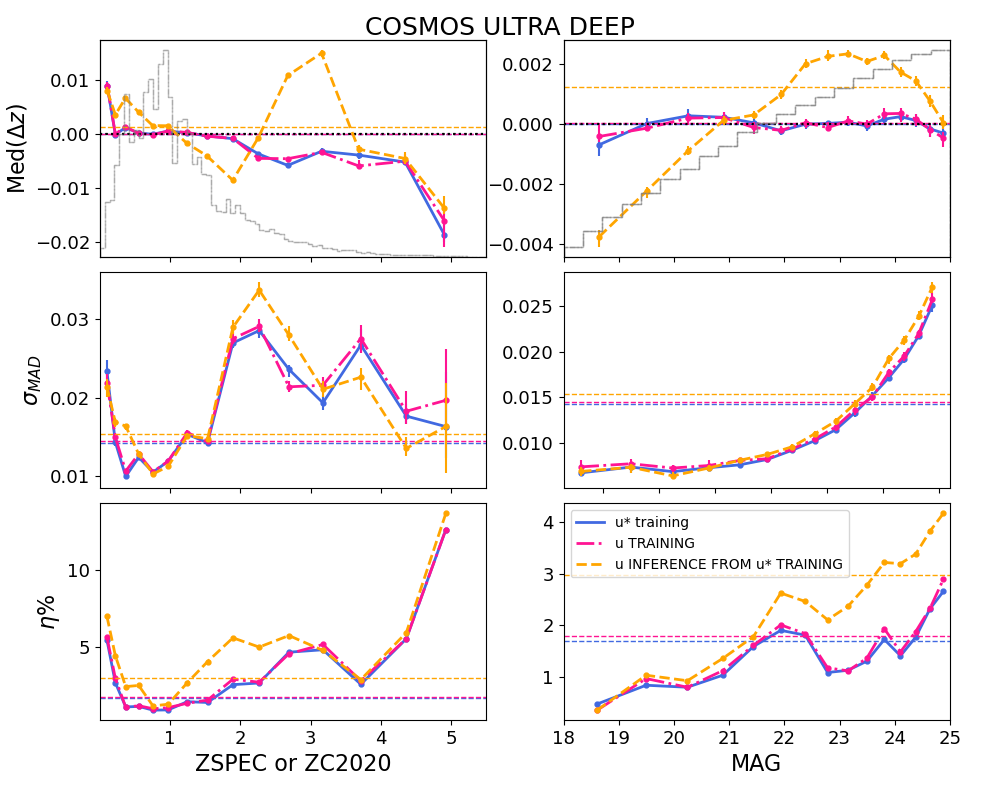}
\caption{The CNN metrics as a function of redshift label and $i$-band magnitude for the $u^\star grizy$ and $ugrizy$ validation sets, and for the $u$ dataset inferred from the $u^\star$-trained model. Using either band for training makes no difference but inferring one from the other significantly degrades the performance. In this and all similar figures, the redshift and magnitude distributions are overlaid in gray in the corresponding upper panels, and the 1$\sigma$ error bars are based on 100 bootstrap samples.}
\label{fig:deep_cosmos_cross}
\end{figure}

 The following metrics are used to quantify the performance of redshift predictions ($z_{\rm CNN}$, $z_{\rm C2020}$, $z_{\rm PHOT}$):

\begin{itemize}
\item the \textbf{normalized residuals} $\Delta z = (z_{\rm PRED}-z_{\rm TRUE})/(1+z_{\rm TRUE})$
\item the \textbf{prediction bias} $b=$ Med($\Delta z $) (median of the residuals)
\item the \textbf{deviation} $\sigma_{\rm MAD}=1.4826\times {\rm MAD}$, where MAD (Median Absolute Deviation) = Med($|\Delta z - \textrm{Med}(\Delta z)|$) 
\item the \textbf{fraction $\eta$  of outliers} with $|\Delta z|$>0.15
\end{itemize}
where $z_{\rm PRED}$ and $z_{\rm TRUE}$ refer to the predicted and "true" redshifts (the labels), respectively.  
We first trained the network on both training datasets ($ugrizy$ and $u^\star grizy$) with a cross-validation protocol (80\% training and 20\% validation). The PDF predictions of each validation fold were averaged over 6 runs. Figure \ref{fig:deep_cosmos_cross} shows the prediction bias, MAD and outlier fraction as a function of redshift label (spectroscopic or C2020) and $i$-band magnitude for the $u^\star grizy$ model, the $ugrizy$ model, and the $u^\star grizy$-trained model applied to the $ugrizy$ data. Using either set of bands makes no difference but inferring one from the other significantly degrades the performance. Therefore we will use whichever band is suited for each of the other fields ($u^\star$ for XMM and COSMOS DEEP, $u$ for ELAIS and DEEP2). As expected from the declining quality of the images (P19), the metrics significantly degrade with magnitude, hence our choice to limit the present analysis to $i=25$. They also do at $z<0.5$, probably due to the paucity of low redshift galaxies in the training set and to the C2020 outliers seen in Fig. \ref{fig:zc2020_zspec}.

\subsection{Inference}
\label{subsec:inference}

\begin{figure}
\centering
\includegraphics[width=\linewidth]{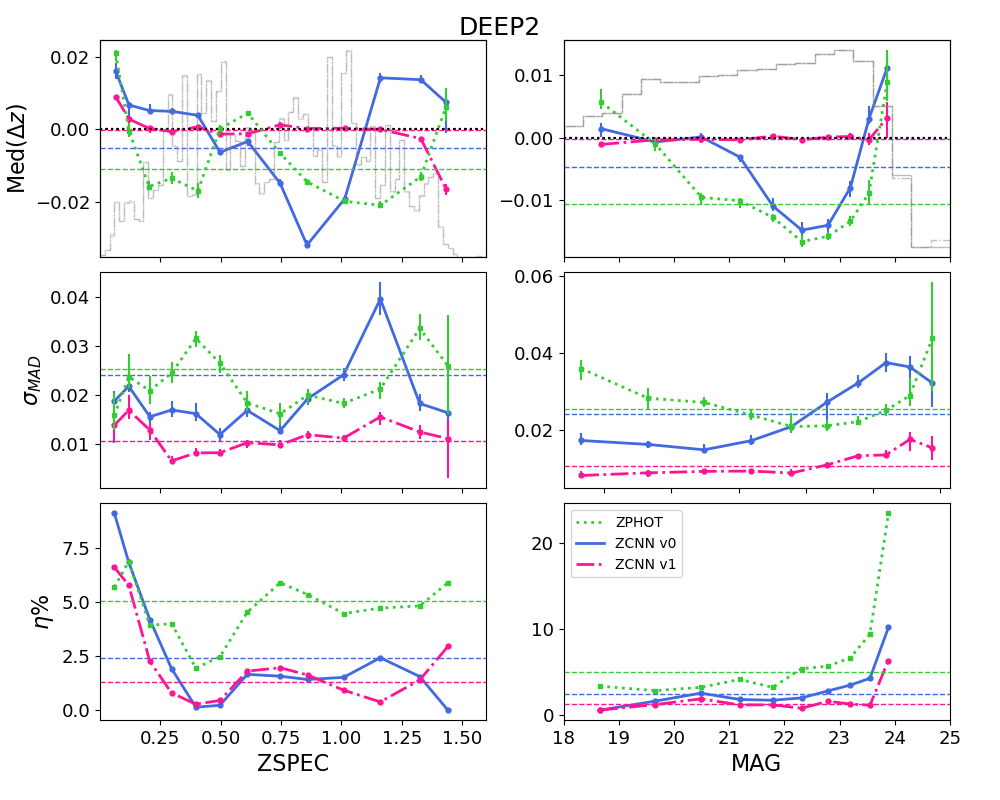}
\caption{The v0 and v1 CNN metrics (blue and pink lines, respectively. See Section \ref{subsec:inference}) for the DEEP2 spectroscopic sample as a function of redshift and magnitude, compared to the $ugrizy$ photometric redshifts (green lines).}
\label{fig:BL_cross_metrics}
\end{figure}

\begin{figure}
\centering
\includegraphics[width=\linewidth]{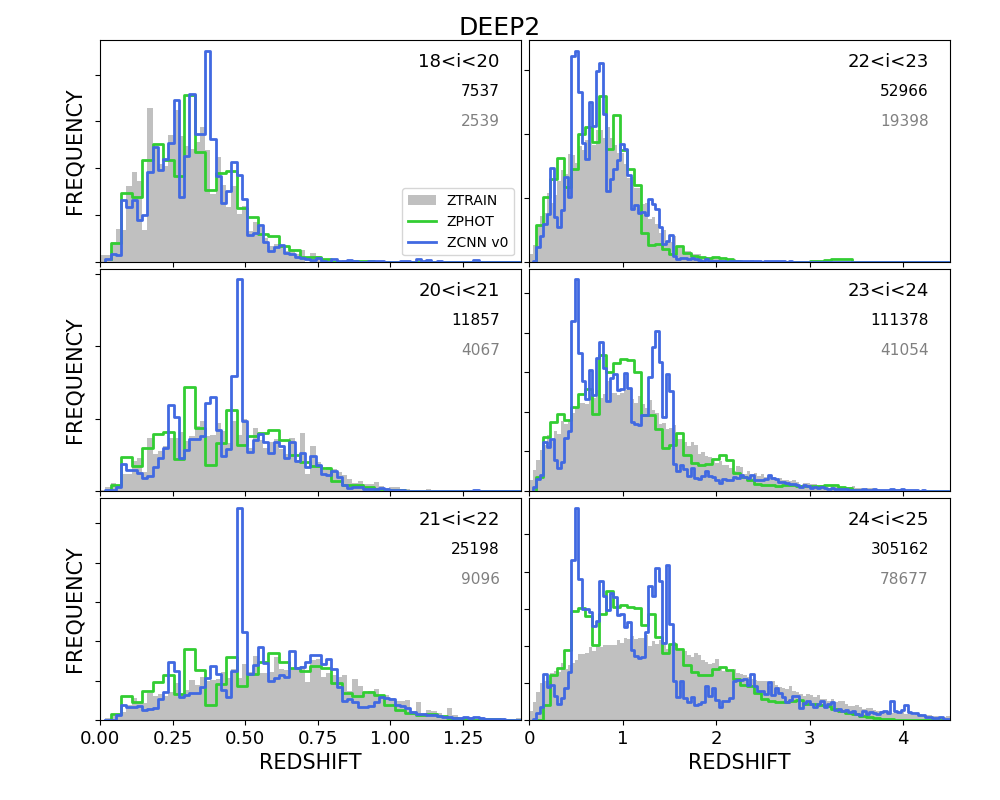}
\includegraphics[width=\linewidth]{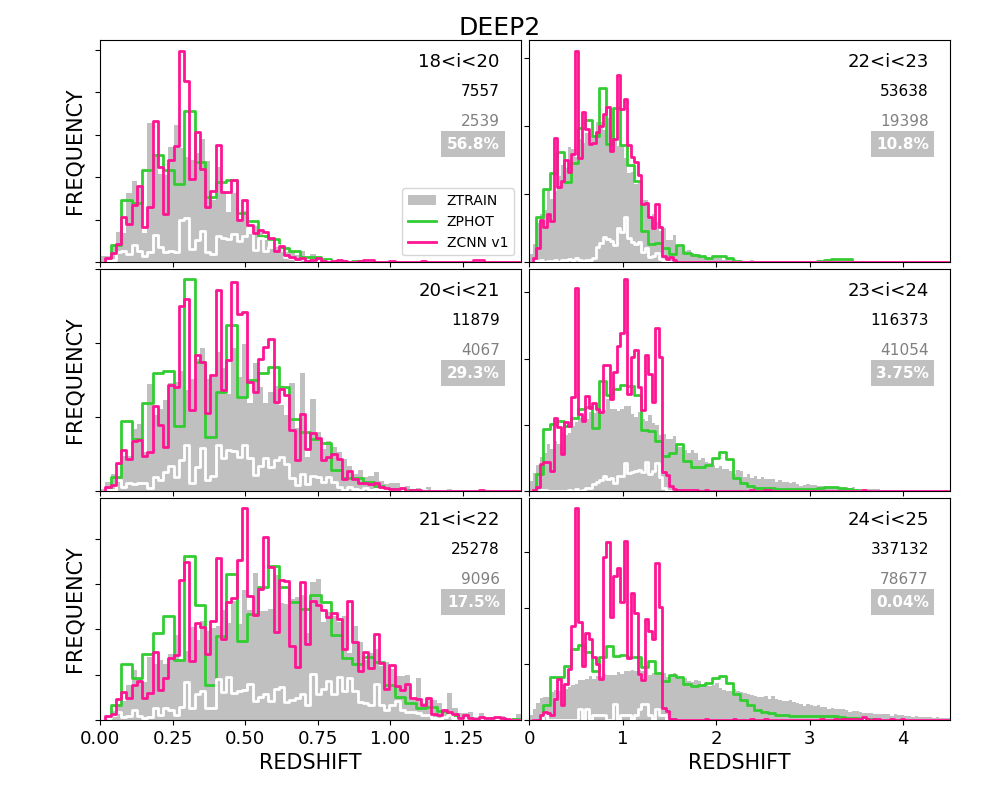}
\caption{The redshift distributions of the complete DEEP2 data inferred from the baseline CNN trained on the COSMOS UDF sample alone (v0, blue histograms) and on the COSMOS UDF data supplemented by the DEEP2 spectroscopic data (v1, pink histograms). The distributions are compared to the $ugrizy$ SED-fitting redshift distributions (green histograms) and to the COSMOS UDF training distributions (shaded gray histograms). All are normalized to 1. The white histograms are the redshift distributions of the DEEP2 spectroscopic sources included in the training set, arbitrarily scaled for visibility. The number of DEEP2 galaxies, of COSMOS UDF training galaxies and the fraction of DEEP2 training galaxies with respect to the latter are quoted in black, gray and white on gray, respectively, in each magnitude range.}
\label{fig:BL_cross_inference}
\end{figure}

From the above baseline model, we inferred baseline redshifts, referred to as v0 redshifts, for the spectroscopic (labeled) population ($\sim 11600$ galaxies) and for the photometric (unlabeled) population  ($\sim 714 000$ galaxies) in the DEEP2 field, as an illustration. In a second experiment, we enriched the COSMOS UDF training sample with the spectroscopic galaxies from the DEEP2 field and retrained the network with a cross-validation protocol (20\% validation, 80\% training, 6 ensembles). We refer to this model as v1. 

Figure \ref{fig:BL_cross_metrics} shows the v0 and v1 CNN metrics for the DEEP2 spectroscopic sample as a function of redshift and magnitude, compared to the $ugrizy$ SED-fitting redshifts. Except for the fraction of outliers, the v0 model is not a significant improvement on those. The deviation is even worse at $22<i<24$. However the gain is unambiguous for all three metrics with the v1 model.

Figure \ref{fig:BL_cross_inference} shows the redshift distributions of the complete, mostly unlabeled, DEEP2 sample in 6 intervals of magnitude inferred from the v0 and v1 models (top and bottom panels respectively). The distributions are normalized and compared the SED-fitting redshift distributions and to the training distributions, which are assumed to be representative and whose numbers of galaxies per magnitude bin are quoted in gray. All the predicted distributions are roughly consistent with the training distributions at $i<22$. The agreement degrades at fainter magnitude where the distributions from both CNN models become significantly, but differently, distorted. 

The v1 redshift distributions are more inadequate than the v0 predictions. The network has classified all DEEP2 sources within the redshift range of the few DEEP2 spectroscopic sources included in the training set (white histograms), however small their contribution (quoted in white on gray). This suggests that it has primarily learned to distinguish between DEEP2 and COSMOS images and that it classifies the redshifts of DEEP2 sources based on DEEP2 images and labels present in the training set rather than on redshift-related patterns that could be learned from the much larger number of COSMOS images. The improvement seen in Fig. \ref{fig:BL_cross_metrics} is therefore very misleading: the network has learnt to estimate the redshift of DEEP2 spectroscopic sources from other DEEP2 spectroscopic sources drawn from the exact same distribution, and to ignore COSMOS sources as irrelevant to the task. 

DEEP2 illustrates what is known as the "domain adaptation problem", the inability of the network to use information from one field to predict adequate results in another. This problem occurs, more or less, in all the other HSC regions (Section \ref{sec:otherfields}).

\section{The adversarial model}
\label{sec:adversarial}

The COSMOS UDF differs from the other fields in a number of ways: S/N (Fig. \ref{fig:hscfields}), Galactic dust extinction (Fig. \ref{fig:EBV_DIS}), seeing, etc. Whatever recipe the network learns from the COSMOS UDF images to estimate redshifts fails to work adequately with images from other sky regions.
To try and blind the CNN to these differences, we developed the adversarial approach described below.

\subsection{Architecture and training procedure}

\begin{figure}
\centering
\includegraphics[width=\linewidth]{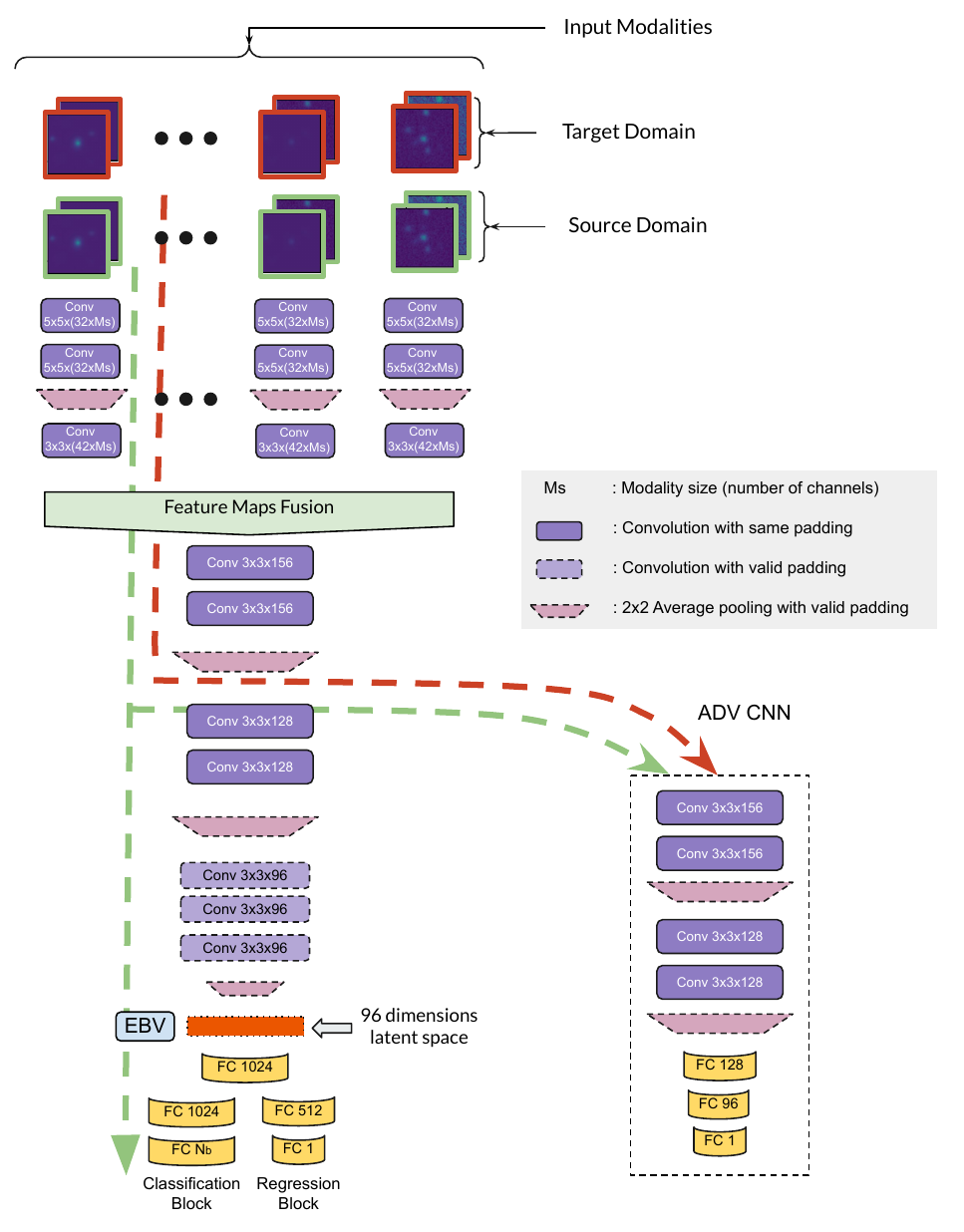}
\caption {The domain adaptation architecture with the added adversarial module.}
\label{fig:adv_archi}
\end{figure}

Figure~\ref{fig:adv_archi} shows how we integrated an adversarial module into the CNN architecture depicted in Fig. \ref{fig:10conv_archi}. This module is composed of four convolutional layers: two with 156 kernels and two with 128 kernels, all of size $3 \times 3$. Every pair of convolutional layers is followed by a pooling layer. The module ends with fully connected layers generating the adversarial classification. 

The network is trained with the source domain (the COSMOS UDF training set) and a target domain (100,000 sources randomly drawn from the photometric datacube of a field other than the COSMOS UDF. No labels are required in the target domain). At each iteration, a mini-batch is sampled from each domain. The source domain batch is used both for the redshift classification training and for the adversarial objective, while the target batch only serves the adversarial objective. 

The adversarial module receives feature maps from a layer of the main network (the 5th in our setup). Its objective is to identify their field of origin - source or target. In a min-max training paradigm, the main network is trained, besides estimating redshift, to produce feature maps that will fool the adversarial network.

The adversarial module is trained with the cross-entropy loss function (Equation~\ref{eq:advloss}), which quantifies the disparity between its predictions $p_{i}$ and the domain labels $d_{i}$ defined as 0 for source and 1 for target:
\begin{equation}
L =  - \sum_{i=1}^{2}  d_{i}\log(p_{i}) 
\label{eq:advloss}
\end{equation}

This loss function updates only the adversarial layers. Following the min-max paradigm, the main layers of the network, which supply the feature map input to the adversarial module, are trained with the inverse of this loss, known as the confusion loss. It is computed by just inverting the domain labels $d_{i}$. Thus, these layers aim to generate feature maps that maximize the adversarial error while enabling accurate redshift predictions. The adversarial and redshift estimation objectives are optimized simultaneously at each iteration, using the same learning rate and the same cross-entropy loss function.

As this network carries a risk of negative domain transfer (by producing indistinguishable feature maps for objects that should be distinguishable, like a faint source domain galaxy and a bright target domain galaxy), we tried two complementary approaches: 

\begin{enumerate}
    \item the selection of the target batch based on magnitude:
    negative domain transfer can be avoided by using pseudo labels guiding the adversarial training to align similar objects in the two domains \citep{pei2018multi}. We chose the $i$-band magnitude, which may help disentangle the imaging conditions from the physical properties that are crucial for redshift estimation. In practice, for each galaxy in the source batch, we randomly select one galaxy from the target domain among the 30 sources that are closest in $i$-band magnitude.
  
    \item pairing the selections: to make the magnitude pairing idea more efficient, we modified the domain adaptation learning objective. Instead of having the adversarial network determine whether feature maps originate from the source or target domain, it must decide if they come from the same field or not. To do so, we concatenate the feature maps along the width axis before feeding them to the adversarial network. The adversarial network produces one estimate per pair of images. We ensure that 50\% of the pairs are from the same field (equally split between target and source) and the remaining 50\% are pairs from different domains.
\end{enumerate}

We found no significant difference with and without these features in the results presented below.

\subsection{Results}
\label{subsec:adv_results}

\begin{figure}
\centering
\includegraphics[width=\linewidth]{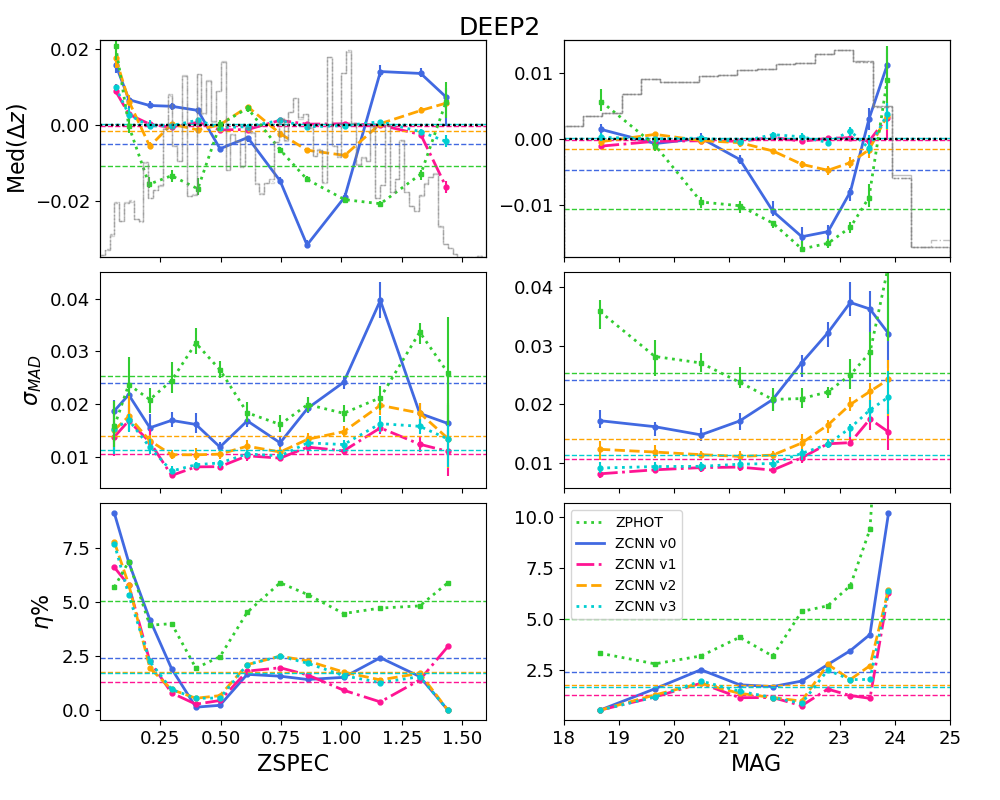}
\caption{The performance of the adversarial model trained with the COSMOS UDF sample alone (v2, orange lines) or with the COSMOS UDF sample supplemented by the DEEP2 spectroscopic sample in cross-validation mode (v3, turquoise lines) for the DEEP2 spectroscopic sample as a function of redshift and magnitude. The dark blue and pink lines are the baseline results (v0 and v1) shown in Fig.\ref{fig:BL_cross_metrics}.  The $ugrizy$ SED-fitting metrics are shown in green.}
\label{fig:ADV_metrics}
\end{figure}

\begin{figure}
\centering
\includegraphics[width=\linewidth]{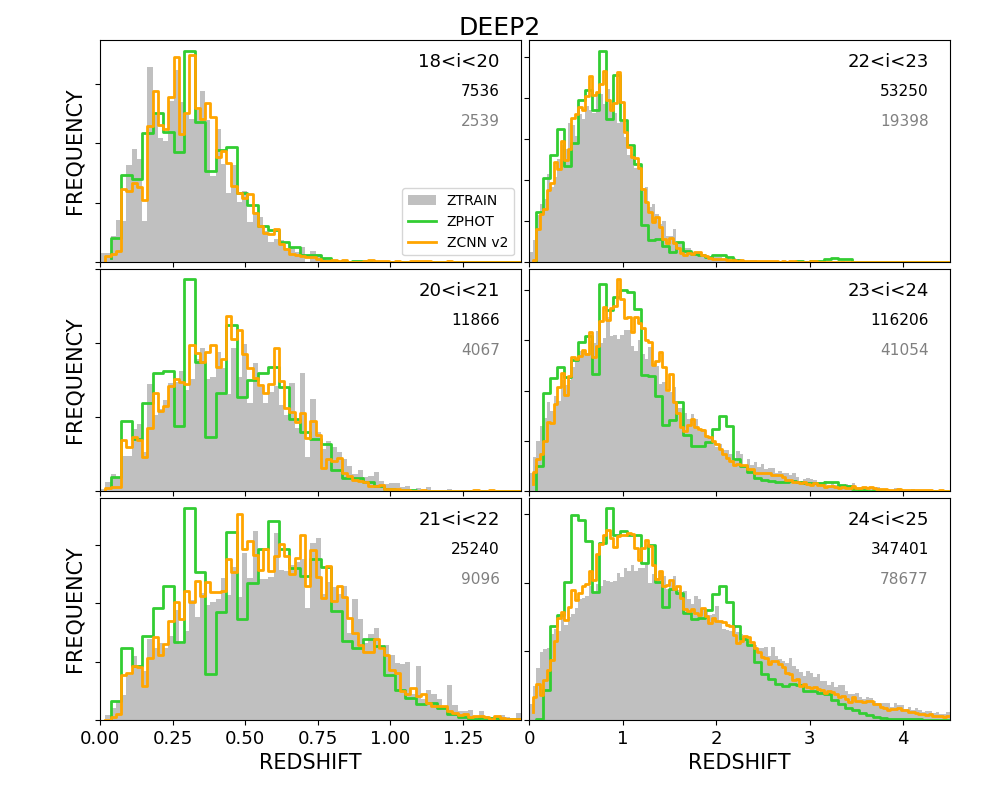}
\includegraphics[width=\linewidth]{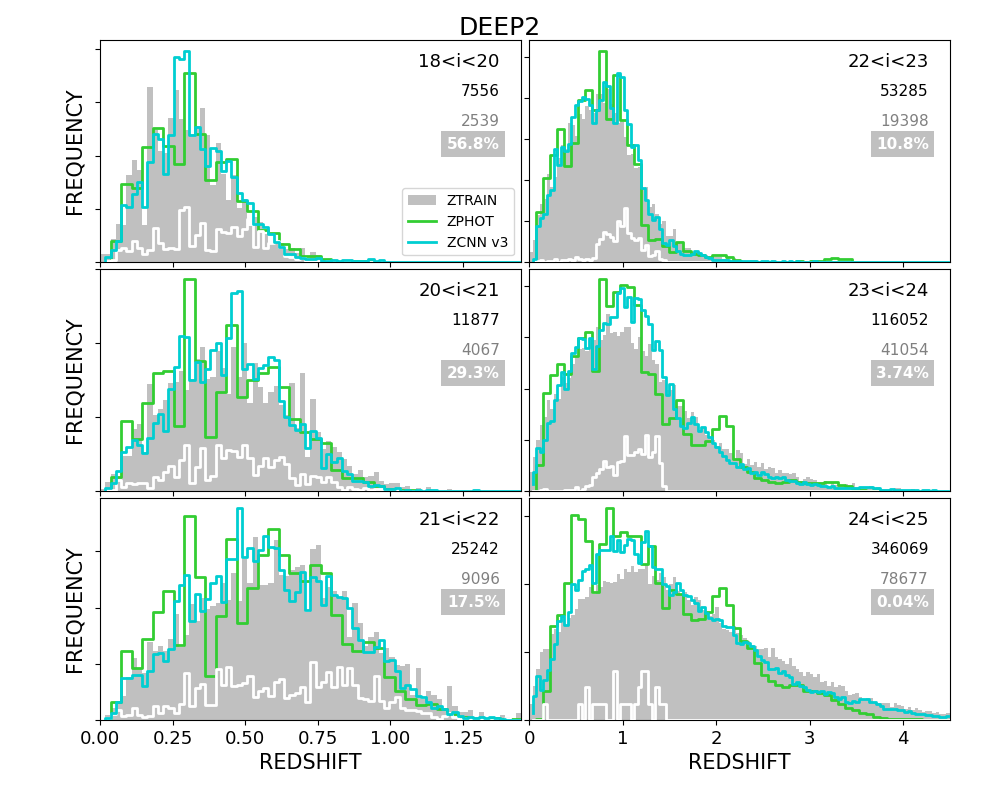 }
\caption{The predicted redshift distributions of the full DEEP2 data inferred from the adversarial network trained on the COSMOS UDF sample alone (v2, orange histograms in the top panels) and on the COSMOS UDF data supplemented by the DEEP2 spectroscopic data (v3, turquoise histograms in the bottom panels). See Fig. \ref{fig:BL_cross_inference} for the full caption.
}
\label{fig:ADV_inference}
\end{figure}

We trained the adversarial CNN with the COSMOS UDF training sample as source data and 100,000 unlabeled DEEP2 sources as target data. The PDFs inferred on DEEP2 are the average of 10 trainings. We refer to this model as v2. We also trained the adversarial model using the COSMOS UDF sample supplemented by the spectroscopic DEEP2 data as training set with a cross-validation protocol. We refer to this last experiment as v3.

Figure \ref{fig:ADV_metrics} shows the v2 and v3 metrics as a function of redshift and magnitude for the DEEP2 spectroscopic sample, compared to the baseline results (v0 and v1) shown in Fig.\ref{fig:BL_cross_metrics} and to the SED-fitting metrics. The v2 versus v0 comparison, the two models trained on COSMOS UDF alone, demonstrates the significant impact of the adversarial module. Adding the DEEP2 spectroscopic data to the training set brings the v3 metrics nearly down to the v1 level. 

Contrary to the baseline results in Fig. \ref{fig:BL_cross_inference}, a dramatic improvement is also seen in the redshift distributions inferred for the unlabeled galaxies, shown in Fig. \ref{fig:ADV_inference}. These distributions are significantly more consistent with the reference sample at faint magnitude, without or with the addition of the DEEP2 spectroscopic data to the training sample. The two models are very similar but the v3 model appears to create more structures, most notably at $i<20$ and at $i>23$. This suggests that the presence of even a small fraction of unrepresentative target sources amid the training data may still induce ill-understood biases despite the adversarial module, and that the improved metrics for the spectroscopic target population remain misleading. The most conspicuous discrepancy with the underlying training distribution for both models is an excess of galaxies around $z\sim 1$, taken away from the low and high redshift tails. This may be a flaw, but it is also conceivable that we over-flattened the COSMOS UDF distribution (Fig. \ref{fig:smoothing}) and that this growth is at least partly real. This point is further discussed in Section \ref{sec:qa}.

\begin{figure}
\centering
\includegraphics[width=\linewidth]{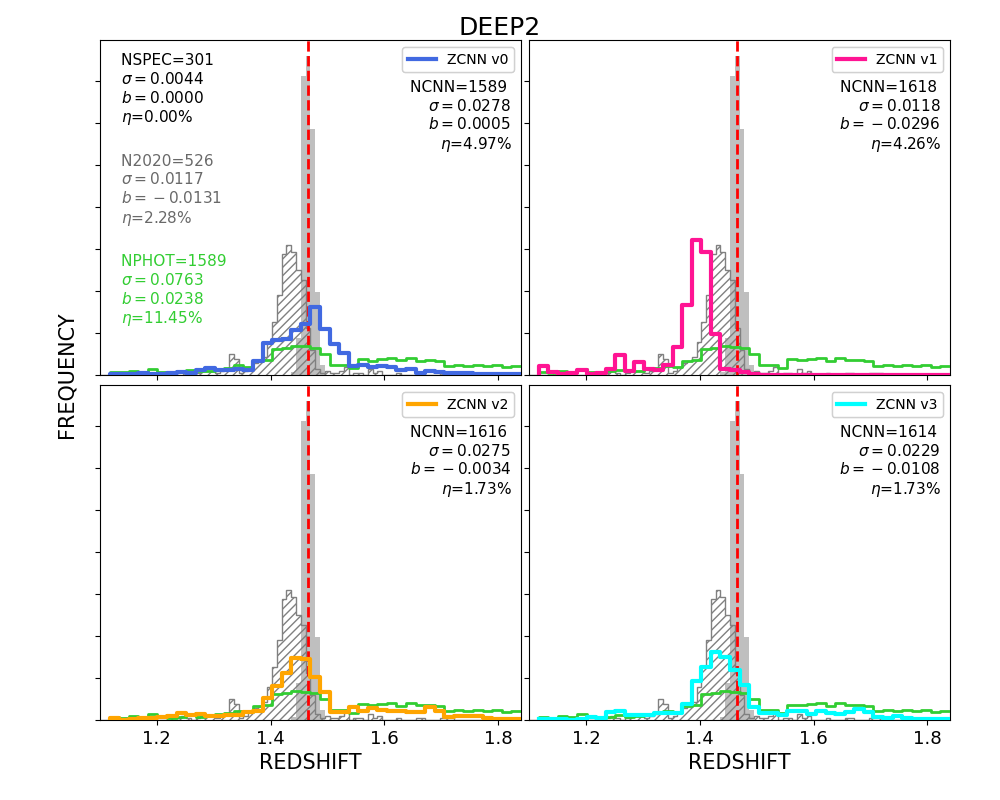}
\caption{Redshifts inferred from the v0, v1, v2, and v3 models for a sample of [OII] emission line galaxy candidates at $z \approx 1.47$ in the DEEP2 field. Their $ugrizy$ SED-fitting redshift distribution is shown in green. The 526 C2020 redshifts available for such galaxies in the COSMOS field are shown as hatched histograms and the spectroscopically confirmed cases in all the fields combined as gray-filled histograms. The corresponding metrics, in the top right corner of each panel for the CNN, and in green and gray for the $ugrizy$ and C2020 redshifts respectively, assume a "true" redshift $z_{EL}=1.4657$ (red vertical line, see text for details). 
}
\label{fig:EL_inference}
\end{figure}

A final test is afforded by a sample of $\sim$15,000 galaxies selected from narrow-band observations designed to detect [OII] emission line galaxies at redshift $z \approx 1.47$  (Section \ref{subsec:spec}). This sample spans a large magnitude range and covers COSMOS, DEEP23, ELAIS and bits of XMM. Figure \ref{fig:EL_inference} shows the redshifts predicted for the $\sim 1600$ such galaxies at $i<25$ in the DEEP2 region by the v0, v1, v2, and v3 models. The distributions are compared to the $ugrizy$ SED-fitting redshift distribution and to that of the 526 C2020 redshifts available for the emission line candidates in the COSMOS UDF. The corresponding metrics quoted in the panels assume a "true" redshift $z_{EL}$ of 1.4657. We computed $z_{EL}$ as the median redshift of the 301 most reliable spectroscopic redshifts available for these galaxies at $i<25$ in all the fields combined, assuming {\tt ZFLAG>2} and redshifts within the $x$-axis range of the plots, $1.1<z<1.84$, which corresponds to $|\Delta z|<0.15$, our definition of acceptable estimates (non catastrophic failures). We note that 5 spectroscopic redshifts with {\tt ZFLAG>2} are in this category, signaling that the narrow-band filter detections are not 100\% [OII] emission line galaxies at $z \approx 1.47$. All values of $\eta$ are, by definition, the fraction of redshift estimates that extend beyond the plots.

Interestingly, the baseline model, v0, although ruled out, does not do such a bad job. The baseline model trained with the addition of the DEEP2 spectroscopic sources, v1, also ruled out, induces a strong negative bias, confirming its inadequacy. Both versions of the adversarial model, v2 and v3, yield significantly more decent results, in the sense that they align with the C2020 histogram, fulfilling the goal of the adversarial module since C2020 redshifts dominate the training sample. The fraction of catastrophic failures are also much reduced. Like v1, v3 whose training also includes the DEEP2 spectroscopic sources, induces a stronger negative bias.

\begin{figure}[h!]
\centering
\includegraphics[width=9cm]{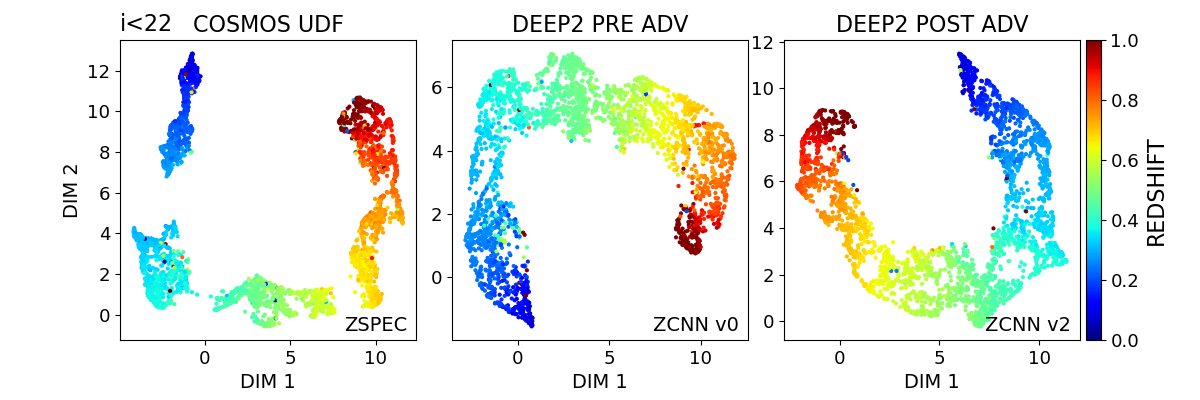}
\includegraphics[width=9cm]{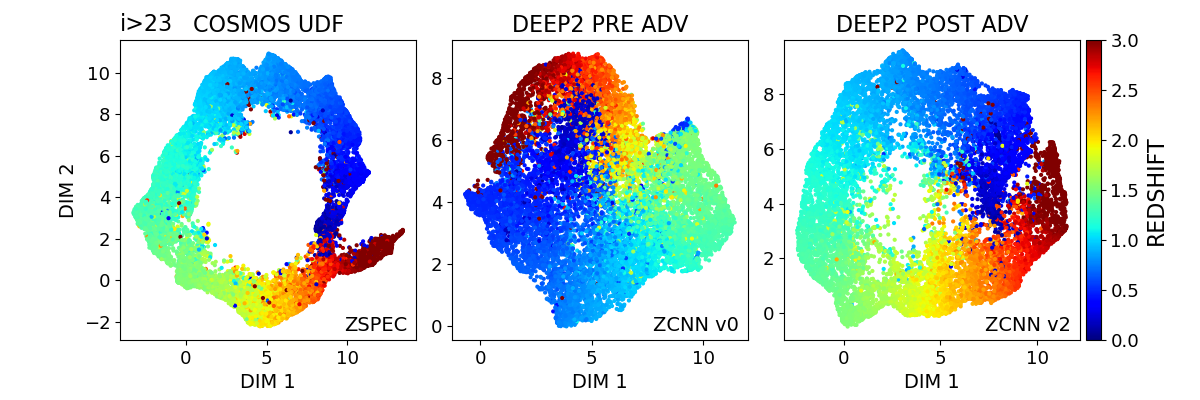}
\caption{UMAP embeddings from a deep convolutional layer, colored by redshift. The leftmost panels displays a COSMOS UDF validation sample from the baseline training, colored by redshift labels. The middle and right panels show the embeddings from the same layer for DEEP2 unlabeled galaxies, colored by redshift predictions, before and after DA (v0 vs v2), respectively. The top and bottom rows make use of galaxy samples with $i<22$ and $i>23$, respectively.}
\label{fig:umap}
\end{figure}

Given this test and the shapes of the overall redshift distributions, we settle for v2 as our preferred model.
UMAP \citep{UMAP} or t-SNE \citep{tSNE} can help us visualize the impact of DA within our network. To select a relevant layer, we used a COSMOS UDF validation sample from the trained baseline network (Section \ref{subsection:training}) to quantify the quality of the internal representations produced at each layer. For every layer, we extracted per-galaxy activation vectors, applied dimensionality reduction using UMAP and t-SNE, and computed silhouette scores with respect to redshift. As expected, the deepest layers exhibit the strongest feature separability in redshift space. The top and bottom left panels of Fig. \ref{fig:umap} show the UMAP embeddings from one of the last convolutional layers (more visually dispersed than its t-SNE equivalent) for validation galaxies with $i<22$ and $i>23$, respectively, colored by redshift (spectroscopic or C2020 labels). The four right-hand panels show UMAP embeddings from the same layer for a random subset of bright and faint unlabeled DEEP2 galaxies colored by predicted redshift, before and after DA (v0 vs. v2). The improvement in redshift encoding at faint magnitude is significant, confirming the results presented in this section. We note that the lowest and highest redshifts are not perfectly separated in the COSMOS UDF validation sample at $i>23$, a weakness that DA cannot improve upon. The fact that labels are predominantly C2020 photometric estimates at faint magnitude, with low and high redshift outliers (visible in Fig. \ref{fig:zc2020_zspec} for the bright-magnitude biased sample with spectroscopic redshifts), contributes to degrading redshift classification as magnitude increases (this is further discussed in Section \ref{sec:qa}).

\section{The other HSC regions}
\label{sec:otherfields}

In the following subsections, we present the results of our adaptation model applied to the 4 other HSC sub-regions\footnote{For each region we performed 10 training runs of 50 epochs using V100 and A100 GPUs on the Jean Zay supercomputer at IDRIS. One training run consumes $\sim$17 hours of V100 or $\sim$7.5 hours of A100, with a memory footprint of $\sim$27 GB (RES).}: ELAIS, XMM ULTRA DEEP, XMM DEEP, and COSMOS DEEP. As described in Section \ref{subsec:phot}, each of these fields has its own photometric characteristics, that more or less differ from the COSMOS UDF, making adaptation more or less critical for redshift estimation based on that particular source field. The quality of the SED-fitting photometric redshifts also differs: in XMM DEEP, XMM ULTRA DEEP, and COSMOS DEEP, they were computed with the addition of 3 near-infrared bands for 51.8\%, 24.6\% and 3\% of the sources, respectively. In the case of COSMOS DEEP, 71.6\% of the "spectroscopic" sample are C2020  redshifts. 

\begin{figure}[h]
\centering
\includegraphics[width=\linewidth]{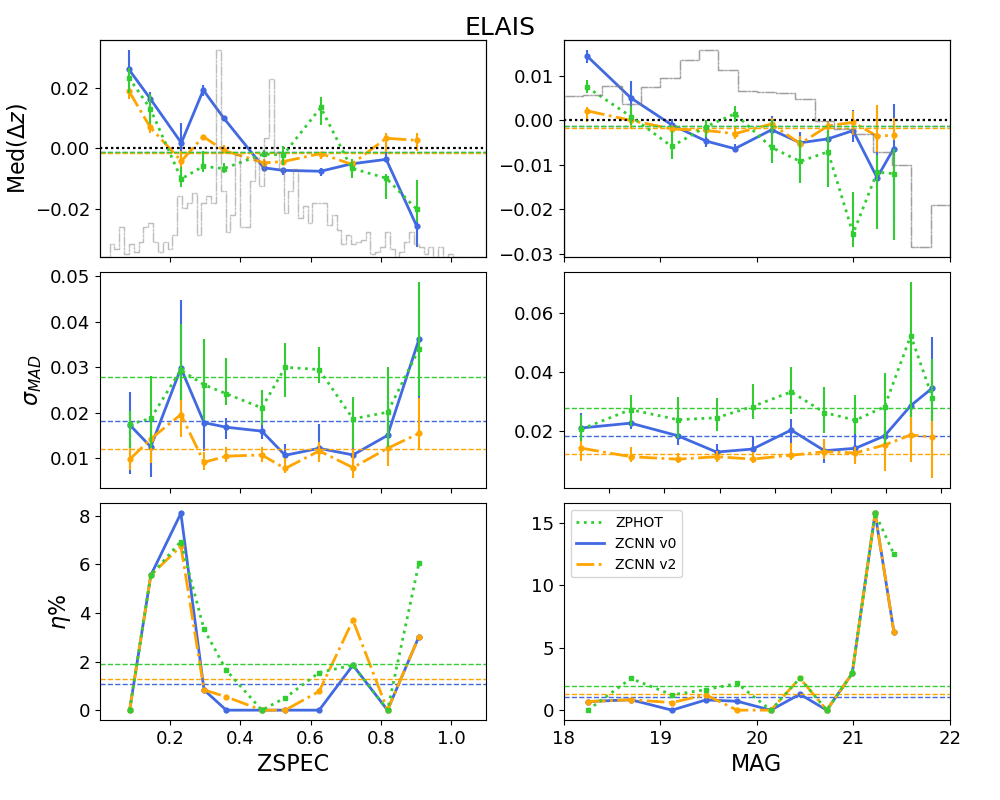}
\centering
\includegraphics[width=\linewidth]{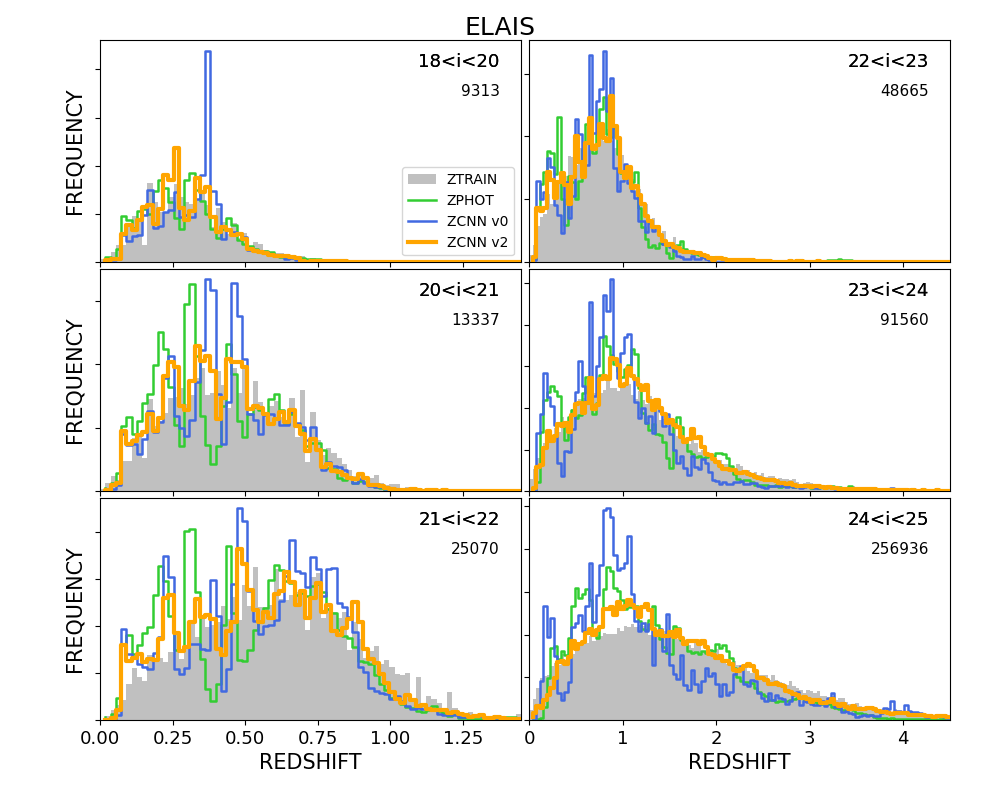}
\centering
\includegraphics[width=\linewidth]{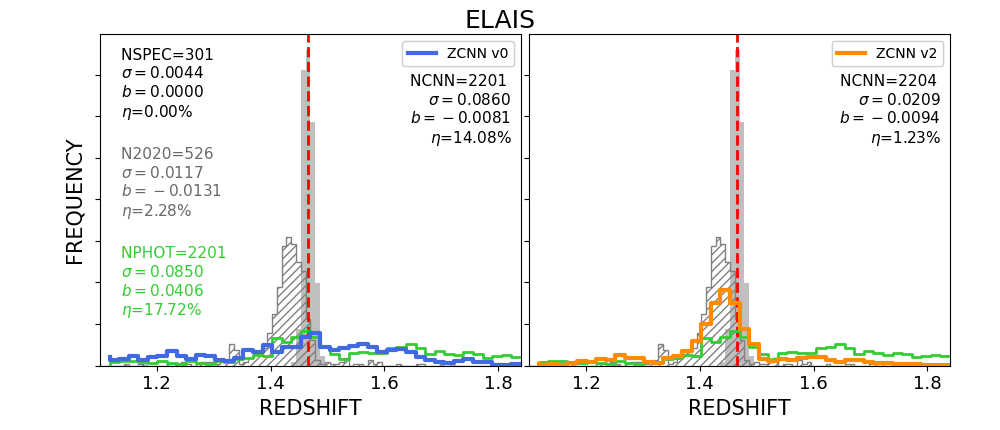}
\caption{The predictions of the baseline CNN (v0 in blue) and adversarial CNN (v2 in orange) trained with the COSMOS UDF sample for the ELAIS data, compared to the $ugrizy$ SED-fitting redshifts (in green). {\bf Top:} The metrics measured for the spectroscopic sample as a function of redshift and magnitude. {\bf Middle:} The predicted redshift distributions of the complete catalog in 6 magnitude bins. See Fig. \ref{fig:BL_cross_inference} for the complete caption. {\bf Bottom:} The redshifts inferred for the [OII] emission line candidates. See Fig. \ref{fig:EL_inference} for the complete caption. }
\label{fig:shallow_elais}
\end{figure}

\begin{figure}
\centering
\includegraphics[width=\linewidth]{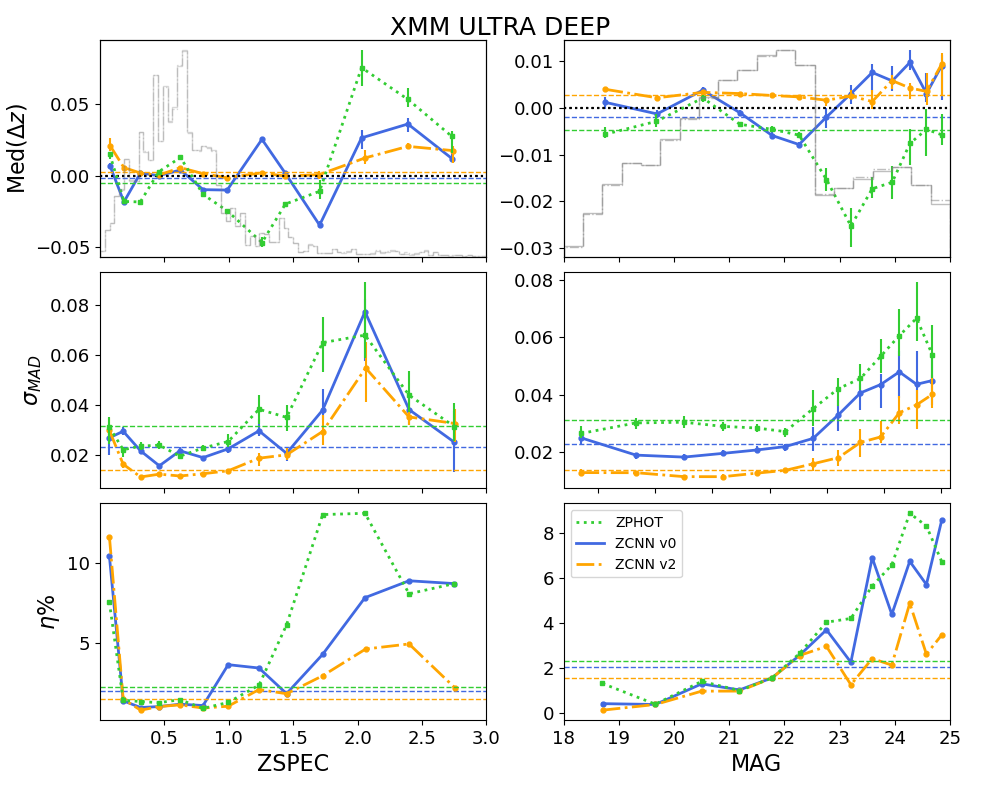}
\includegraphics[width=\linewidth]{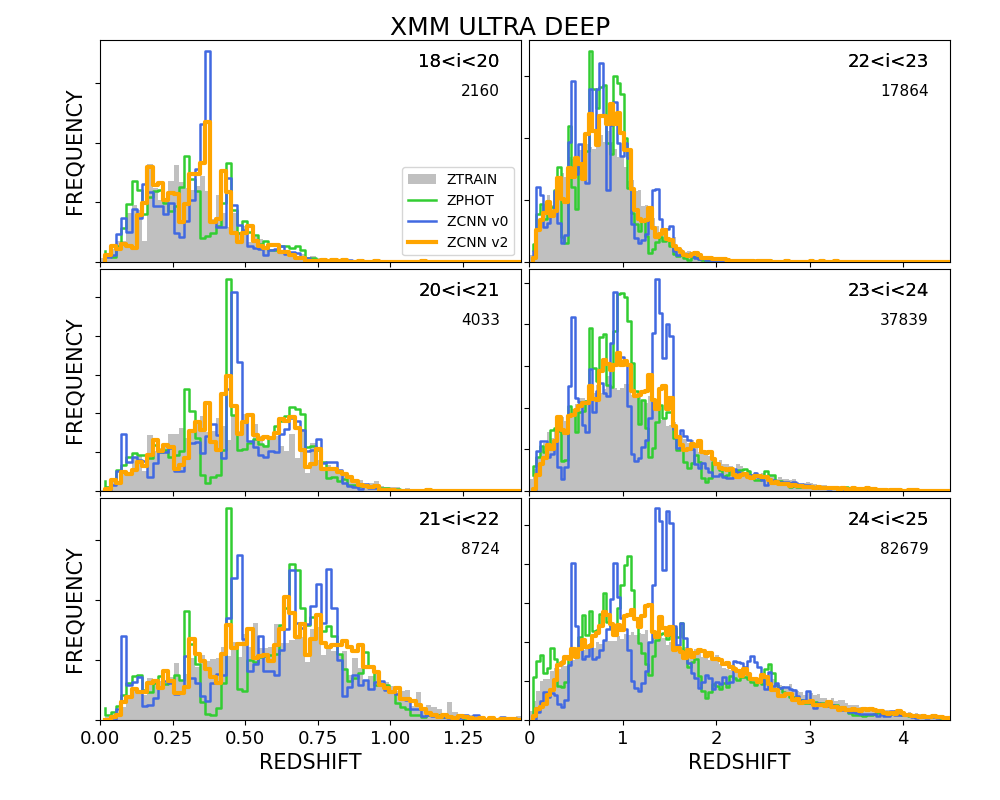}
\includegraphics[width=\linewidth]{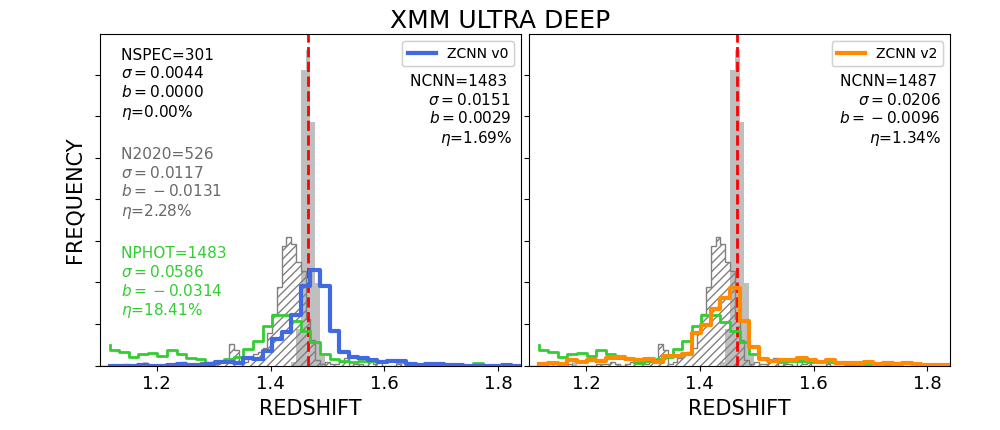}
\caption{Same as Fig. \ref{fig:shallow_elais} for the XMM UDF.}
\label{fig:deep_xmm}
\end{figure}

\begin{figure}
\centering
\includegraphics[width=\linewidth]{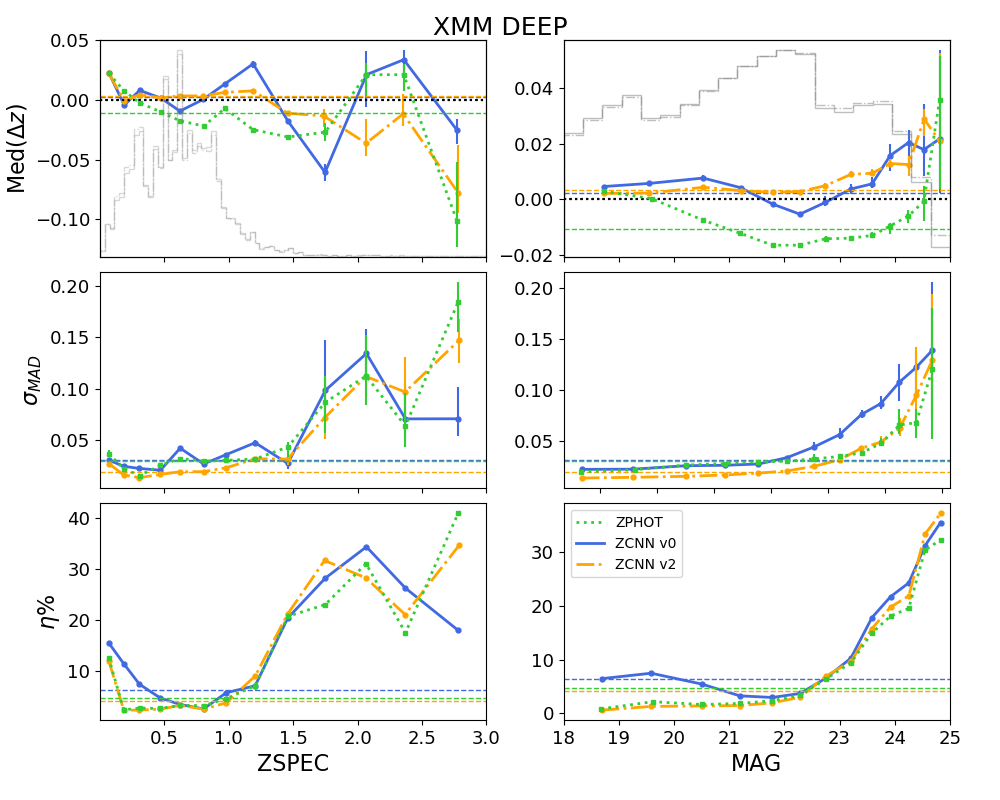}
\includegraphics[width=\linewidth]{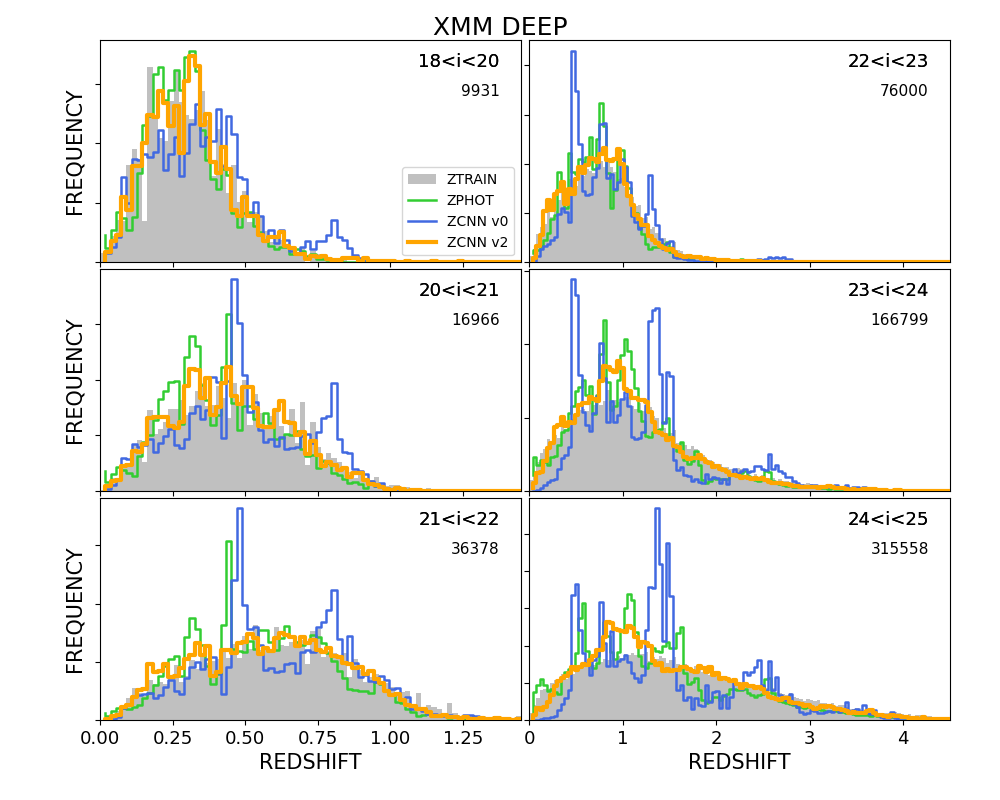}
\includegraphics[width=\linewidth]{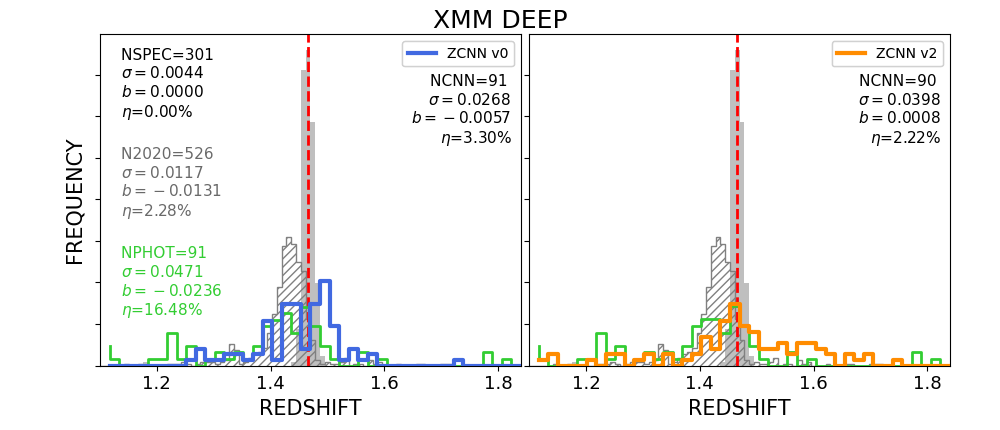}
\caption{Same as Fig. \ref{fig:shallow_elais} for the XMM DEEP field.} 
\label{fig:shallow_xmm}
\end{figure}

\begin{figure}
\centering
\includegraphics[width=\linewidth]{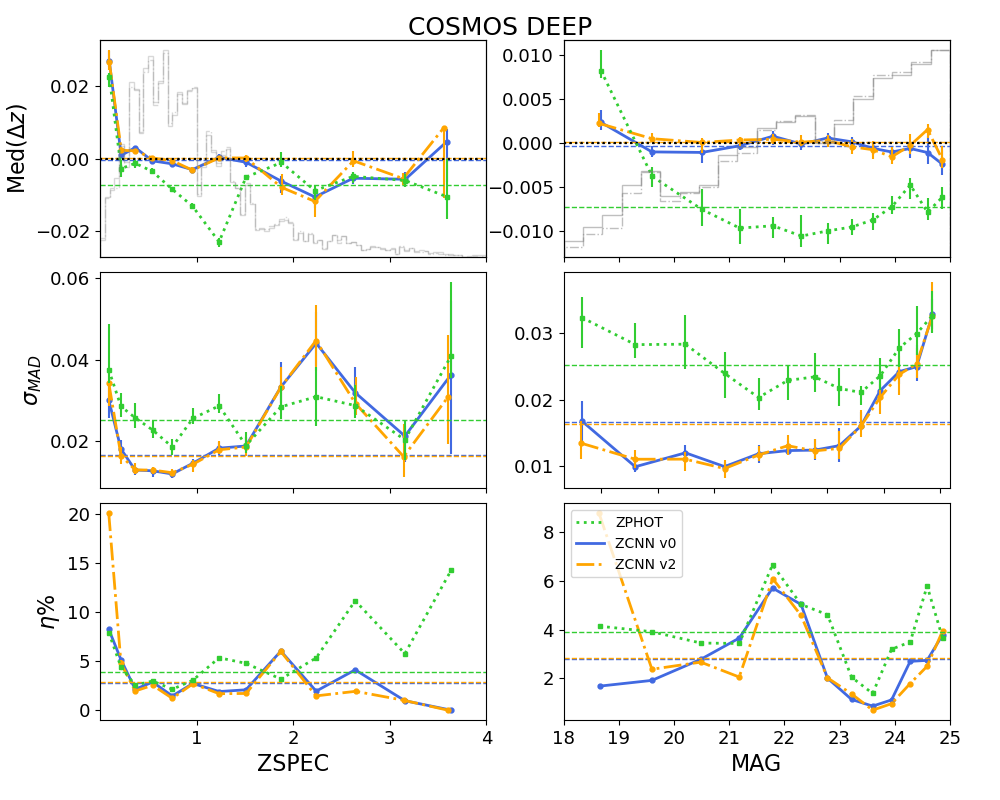}
\includegraphics[width=\linewidth]{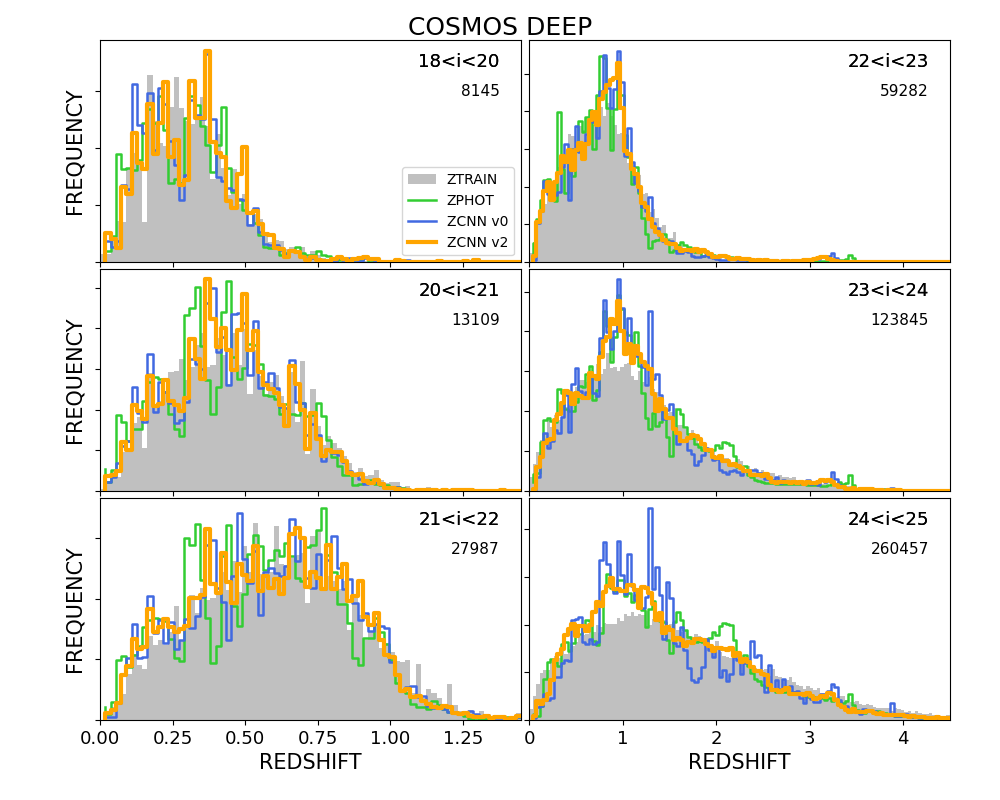}
\includegraphics[width=\linewidth]{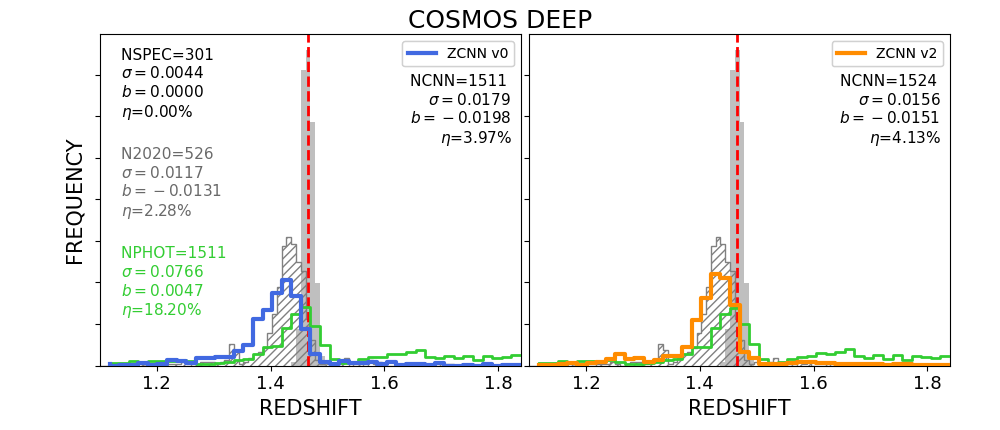}
\caption{Same as Fig. \ref{fig:shallow_elais} for the COSMOS DEEP field.}
\label{fig:shallow_cosmos}
\end{figure}

The results of the adaptation model (v2) for these 4 sub-regions are compared to the inferences from the baseline CNN (v0) and to the SED-fitting redshifts, in Figs. \ref{fig:shallow_elais}, \ref{fig:deep_xmm},  \ref{fig:shallow_xmm} and \ref{fig:shallow_cosmos}. These figures show, from top to bottom, the performance of the 3 methods for the spectroscopic samples, their predicted redshift distributions for the complete samples in 6 magnitude bins, and their predicted redshift distributions for the [OII] emission line galaxy candidates. 

\begin{itemize}
\item ELAIS (Fig. \ref{fig:shallow_elais}) :
The ELAIS field is very poor in spectroscopy, with only $\sim 1200$ redshifts smaller than 1 in this field. The adversarial model is not a dramatic improvement on the baseline CNN or even on the $ugrizy$ SED-fitting reshifts for those few essentially bright, low redshift sources. The redshift distributions inferred from the v0 and v2 models for the full sample at bright magnitude are also similarly distorted by unlikely prominent redshift structures. Only at $i>23$ does the v2 model appear to be a significant improvement on v0 and on the SED-fitting reshift. An excess of faint galaxies around $z\sim 1$ with respect to the training distribution is present as in DEEP2 but less prominently. The emission line galaxy predictions are brought considerably closer to the C2020 values.

\item XMM ULTRA DEEP (Fig. \ref{fig:deep_xmm}): 
The improvement from SED-fitting to v0 to v2 is readily visible in this field. The v2 redshift distributions are still quite distorted at bright magnitude but there are 3 times fewer galaxies than in ELAIS and DEEP2. The SED-fitting photometric redshifts were computed with the addition of 3 NIR bands (Section \ref{subsec:phot}) for a small quarter of the sources, which, strangely, seems to degrade the redshift distributions at faint magnitude and to cause the emission line galaxy predictions to be more negatively biased than in the DEEP2 and ELAIS cases. The v0 model is more aligned with spectroscopically confirmed emission line galaxies than v2 but the latter has learned what it was designed for.

\item XMM DEEP (Fig. \ref{fig:shallow_xmm}):
The difference between v0 and v2 is less significant in this field for the spectroscopic sources, but the gain from v2 is once again obvious in the redshift distributions of the complete sample. 
The distributions at $i>23$ show the largest excess of galaxies around $z \sim 1$ of all the HSC regions. The emission line galaxy predictions are also the worst of all fields but the sample is much smaller. XMM DEEP has the lowest S/N of all regions (Fig.~\ref{fig:hscfields}), which probably contributes to the degraded results (P19). We also found that its $r-i$ color distribution is shifted by $\sim -0.2$ compared to the other fields, while all other colors align. Some flaw in the calibrations may also contribute.

\item COSMOS DEEP (Fig. \ref{fig:shallow_cosmos}): This region surrounds the COSMOS UDF. Its labeled sample contains 8561 spectroscopic redshifts and 6359 C2020 redshifts in the high S/N outskirts of the UDF. The S/N remains quite high throughout most of the region (Fig. \ref{fig:hscfields}). In this particular case, the difference between v0 and v2 is understandably minimal compared to the other regions, including for the redshift distributions of the complete sample. Yet the adaptation module still provides a noticeable improvement at the faintest magnitudes and for [OII] emission line candidates. 
\end{itemize}

\begin{figure}
\includegraphics[width=\linewidth]{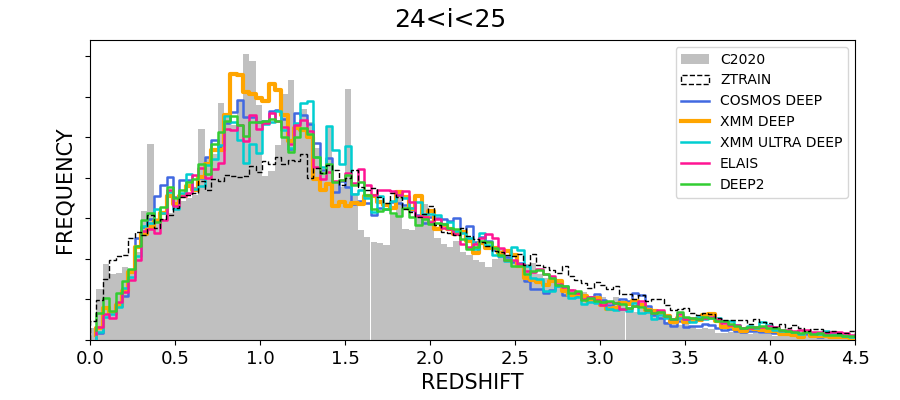}
\caption{The predicted (v2) redshift distributions in the 5 HSC regions at $24<i<25$, compared to the original COSMOS UDF distribution (gray shaded histogram) and its smoothed training version (dashed line). All histograms are normalized to 1.}
\label{fig:zdis_PHOT_ALLFIELDS}
\end{figure}

The most notable discrepancy between the predicted redshift distributions of the complete samples in all regions and the training distributions is an excess of galaxies around $z\sim 1$ at faint magnitude, at the expense of the low and high redshift tails. This excess is a flawed result of the model insofar as we consider the training distributions to be representative, but, as noted in Section \ref{subsec:adv_results}, our smoothing of the COSMOS UDF redshift distribution (Fig. \ref{fig:smoothing}) may not be the best representation of the sky-averaged redshift distribution of galaxies. Figure \ref{fig:zdis_PHOT_ALLFIELDS} overlays the predicted redshift distributions of the 5 regions in the faintest magnitude bin, $24<i<25$ (shown in previous figures), with the original COSMOS UDF distribution and its smoothed training version. Even if other regions of the sky have no reason to exhibit such big double structure around $z \sim 1$, our smoothing procedure could very well underestimate the actual frequency of galaxies in that redshift range, in which case the network's predictions may not be so far from the ground truth. It is actually a success not to reproduce the training distribution, which should ideally be completely flat.

\section{Quality assessment}
\label{sec:qa}

\begin{figure}
\includegraphics[width=\linewidth]{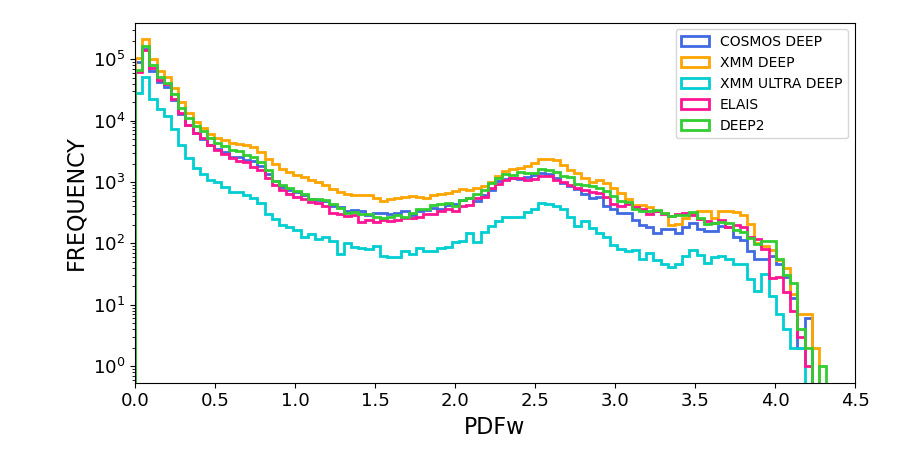}
\caption{The PDF width distribution of the complete samples in the 5 HSC regions (v2 model).}
\label{fig:PDFW_DIS}
\end{figure}

\begin{figure}
\centering
\includegraphics[width=\linewidth]{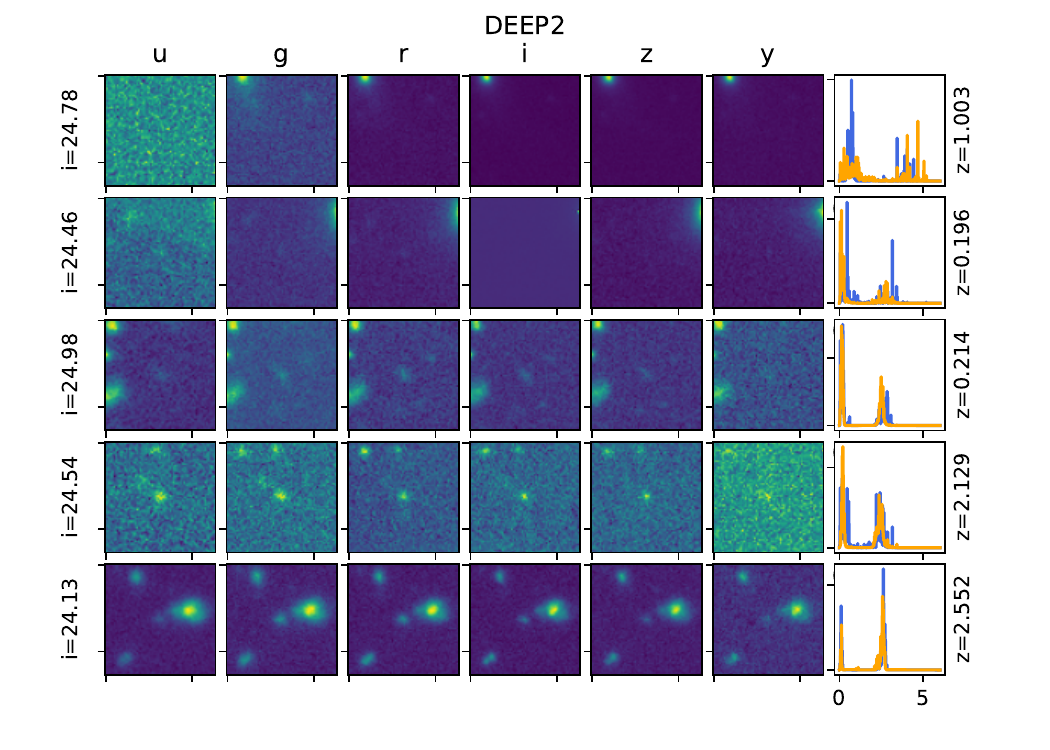}
\caption{A random sample of $ugrizy$ images of DEEP2 galaxies at $24<i<25$ with v2 PDFw $>1.5$, to be compared to samples with PDFw $<1.5$ in Fig. \ref{fig:images_pdf}. The right most panels show the PDFs from the v0 and v2 models in blue and orange, respectively. The redshift values are the final v2 estimates.}
\label{fig:deep2_badpdf}
\end{figure}

A good assessment of redshift reliability is provided by the width of the CNN classification, PDFw, defined as the redshift interval underlying the central 68\% of the PDF \citep{P19,treyer2024}. Figure \ref{fig:PDFW_DIS} shows the v2 PDFw distribution of the complete samples at $i<25$ in the 5 HSC regions analyzed in the previous sections. All fields show a resurgence of very insecure predictions at ${\rm PDFw}>1.5$, emphasized by the log scale on the $y$-axis. These cases increase with magnitude, reaching between 7.3\% and 9.5\% at $24<i<25$. The PDFw distributions of the v0 model have a similar shape, but with fractions of $\rm{PDFw} >1.5$ ranging from 10.7\% to 22.7\% at $24<i<25$, bounded by XMM UDF and XMM DEEP. Figure \ref{fig:deep2_badpdf} shows a random sample of DEEP2 galaxies with v2 PDFw $>1.5$, to be compared to samples with v2 $\rm{PDFw} <1.5$ shown in Appendix \ref{sec:images_pdf}. The images appear of poor quality and/or have multiple sources brighter than the central target, but that is not unusual in Fig. \ref{fig:images_pdf} either. Both the v0 and v2 PDFs are wide due to their multimodality. In the examples displayed in Appendix (v2 $\rm{PDFw} <1.5$), v2 seems to have resolved the v0 multimodal conflicts. In all that follows, we restrict the statistics to ${\rm PDFw}<1.5$.

\begin{figure*}
\includegraphics[width=\textwidth]{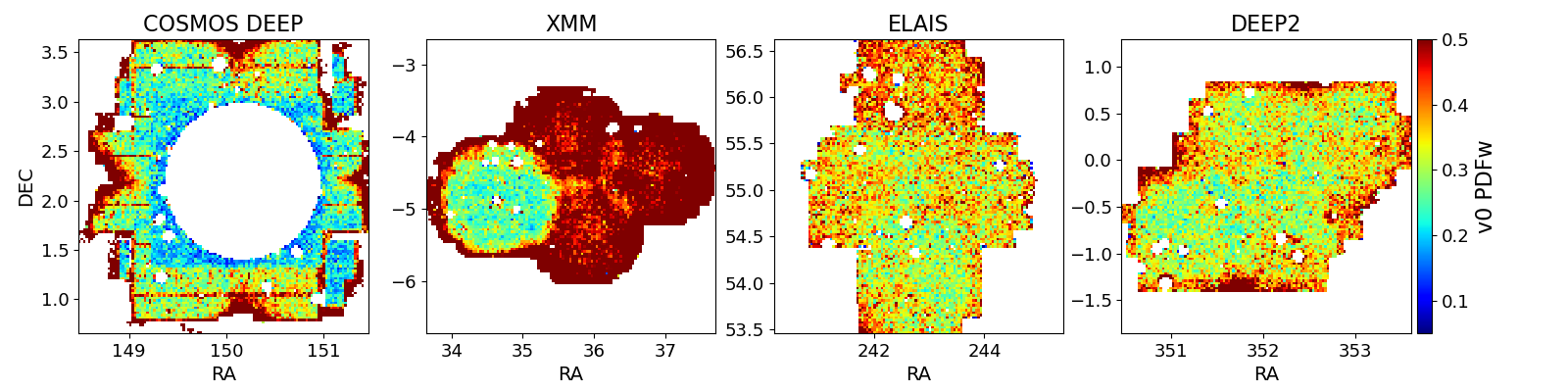}
\includegraphics[width=\textwidth]{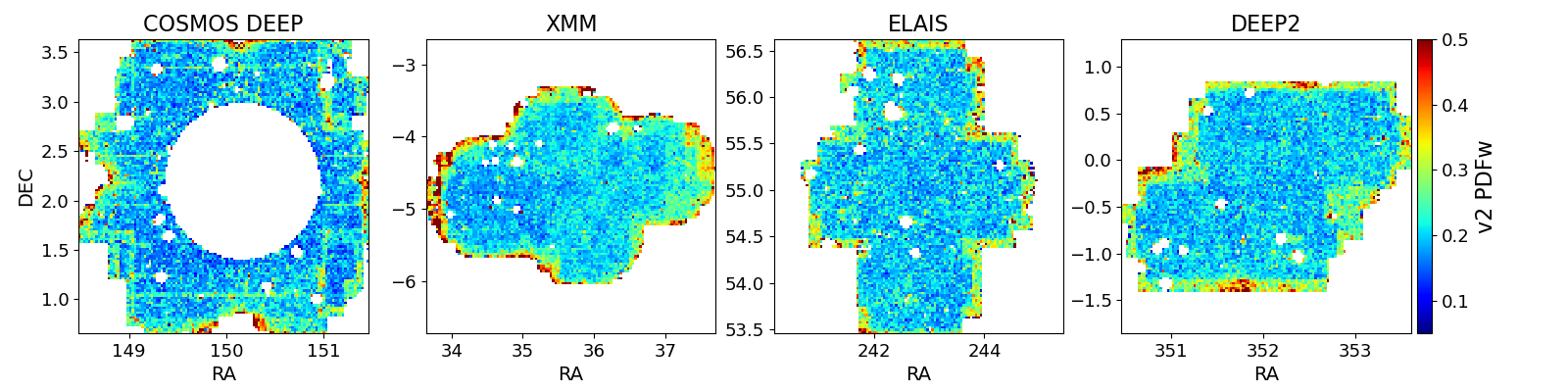}
\caption{The four HSC regions color-coded by the PDF width (PDFw $<1.5$) of sources at $24<i<25$ for the v0 and v2 models in the upper and lower panels, respectively. The color scale is the same for all.}
\label{fig:hscfields_pdfw}
\end{figure*}

Figure \ref{fig:hscfields_pdfw} shows the gain in PDFw per region from the baseline CNN to the adversarial model at $24<i<25$. Note that since PDFw larger than 1.5 have been excluded in both cases, more poor values have been discarded in the upper panels than in the lower ones (more than twice as many in the case of XMM DEEP). The gain is therefore larger than it looks. XMM DEEP clearly has its own specific issues compared to the other regions. As mentioned in Section \ref{sec:otherfields}, we found a color shift in $r-i$ color compared to the other fields, and to COSMOS UDF in particular, which might explain why the baseline inference fails more dramatically than in other regions. The adversarial model largely remedies this problem. The XMM UDF, which is closest to the COSMOS UDF in S/N and E(B-V), fares consistently better. Poor results persist in the outskirts of the fields.

\begin{figure}
\includegraphics[width=\linewidth]{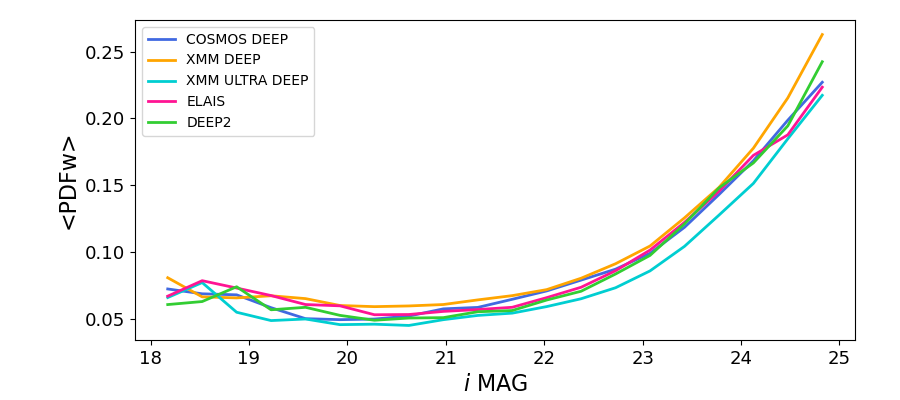}
\caption{The mean PDF width (PDFw $<1.5$) as a function of magnitude in the 5 HSC regions (v2 model). The range is bounded by XMM ULTRA DEEP and XMM DEEP, which have the lowest and largest values, respectively.
}
\label{fig:MEANPDFW_MAG}
\end{figure}

\begin{figure*}
\includegraphics[width=\linewidth]{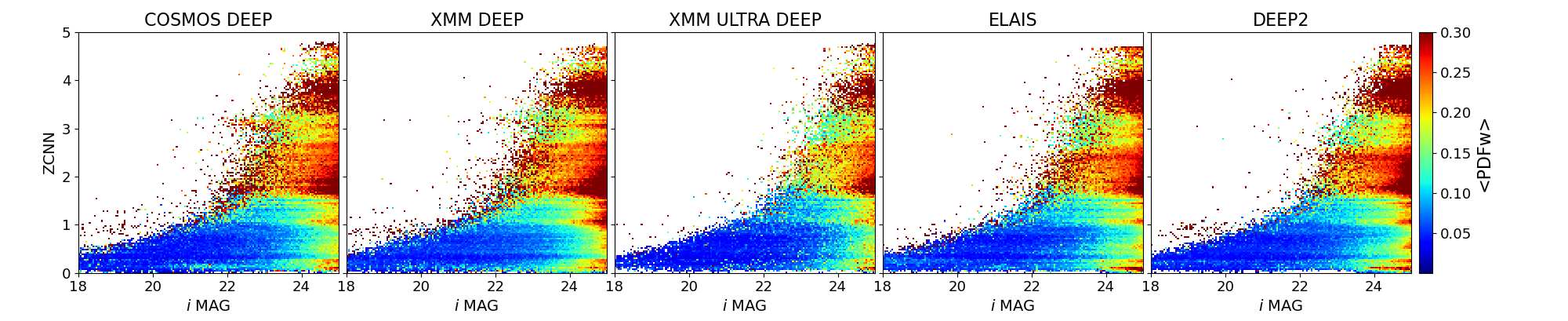}
\caption{The distribution of PDF uncertainties (mean PDF width per pixel) in magnitude-redshift space is remarkably similar in all the fields, with similarly uneven patches of particularly degraded predictions at $i\gtrsim 23$.}
\label{fig:HSCfields_magz_pdfw}
\end{figure*}

Although the mean PDF width increases smoothly with magnitude (Fig. \ref{fig:MEANPDFW_MAG}), the degradation in redshift accuracy exhibits some complexity. Figure \ref{fig:HSCfields_magz_pdfw} shows the mean PDF width per pixel in magnitude-redshift space in the 5 regions. The patterns are remarkably similar in all the fields. At $i\gtrsim 23$, the widening of the PDFs (essentially due to multimodality) varies unevenly with redshift (their median values), with 2 extended patches of particularly degraded predictions around $z_{\rm CNN}\approx 1.8$ and $z_{\rm CNN}\approx 3.8$. In contrast, the network appears much less hesitant about piling up galaxies around $z\sim 1$, which aligns with the earlier argument that the predicted redshift frequency in this range may not be a classification failure. XMM DEEP, whose "excess" around $z\sim 1$ is more extreme than in the other regions (Fig. \ref{fig:zdis_PHOT_ALLFIELDS}) is more degraded at this particular redshift (Fig. \ref{fig:HSCfields_magz_pdfw}).

\begin{figure}
\includegraphics[width=8.5cm]{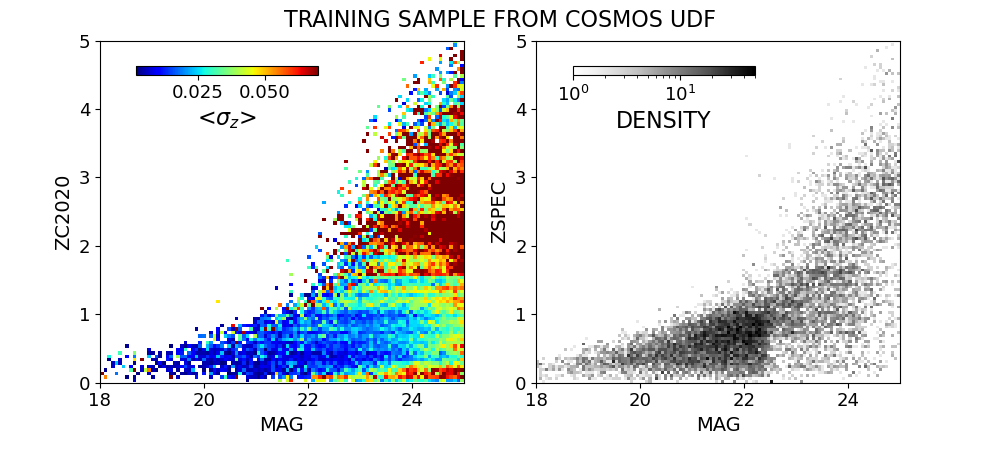}
\caption{{\bf Left:} The $\sigma_z$ distribution of the C2020 labels (see Section \ref{subsec:spec}) in magnitude-redshift space. {\bf Right:}  Spectroscopic redshifts replacing C2020 values in the training sample. In the red regions on the left, the CNN relies essentially on dubious labels.}
\label{fig:COSMOS2020_magz_sigz}
\end{figure}

As mentioned in Section \ref{subsec:spec}, the C2020 redshift we use as labels are the means of 4 estimates, with dispersion $\sigma_z\le 0.1(1+\bar z)$. The left panel of Fig. \ref{fig:COSMOS2020_magz_sigz} shows the mean $\sigma_z$ per pixel in magnitude-redshift space. It appears that $\sigma_z$ also varies unevenly with redshift at faint magnitude. Although the red patches are not centered around the same redshifts as in Fig. \ref{fig:HSCfields_magz_pdfw}, the topological similarity between the distributions in both figures is noteworthy. It suggests that the CNN detects the more or less blurry nature of the redshift labels depending on their location in the parameter space and that it affects its classification ability, reflected by the PDF widths. The right panel shows the magnitude-redshift distribution of the spectroscopic redshifts we substituted to the C2020 values in the training sample, when available. At $i>23$, they account for less than 8\% of the sample. They may be too few to have much stabilizing influence on the classification or, as they tend to be larger than the C2020 estimates (the small negative bias in Fig. \ref{fig:zc2020_zspec} worsens with redshift), they may add to the CNN's confusion around these galaxies. In any case, we may assume that the uneven consistency of the labels reflected by the $\sigma_z$ distribution is at least partly responsible for the uneven PDFw distribution in Fig.~\ref{fig:HSCfields_magz_pdfw}, even if not straightforwardly. 

All distributions in Fig. \ref{fig:HSCfields_magz_pdfw} plateau at $z\sim4.8$, above which no point estimate is ever predicted despite training the model to $z=6$. The latter phenomenon was also observed when deriving photometric redshifts in the SDSS via a CNN classification (P19). It appears to be due to the shortage of galaxies in the high redshift tail of the training distribution, which does not adequately populate the last bins of the classification. Wider bins may be used at the highest redshifts to alleviate the problem \citep{treyer2024}. At faint magnitude, the network is also plagued by other weaknesses of the training sample, in particular the lower S/N, and of the inference samples, whose S/N are even lower, except in the XMM UDF.

\section{Conclusion}
\label{sec:conclusion}

We used a representative sample of $\sim 180,000$ redshifts available in the COSMOS UDF as labels to train a CNN with the aim of predicting redshifts for the remaining $\sim$ 4 millions galaxies in the HSC-CLAUDS survey at $i<25$ \citep{Desprez2023}. For most of the brightest galaxies in this sample ($i<22.5$), the redshifts are spectroscopic; for most of the faintest galaxies, they are photometric, based on 30-band SED-fitting from the COSMOS2020 sample \citep{Weaver2022}. The input data supplied to the CNN consist of 6 stamp images in the $ugrizy$ bands centered on each galaxy of the HSC survey. The CNN architecture is from \citet{aitouahmed2024}. Its output is a classification into contiguous bins that we use as redshift PDF, whose median and width are our point estimate and uncertainty measure, respectively.   

Training the CNN with the COSMOS-UDF labeled sample revealed a very strong domain mismatch problem when inferring redshifts for sources from other HSC regions. To alleviate this problem, we developed an unsupervised adversarial-based domain adaptation network. This approach aims to transfer the redshift classification capability that the CNN learns from the COSMOS UDF data - the source field - to another, so called target field. Its effectiveness was tested with: 1/ the metrics that can be measured for the spectroscopic sample available in each target field (mostly bright galaxies) 2/ the shape of the magnitude-dependent redshift distributions predicted for the entire target population, that are expected to resemble those of the presumably representative training sample 3/ the redshift predictions for a sample of galaxies detected by deep, narrow-band filter observations expected to catch strong [OII] emitters at $z \sim 1.47$. All 3 tests demonstrate that the adversarial network is able to transfer its ability to estimate redshifts from the source training set to the other regions. It has yielded a unified representation that captures the underlying data structure. We find the PDF widths to be good measures of redshift reliability. In particular, they are able to point back to regions where labels are most uncertain in magnitude-redshift space.

The limited quality of the training sample remains a significant downside. The training redshifts from COSMOS2020, or COSMOS2025 \citep{Shuntov2025} for a future analysis at fainter magnitudes, may be better represented by soft instead of one-hot labels. The method could also be combined with self-supervised techniques further ensuring that the resulting latent space does not discriminate between the different fields. New self-supervised methods for extracting robust features from large databases have emerged in astrophysics, often relying on foundation models \citep{Parker2024,Lastufka2024}. However, these models remain biased, and t-SNE projections can end up separating data by field or survey rather than revealing true intrinsic structure. Despite their flaws, our results show that adversarial-based domain adaptation networks are capable of remedying this bias by aligning representations across different regions. Our approach offers promising perspectives for redshift estimation in future large imaging surveys where little spectroscopy will be available for training, such as those expected from, e.g., the Vera Rubin telescope. Pending improvements, the redshift catalogs we generated for the COSMOS, XMM, DEEP2 and ELAIS fields are available at deepdip.iap.fr.

\begin{acknowledgements}
This work was carried out using computing and storage resources at IDRIS thanks to grants 2024-AD010414147R1 and 2025-AD010414147R2 awarded by GENCI on the V100 and A100 partitions of the Jean Zay supercomputer. It benefited from the support of the French National Research Agency (ANR) as part of the DEEPDIP project (ANR-19-CE31-0023). 
\end{acknowledgements}

\bibliographystyle{aa}
\bibliography{biblio.bib}

@ARTICLE{Wei2025PASA,
       author = {{Wei}, Shirui and {Li}, Changhua and {Zhang}, Yanxia and {Cui}, Chenzhou and {Tang}, Chao and {Zhang}, Jingyi and {Zhao}, Yongheng and {Wu}, Xue-Bing and {Tao}, Yihan and {Fan}, Dongwei and {Li}, Shanshan and {Xu}, Yunfei and {Huang}, Maoyuan and {Yang}, Xingyu and {Kang}, Zihan and {Shi}, Jinghang},
        title = "{Photometric redshift estimation for emission line galaxies of DESI Legacy Imaging Surveys by CNN-MLP}",
      journal = {\pasa},
         year = 2025,
        month = jun,
       volume = {42},
          eid = {e092},
        pages = {e092},
          doi = {10.1017/pasa.2025.10056},
archivePrefix = {arXiv},
       eprint = {2505.24175},
 primaryClass = {astro-ph.IM},
       adsurl = {https://ui.adsabs.harvard.edu/abs/2025PASA...42...92W}
}

@ARTICLE{Picouet2023,
       author = {{Picouet}, V. and {Arnouts}, S. and {Le Floc'h}, E. and {Moutard}, T. and {Kraljic}, K. and {Ilbert}, O. and {Sawicki}, M. and {Desprez}, G. and {Laigle}, C. and {Schiminovich}, D. and {de la Torre}, S. and {Gwyn}, S. and {McCracken}, H.~J. and {Dubois}, Y. and {Dav{\'e}}, R. and {Toft}, S. and {Weaver}, J.~R. and {Shuntov}, M. and {Kauffmann}, O.~B.},
        title = "{HSC-CLAUDS survey: The star formation rate functions since z {\ensuremath{\sim}} 2 and comparison with hydrodynamical simulations}",
      journal = {\aap},
         year = 2023,
        month = jul,
       volume = {675},
          eid = {A164},
        pages = {A164},
          doi = {10.1051/0004-6361/202245756},
archivePrefix = {arXiv},
       eprint = {2305.05504},
 primaryClass = {astro-ph.GA},
       adsurl = {https://ui.adsabs.harvard.edu/abs/2023A&A...675A.164P}
}

@ARTICLE{Roster2024,
       author = {{Roster}, W. and {Salvato}, M. and {Krippendorf}, S. and {Saxena}, A. and {Shirley}, R. and {Buchner}, J. and {Wolf}, J. and {Dwelly}, T. and {Bauer}, F.~E. and {Aird}, J. and {Ricci}, C. and {Assef}, R.~J. and {Anderson}, S.~F. and {Liu}, X. and {Merloni}, A. and {Weller}, J. and {Nandra}, K.},
        title = "{PICZL: Image-based photometric redshifts for AGN}",
      journal = {\aap},
         year = 2024,
        month = dec,
       volume = {692},
          eid = {A260},
        pages = {A260},
          doi = {10.1051/0004-6361/202452361},
archivePrefix = {arXiv},
       eprint = {2411.07305},
 primaryClass = {astro-ph.GA},
       adsurl = {https://ui.adsabs.harvard.edu/abs/2024A&A...692A.260R}
}

@ARTICLE{Zhang2024,
       author = {{Zhang}, Chen and {Wang}, Wenyu and {Qu}, Meixia and {Jiang}, Bin and {Zhang}, YanXia},
        title = "{Photometric Redshift Estimation of Quasars by a Cross-modal Contrast Learning Method}",
      journal = {\aj},
         year = 2024,
        month = dec,
       volume = {168},
       number = {6},
          eid = {244},
        pages = {244},
          doi = {10.3847/1538-3881/ad79f9},
       adsurl = {https://ui.adsabs.harvard.edu/abs/2024AJ....168..244Z}
}

@ARTICLE{Yao2023,
       author = {{Yao}, Lin and {Qiu}, Bo and {Luo}, A. -Li and {Zhou}, Jianwei and {Wu}, Kuang and {Kong}, Xiao and {Liu}, Yuanbo and {Zhao}, Guiyu and {Wang}, Kun},
        title = "{Photometric redshift estimation of quasars with fused features from photometric data and images}",
      journal = {\mnras},
         year = 2023,
        month = aug,
       volume = {523},
       number = {4},
        pages = {5799-5811},
          doi = {10.1093/mnras/stad1842},
       adsurl = {https://ui.adsabs.harvard.edu/abs/2023MNRAS.523.5799Y}
}

@ARTICLE{Li2024CatBoost,
       author = {{Li}, Changhua and {Zhang}, Yanxia and {Cui}, Chenzhou and {Wei}, Shirui and {Zhang}, Jingyi and {Zhao}, Yongheng and {Wu}, Xue-Bing and {Tao}, Yihan and {Li}, Shanshan and {Wang}, Youfen and {Kang}, Zihan},
        title = "{A Photometric Redshift Catalogue of Galaxies from the DESI Legacy Imaging Surveys DR10}",
      journal = {\aj},
         year = 2024,
        month = dec,
       volume = {168},
       number = {6},
          eid = {233},
        pages = {233},
          doi = {10.3847/1538-3881/ad7c52},
       adsurl = {https://ui.adsabs.harvard.edu/abs/2024AJ....168..233L}
}

@ARTICLE{Li2022XGBoost,
       author = {{Li}, Changhua and {Zhang}, Yanxia and {Cui}, Chenzhou and {Fan}, Dongwei and {Zhao}, Yongheng and {Wu}, Xue-Bing and {Zhang}, Jing-Yi and {Han}, Jun and {Xu}, Yunfei and {Tao}, Yihan and {Li}, Shanshan and {He}, Boliang},
        title = "{Photometric redshift estimation of BASS DR3 quasars by machine learning}",
      journal = {\mnras},
         year = 2022,
        month = jan,
       volume = {509},
       number = {2},
        pages = {2289-2303},
          doi = {10.1093/mnras/stab3165},
archivePrefix = {arXiv},
       eprint = {2110.14951},
 primaryClass = {astro-ph.IM},
       adsurl = {https://ui.adsabs.harvard.edu/abs/2022MNRAS.509.2289L}
}

@ARTICLE{Zhang2013kNN,
       author = {{Zhang}, Yanxia and {Ma}, He and {Peng}, Nanbo and {Zhao}, Yongheng and {Wu}, Xue-bing},
        title = "{Estimating Photometric Redshifts of Quasars via the k-nearest Neighbor Approach Based on Large Survey Databases}",
      journal = {\aj},
         year = 2013,
        month = aug,
       volume = {146},
       number = {2},
          eid = {22},
        pages = {22},
          doi = {10.1088/0004-6256/146/2/22},
archivePrefix = {arXiv},
       eprint = {1305.5023},
 primaryClass = {astro-ph.IM},
       adsurl = {https://ui.adsabs.harvard.edu/abs/2013AJ....146...22Z}
}

@ARTICLE{Csabai2003,
       author = {{Csabai}, Istv{\'a}n and {Budav{\'a}ri}, Tam{\'a}s and {Connolly}, Andrew J. and {Szalay}, Alexander S. and {Gy{\H{o}}ry}, Zsuzsanna and {Ben{\'\i}tez}, Narciso and {Annis}, Jim and {Brinkmann}, Jon and {Eisenstein}, Daniel and {Fukugita}, Masataka and {Gunn}, Jim and {Kent}, Stephen and {Lupton}, Robert and {Nichol}, Robert C. and {Stoughton}, Chris},
        title = "{The Application of Photometric Redshifts to the SDSS Early Data Release}",
      journal = {\aj},
         year = 2003,
        month = feb,
       volume = {125},
       number = {2},
        pages = {580-592},
          doi = {10.1086/345883},
archivePrefix = {arXiv},
       eprint = {astro-ph/0211080},
 primaryClass = {astro-ph},
       adsurl = {https://ui.adsabs.harvard.edu/abs/2003AJ....125..580C}
}

@ARTICLE{Henghes2021,
       author = {{Henghes}, Ben and {Pettitt}, Connor and {Thiyagalingam}, Jeyan and {Hey}, Tony and {Lahav}, Ofer},
        title = "{Benchmarking and scalability of machine-learning methods for photometric redshift estimation}",
      journal = {\mnras},
         year = 2021,
        month = aug,
       volume = {505},
       number = {4},
        pages = {4847-4856},
          doi = {10.1093/mnras/stab1513},
archivePrefix = {arXiv},
       eprint = {2104.01875},
 primaryClass = {astro-ph.IM},
       adsurl = {https://ui.adsabs.harvard.edu/abs/2021MNRAS.505.4847H}
}

@ARTICLE{Henghes2022,
       author = {{Henghes}, Ben and {Thiyagalingam}, Jeyan and {Pettitt}, Connor and {Hey}, Tony and {Lahav}, Ofer},
        title = "{Deep learning methods for obtaining photometric redshift estimations from images}",
      journal = {\mnras},
         year = 2022,
        month = may,
       volume = {512},
       number = {2},
        pages = {1696-1709},
          doi = {10.1093/mnras/stac480},
archivePrefix = {arXiv},
       eprint = {2109.02503},
 primaryClass = {astro-ph.IM},
       adsurl = {https://ui.adsabs.harvard.edu/abs/2022MNRAS.512.1696H}
}

@ARTICLE{Menou2019,
       author = {{Menou}, Kristen},
        title = "{Morpho-photometric redshifts}",
      journal = {\mnras},
         year = 2019,
        month = nov,
       volume = {489},
       number = {4},
        pages = {4802-4808},
          doi = {10.1093/mnras/stz2477},
archivePrefix = {arXiv},
       eprint = {1811.06374},
 primaryClass = {astro-ph.IM},
       adsurl = {https://ui.adsabs.harvard.edu/abs/2019MNRAS.489.4802M}
}

@ARTICLE{Vanzella2004,
       author = {{Vanzella}, E. and {Cristiani}, S. and {Fontana}, A. and {Nonino}, M. and {Arnouts}, S. and {Giallongo}, E. and {Grazian}, A. and {Fasano}, G. and {Popesso}, P. and {Saracco}, P. and {Zaggia}, S.},
        title = "{Photometric redshifts with the Multilayer Perceptron Neural Network: Application to the HDF-S and SDSS}",
      journal = {\aap},
         year = 2004,
        month = aug,
       volume = {423},
        pages = {761-776},
          doi = {10.1051/0004-6361:20040176},
archivePrefix = {arXiv},
       eprint = {astro-ph/0312064},
 primaryClass = {astro-ph},
       adsurl = {https://ui.adsabs.harvard.edu/abs/2004A&A...423..761V}
}

@ARTICLE{Wadadekar2005,
       author = {{Wadadekar}, Yogesh},
        title = "{Estimating Photometric Redshifts Using Support Vector Machines}",
      journal = {\pasp},
         year = 2005,
        month = jan,
       volume = {117},
       number = {827},
        pages = {79-85},
          doi = {10.1086/427710},
archivePrefix = {arXiv},
       eprint = {astro-ph/0412005},
 primaryClass = {astro-ph},
       adsurl = {https://ui.adsabs.harvard.edu/abs/2005PASP..117...79W}
}

@ARTICLE{Lastufka2024,
       author = {{Lastufka}, E. and {Bait}, O. and {Taran}, O. and {Drozdova}, M. and {Kinakh}, V. and {Piras}, D. and {Audard}, M. and {Dessauges-Zavadsky}, M. and {Holotyak}, T. and {Schaerer}, D. and {Voloshynovskiy}, S.},
        title = "{Self-supervised learning on MeerKAT wide-field continuum images}",
      journal = {\aap},
         year = 2024,
        month = oct,
       volume = {690},
          eid = {A310},
        pages = {A310},
          doi = {10.1051/0004-6361/202449964},
archivePrefix = {arXiv},
       eprint = {2408.06147},
 primaryClass = {astro-ph.IM},
       adsurl = {https://ui.adsabs.harvard.edu/abs/2024A&A...690A.310L}
}

@ARTICLE{Parker2024,
       author = {{Parker}, Liam and {Lanusse}, Francois and {Golkar}, Siavash and {Sarra}, Leopoldo and {Cranmer}, Miles and {Bietti}, Alberto and {Eickenberg}, Michael and {Krawezik}, Geraud and {McCabe}, Michael and {Morel}, Rudy and {Ohana}, Ruben and {Pettee}, Mariel and {R{\'e}galdo-Saint Blancard}, Bruno and {Cho}, Kyunghyun and {Ho}, Shirley and {Polymathic AI Collaboration}},
        title = "{AstroCLIP: a cross-modal foundation model for galaxies}",
      journal = {\mnras},
         year = 2024,
        month = jul,
       volume = {531},
       number = {4},
        pages = {4990-5011},
          doi = {10.1093/mnras/stae1450},
archivePrefix = {arXiv},
       eprint = {2310.03024},
 primaryClass = {astro-ph.IM},
       adsurl = {https://ui.adsabs.harvard.edu/abs/2024MNRAS.531.4990P}
}

@ARTICLE{Shuntov2025,
       author = {{Shuntov}, Marko and {Akins}, Hollis B. and {Paquereau}, Louise and {Casey}, Caitlin M. and {Ilbert}, Olivier and {Arango-Toro}, Rafael C. and {McCracken}, Henry Joy and {Franco}, Maximilien and {Harish}, Santosh and {Kartaltepe}, Jeyhan S. and {Koekemoer}, Anton M. and {Yang}, Lilan and {Huertas-Company}, Marc and {Berman}, Edward M. and {McCleary}, Jacqueline E. and {Toft}, Sune and {Gavazzi}, Rapha{\"e}l and {Achenbach}, Mark J. and {Bertin}, Emmanuel and {Brinch}, Malte and {Champagne}, Jackie and {Chartab}, Nima and {Drakos}, Nicole E. and {Egami}, Eiichi and {Endsley}, Ryan and {Faisst}, Andreas L. and {Fan}, Xiaohui and {Flayhart}, Carter and {Hartley}, William G. and {Hatamnia}, Hossein and {Gozaliasl}, Ghassem and {Gentile}, Fabrizio and {Jermann}, Iris and {Jin}, Shuowen and {Kakiichi}, Koki and {Khostovan}, Ali Ahmad and {K{\"u}mmel}, Martin and {Laigle}, Clotilde and {Laishram}, Ronaldo and {Lambrides}, Erini and {Liu}, Daizhong and {Lyu}, Jianwei and {Magdis}, Georgios and {Mobasher}, Bahram and {Moutard}, Thibaud and {Renzini}, Alvio and {Robertson}, Brant E. and {Schefer}, Marc and {Scognamiglio}, Diana and {Scoville}, Nick and {Sattari}, Zahra and {Sanders}, David B. and {Taamoli}, Sina and {Trakhtenbrot}, Benny and {Valentino}, Francesco and {Wang}, Feige and {Weaver}, John R. and {Yang}, Jinyl},
        title = "{COSMOS2025: The COSMOS-Web galaxy catalog of photometry, morphology, redshifts, and physical parameters from JWST, HST, and ground-based imaging}",
      journal = {arXiv e-prints},
         year = 2025,
        month = jun,
          eid = {arXiv:2506.03243},
        pages = {arXiv:2506.03243},
          doi = {10.48550/arXiv.2506.03243},
archivePrefix = {arXiv},
       eprint = {2506.03243},
 primaryClass = {astro-ph.GA},
       adsurl = {https://ui.adsabs.harvard.edu/abs/2025arXiv250603243S}
}

@ARTICLE{Khostovan2025,
       author = {{Khostovan}, Ali Ahmad and {Kartaltepe}, Jeyhan S. and {Salvato}, Mara and {Ilbert}, Olivier and {Casey}, Caitlin M. and {Algera}, Hiddo and {Antwi-Danso}, Jacqueline and {Battisti}, Andrew and {Brinch}, Malte and {Brusa}, Marcella and {Calabro}, Antonello and {Capak}, Peter L. and {Chartab}, Nima and {Cooper}, Olivia R. and {Cox}, Isa G. and {Darvish}, Behnam and {Drakos}, Nicole E. and {Faisst}, Andreas L. and {George}, Matthew R. and {Gozaliasl}, Ghassem and {Harish}, Santosh and {Hasinger}, Gunther and {Hatamnia}, Hossein and {Iovino}, Angela and {Jin}, Shuowen and {Kashino}, Daichi and {Koekemoer}, Anton M. and {Laishram}, Ronaldo and {Lee}, Khee-Gan and {Lertprasertpong}, Jitrapon and {Lilly}, Simon J. and {Masters}, Daniel C. and {Mobasher}, Bahram and {Nagao}, Tohru and {Onodera}, Masato and {Peng}, Yingjie and {Sanders}, David B. and {Sanders}, Ryan L. and {Sattari}, Zahra and {Scoville}, Nick and {Shah}, Ekta A. and {Silverman}, John D. and {Suzuki}, Nao and {Tanaka}, Masayuki and {Toft}, Sune and {Trakhtenbrot}, Benny and {Trump}, Jonathan R. and {Vaccari}, Mattia and {Valentino}, Francesco and {Vanderhoof}, Brittany N. and {Weaver}, John R. and {Yun}, Min S. and {Zavala}, Jorge A.},
        title = "{COSMOS Spectroscopic Redshift Compilation (First Data Release): 165k Redshifts Encompassing Two Decades of Spectroscopy}",
      journal = {arXiv e-prints},
         year = 2025,
        month = feb,
          eid = {arXiv:2503.00120},
        pages = {arXiv:2503.00120},
          doi = {10.48550/arXiv.2503.00120},
archivePrefix = {arXiv},
       eprint = {2503.00120},
 primaryClass = {astro-ph.GA},
       adsurl = {https://ui.adsabs.harvard.edu/abs/2025arXiv250300120K}
}

@ARTICLE{aitouahmed2024,
       author = {{Ait Ouahmed}, R. and {Arnouts}, S. and {Pasquet}, J. and {Treyer}, M. and {Bertin}, E.},
        title = "{Multimodality for improved CNN photometric redshifts}",
      journal = {\aap},
         year = 2024,
        month = mar,
       volume = {683},
          eid = {A26},
        pages = {A26},
          doi = {10.1051/0004-6361/202347395},
       adsurl = {https://ui-adsabs-harvard-edu.insu.bib.cnrs.fr/abs/2024A&A...683A..26A}
}

@ARTICLE{treyer2024,
       author = {{Treyer}, M. and {Ait Ouahmed}, R. and {Pasquet}, J. and {Arnouts}, S. and {Bertin}, E. and {Fouchez}, D.},
        title = "{CNN photometric redshifts in the SDSS at r {\ensuremath{\leq}} 20}",
      journal = {\mnras},
         year = 2024,
        month = jan,
       volume = {527},
       number = {1},
        pages = {651-671},
          doi = {10.1093/mnras/stad3171},
archivePrefix = {arXiv},
       eprint = {2310.02173},
 primaryClass = {astro-ph.CO},
       adsurl = {https://ui-adsabs-harvard-edu.insu.bib.cnrs.fr/abs/2024MNRAS.527..651T}
}

@ARTICLE{P19,
       author = {{Pasquet}, Johanna and {Bertin}, E. and {Treyer}, M. and {Arnouts}, S. and
         {Fouchez}, D.},
        title = "{Photometric redshifts from SDSS images using a convolutional neural network}",
      journal = {\aap},
         year = "2019",
        month = "Jan",
       volume = {621},
          eid = {A26},
        pages = {A26},
          doi = {10.1051/0004-6361/201833617},
archivePrefix = {arXiv},
       eprint = {1806.06607},
 primaryClass = {astro-ph.IM},
       adsurl = {https://ui.adsabs.harvard.edu/abs/2019A&A...621A..26P}
}

@ARTICLE{Hayashi2020,
       author = {{Hayashi}, Masao and {Shimakawa}, Rhythm and {Tanaka}, Masayuki and {Onodera}, Masato and {Koyama}, Yusei and {Inoue}, Akio K. and {Komiyama}, Yutaka and {Lee}, Chien-Hsiu and {Lin}, Yen-Ting and {Yabe}, Kiyoto},
        title = "{A 16 deg$^{2}$ survey of emission-line galaxies at z < 1.6 from HSC-SSP PDR2 and CHORUS}",
      journal = {\pasj},
         year = 2020,
        month = oct,
       volume = {72},
       number = {5},
          eid = {86},
        pages = {86},
          doi = {10.1093/pasj/psaa076},
archivePrefix = {arXiv},
       eprint = {2007.07413},
 primaryClass = {astro-ph.GA},
       adsurl = {https://ui-adsabs-harvard-edu.insu.bib.cnrs.fr/abs/2020PASJ...72...86H}
}

@ARTICLE{2020SciPy-NMeth,
  author  = {Virtanen, Pauli and Gommers, Ralf and Oliphant, Travis E. and
            Haberland, Matt and Reddy, Tyler and Cournapeau, David and
            Burovski, Evgeni and Peterson, Pearu and Weckesser, Warren and
            Bright, Jonathan and {van der Walt}, St{\'e}fan J. and
            Brett, Matthew and Wilson, Joshua and Millman, K. Jarrod and
            Mayorov, Nikolay and Nelson, Andrew R. J. and Jones, Eric and
            Kern, Robert and Larson, Eric and Carey, C J and
            Polat, {\.I}lhan and Feng, Yu and Moore, Eric W. and
            {VanderPlas}, Jake and Laxalde, Denis and Perktold, Josef and
            Cimrman, Robert and Henriksen, Ian and Quintero, E. A. and
            Harris, Charles R. and Archibald, Anne M. and
            Ribeiro, Ant{\^o}nio H. and Pedregosa, Fabian and
            {van Mulbregt}, Paul and {SciPy 1.0 Contributors}},
  title   = {{{SciPy} 1.0: Fundamental Algorithms for Scientific
            Computing in Python}},
  journal = {Nature Methods},
  year    = {2020},
  volume  = {17},
  pages   = {261--272},
  adsurl  = {https://rdcu.be/b08Wh},
  doi     = {10.1038/s41592-019-0686-2},
}

@article{hoyle2016,
  title={Measuring photometric redshifts using galaxy images and Deep Neural Networks},
  author={Hoyle, Ben},
  journal={Astronomy and Computing},
  volume={16},
  pages={34--40},
  year={2016},
  publisher={Elsevier}
}

@preamble{ " \newcommand{\noop}[1]{} " }

@ARTICLE{Dey2022,
       author = {{Dey}, Biprateep and {Andrews}, Brett H. and {Newman}, Jeffrey A. and {Mao}, Yao-Yuan and {Rau}, Markus Michael and {Zhou}, Rongpu},
        title = "{Photometric redshifts from SDSS images with an interpretable deep capsule network}",
      journal = {\mnras},
         year = 2022,
        month = oct,
       volume = {515},
       number = {4},
        pages = {5285-5305},
          doi = {10.1093/mnras/stac2105},
archivePrefix = {arXiv},
       eprint = {2112.03939},
 primaryClass = {astro-ph.IM},
       adsurl = {https://ui.adsabs.harvard.edu/abs/2022MNRAS.515.5285D}
}

@ARTICLE{Hayat2021, 
       author = {{Hayat}, Md Abul and {Stein}, George and {Harrington}, Peter and {Luki{\'c}}, Zarija and {Mustafa}, Mustafa},
        title = "{Self-supervised Representation Learning for Astronomical Images}",
      journal = {\apjl},
         year = 2021,
        month = apr,
       volume = {911},
       number = {2},
          eid = {L33},
        pages = {L33},
          doi = {10.3847/2041-8213/abf2c7},
       adsurl = {https://ui-adsabs-harvard-edu.insu.bib.cnrs.fr/abs/2021ApJ...911L..33H}
}

@ARTICLE{Weaver2022,
       author = {{Weaver}, J.~R. and {Kauffmann}, O.~B. and {Ilbert}, O. and {McCracken}, H.~J. and {Moneti}, A. and {Toft}, S. and {Brammer}, G. and {Shuntov}, M. and {Davidzon}, I. and {Hsieh}, B.~C. and {Laigle}, C. and {Anastasiou}, A. and {Jespersen}, C.~K. and {Vinther}, J. and {Capak}, P. and {Casey}, C.~M. and {McPartland}, C.~J.~R. and {Milvang-Jensen}, B. and {Mobasher}, B. and {Sanders}, D.~B. and {Zalesky}, L. and {Arnouts}, S. and {Aussel}, H. and {Dunlop}, J.~S. and {Faisst}, A. and {Franx}, M. and {Furtak}, L.~J. and {Fynbo}, J.~P.~U. and {Gould}, K.~M.~L. and {Greve}, T.~R. and {Gwyn}, S. and {Kartaltepe}, J.~S. and {Kashino}, D. and {Koekemoer}, A.~M. and {Kokorev}, V. and {Le F{\`e}vre}, O. and {Lilly}, S. and {Masters}, D. and {Magdis}, G. and {Mehta}, V. and {Peng}, Y. and {Riechers}, D.~A. and {Salvato}, M. and {Sawicki}, M. and {Scarlata}, C. and {Scoville}, N. and {Shirley}, R. and {Silverman}, J.~D. and {Sneppen}, A. and {Smolc̆i{\'c}}, V. and {Steinhardt}, C. and {Stern}, D. and {Tanaka}, M. and {Taniguchi}, Y. and {Teplitz}, H.~I. and {Vaccari}, M. and {Wang}, W. -H. and {Zamorani}, G.},
        title = "{COSMOS2020: A Panchromatic View of the Universe to z 10 from Two Complementary Catalogs}",
      journal = {\apjs},
         year = 2022,
        month = jan,
       volume = {258},
       number = {1},
          eid = {11},
        pages = {11},
          doi = {10.3847/1538-4365/ac3078},
archivePrefix = {arXiv},
       eprint = {2110.13923},
 primaryClass = {astro-ph.GA},
       adsurl = {https://ui.adsabs.harvard.edu/abs/2022ApJS..258...11W}
}

@article{York2000,
doi = {10.1086/301513},
url = {https://dx.doi.org/10.1086/301513},
year = {2000},
month = {sep},
publisher = {},
volume = {120},
number = {3},
pages = {1579},
author = {York, Donald G. and Adelman, J. and Anderson, Jr., John E. and Anderson, Scott F. and Annis, James and Bahcall, Neta A. and Bakken, J. A. and Barkhouser, Robert and Bastian, Steven and Berman, Eileen and Boroski, William N. and Bracker, Steve and Briegel, Charlie and Briggs, John W. and Brinkmann, J. and Brunner, Robert and Burles, Scott and Carey, Larry and Carr, Michael A. and Castander, Francisco J. and Chen, Bing and Colestock, Patrick L. and Connolly, A. J. and Crocker, J. H. and Csabai, Istv√°n and Czarapata, Paul C. and Davis, John Eric and Doi, Mamoru and Dombeck, Tom and Eisenstein, Daniel and Ellman, Nancy and Elms, Brian R. and Evans, Michael L. and Fan, Xiaohui and Federwitz, Glenn R. and Fiscelli, Larry and Friedman, Scott and Frieman, Joshua A. and Fukugita, Masataka and Gillespie, Bruce and Gunn, James E. and Gurbani, Vijay K. and de Haas, Ernst and Haldeman, Merle and Harris, Frederick H. and Hayes, J. and Heckman, Timothy M. and Hennessy, G. S. and Hindsley, Robert B. and Holm, Scott and Holmgren, Donald J. and Huang, Chi-hao and Hull, Charles and Husby, Don and Ichikawa, Shin-Ichi and Ichikawa, Takashi and Iveziƒá, ≈Ωeljko and Kent, Stephen and Kim, Rita S. J. and Kinney, E. and Klaene, Mark and Kleinman, A. N. and Kleinman, S. and Knapp, G. R. and Korienek, John and Kron, Richard G. and Kunszt, Peter Z. and Lamb, D. Q. and Lee, B. and Leger, R. French and Limmongkol, Siriluk and Lindenmeyer, Carl and Long, Daniel C. and Loomis, Craig and Loveday, Jon and Lucinio, Rich and Lupton, Robert H. and MacKinnon, Bryan and Mannery, Edward J. and Mantsch, P. M. and Margon, Bruce and McGehee, Peregrine and McKay, Timothy A. and Meiksin, Avery and Merelli, Aronne and Monet, David G. and Munn, Jeffrey A. and Narayanan, Vijay K. and Nash, Thomas and Neilsen, Eric and Neswold, Rich and Newberg, Heidi Jo and Nichol, R. C. and Nicinski, Tom and Nonino, Mario and Okada, Norio and Okamura, Sadanori and Ostriker, Jeremiah P. and Owen, Russell and Pauls, A. George and Peoples, John and Peterson, R. L. and Petravick, Donald and Pier, Jeffrey R. and Pope, Adrian and Pordes, Ruth and Prosapio, Angela and Rechenmacher, Ron and Quinn, Thomas R. and Richards, Gordon T. and Richmond, Michael W. and Rivetta, Claudio H. and Rockosi, Constance M. and Ruthmansdorfer, Kurt and Sandford, Dale and Schlegel, David J. and Schneider, Donald P. and Sekiguchi, Maki and Sergey, Gary and Shimasaku, Kazuhiro and Siegmund, Walter A. and Smee, Stephen and Smith, J. Allyn and Snedden, S. and Stone, R. and Stoughton, Chris and Strauss, Michael A. and Stubbs, Christopher and SubbaRao, Mark and Szalay, Alexander S. and Szapudi, Istvan and Szokoly, Gyula P. and Thakar, Anirudda R. and Tremonti, Christy and Tucker, Douglas L. and Uomoto, Alan and Vanden Berk, Dan and Vogeley, Michael S. and Waddell, Patrick and Wang, Shu-i and Watanabe, Masaru and Weinberg, David H. and Yanny, Brian and Yasuda, Naoki},
title = {The Sloan Digital Sky Survey: Technical Summary},
journal = {The Astronomical Journal},
}

@inproceedings{ghifary2015domain,
  title={Domain generalization for object recognition with multi-task autoencoders},
  author={Ghifary, Muhammad and Kleijn, W Bastiaan and Zhang, Mengjie and Balduzzi, David},
  booktitle={Proceedings of the IEEE international conference on computer vision},
  pages={2551--2559},
  year={2015}
}

@inproceedings{long2015learning,
  title={Learning transferable features with deep adaptation networks},
  author={Long, Mingsheng and Cao, Yue and Wang, Jianmin and Jordan, Michael},
  booktitle={International conference on machine learning},
  pages={97--105},
  year={2015},
  organization={PMLR}
}

@article{goodfellow2014generative,
  title={Generative adversarial nets},
  author={Goodfellow, Ian and Pouget-Abadie, Jean and Mirza, Mehdi and Xu, Bing and Warde-Farley, David and Ozair, Sherjil and Courville, Aaron and Bengio, Yoshua},
  journal={Advances in neural information processing systems},
  volume={27},
  year={2014}
}

@article{ganin2016domain,
  title={Domain-adversarial training of neural networks},
  author={Ganin, Yaroslav and Ustinova, Evgeniya and Ajakan, Hana and Germain, Pascal and Larochelle, Hugo and Laviolette, Fran{\c{c}}ois and Marchand, Mario and Lempitsky, Victor},
  journal={The journal of machine learning research},
  volume={17},
  number={1},
  pages={2096--2030},
  year={2016},
  publisher={JMLR. org}
}

@misc{pei2018multi,
      title={Multi-Adversarial Domain Adaptation}, 
      author={Zhongyi Pei and Zhangjie Cao and Mingsheng Long and Jianmin Wang},
      year={2018},
      eprint={1809.02176},
      archivePrefix={arXiv},
      primaryClass={cs.CV},
      url={https://arxiv.org/abs/1809.02176}, 
}

@inproceedings{tzeng2017adversarial,
  title={Adversarial discriminative domain adaptation},
  author={Tzeng, Eric and Hoffman, Judy and Saenko, Kate and Darrell, Trevor},
  booktitle={Proceedings of the IEEE conference on computer vision and pattern recognition},
  pages={7167--7176},
  year={2017}
}

@inproceedings{bousmalis2017unsupervised,
  title={Unsupervised pixel-level domain adaptation with generative adversarial networks},
  author={Bousmalis, Konstantinos and Silberman, Nathan and Dohan, David and Erhan, Dumitru and Krishnan, Dilip},
  booktitle={Proceedings of the IEEE conference on computer vision and pattern recognition},
  pages={3722--3731},
  year={2017}
}

@article{Carliles2010RF,
  title={Random forests for photometric redshifts},
  author={Carliles, Samuel and Budav{\'a}ri, Tam{\'a}s and Heinis, S{\'e}bastien and Priebe, Carey and Szalay, Alexander S},
  journal={The Astrophysical Journal},
  volume={712},
  number={1},
  pages={511},
  year={2010},
  publisher={IOP Publishing}
}

@ARTICLE{Desprez2023,
       author = {{Desprez}, G. and {Picouet}, V. and {Moutard}, T. and {Arnouts}, S. and {Sawicki}, M. and {Coupon}, J. and {Gwyn}, S. and {Chen}, L. and {Huang}, J. and {Golob}, A. and {Furusawa}, H. and {Ikeda}, H. and {Paltani}, S. and {Cheng}, C. and {Hartley}, W. and {Hsieh}, B.~C. and {Ilbert}, O. and {Kauffmann}, O.~B. and {McCracken}, H.~J. and {Shuntov}, M. and {Tanaka}, M. and {Toft}, S. and {Tresse}, L. and {Weaver}, J.~R.},
        title = "{Combining the CLAUDS and HSC-SSP surveys. U + grizy(+YJHK$_{s}$) photometry and photometric redshifts for 18M galaxies in the 20 deg$^{2}$ of the HSC-SSP Deep and ultraDeep fields}",
      journal = {\aap},
         year = 2023,
        month = feb,
       volume = {670},
          eid = {A82},
        pages = {A82},
          doi = {10.1051/0004-6361/202243363},
archivePrefix = {arXiv},
       eprint = {2301.13750},
 primaryClass = {astro-ph.GA},
       adsurl = {https://ui-adsabs-harvard-edu.insu.bib.cnrs.fr/abs/2023A&A...670A..82D}
}

@ARTICLE{Aihara2018a,
   author = {{Aihara}, H. and {Arimoto}, N. and {Armstrong}, R. and {Arnouts}, S. and 
	{Bahcall}, N.~A. and {Bickerton}, S. and {Bosch}, J. and {Bundy}, K. and 
	{Capak}, P.~L. and {Chan}, J.~H.~H. and {Chiba}, M. and {Coupon}, J. and 
	{Egami}, E. and {Enoki}, M. and {Finet}, F. and {Fujimori}, H. and 
	{Fujimoto}, S. and {Furusawa}, H. and {Furusawa}, J. and {Goto}, T. and 
	{Goulding}, A. and {Greco}, J.~P. and {Greene}, J.~E. and {Gunn}, J.~E. and 
	{Hamana}, T. and {Harikane}, Y. and {Hashimoto}, Y. and {Hattori}, T. and 
	{Hayashi}, M. and {Hayashi}, Y. and {He{\l}miniak}, K.~G. and 
	{Higuchi}, R. and {Hikage}, C. and {Ho}, P.~T.~P. and {Hsieh}, B.-C. and 
	{Huang}, K. and {Huang}, S. and {Ikeda}, H. and {Imanishi}, M. and 
	{Inoue}, A.~K. and {Iwasawa}, K. and {Iwata}, I. and {Jaelani}, A.~T. and 
	{Jian}, H.-Y. and {Kamata}, Y. and {Karoji}, H. and {Kashikawa}, N. and 
	{Katayama}, N. and {Kawanomoto}, S. and {Kayo}, I. and {Koda}, J. and 
	{Koike}, M. and {Kojima}, T. and {Komiyama}, Y. and {Konno}, A. and 
	{Koshida}, S. and {Koyama}, Y. and {Kusakabe}, H. and {Leauthaud}, A. and 
	{Lee}, C.-H. and {Lin}, L. and {Lin}, Y.-T. and {Lupton}, R.~H. and 
	{Mandelbaum}, R. and {Matsuoka}, Y. and {Medezinski}, E. and 
	{Mineo}, S. and {Miyama}, S. and {Miyatake}, H. and {Miyazaki}, S. and 
	{Momose}, R. and {More}, A. and {More}, S. and {Moritani}, Y. and 
	{Moriya}, T.~J. and {Morokuma}, T. and {Mukae}, S. and {Murata}, R. and 
	{Murayama}, H. and {Nagao}, T. and {Nakata}, F. and {Niida}, M. and 
	{Niikura}, H. and {Nishizawa}, A.~J. and {Obuchi}, Y. and {Oguri}, M. and 
	{Oishi}, Y. and {Okabe}, N. and {Okamoto}, S. and {Okura}, Y. and 
	{Ono}, Y. and {Onodera}, M. and {Onoue}, M. and {Osato}, K. and 
	{Ouchi}, M. and {Price}, P.~A. and {Pyo}, T.-S. and {Sako}, M. and 
	{Sawicki}, M. and {Shibuya}, T. and {Shimasaku}, K. and {Shimono}, A. and 
	{Shirasaki}, M. and {Silverman}, J.~D. and {Simet}, M. and {Speagle}, J. and 
	{Spergel}, D.~N. and {Strauss}, M.~A. and {Sugahara}, Y. and 
	{Sugiyama}, N. and {Suto}, Y. and {Suyu}, S.~H. and {Suzuki}, N. and 
	{Tait}, P.~J. and {Takada}, M. and {Takata}, T. and {Tamura}, N. and 
	{Tanaka}, M.~M. and {Tanaka}, M. and {Tanaka}, M. and {Tanaka}, Y. and 
	{Terai}, T. and {Terashima}, Y. and {Toba}, Y. and {Tominaga}, N. and 
	{Toshikawa}, J. and {Turner}, E.~L. and {Uchida}, T. and {Uchiyama}, H. and 
	{Umetsu}, K. and {Uraguchi}, F. and {Urata}, Y. and {Usuda}, T. and 
	{Utsumi}, Y. and {Wang}, S.-Y. and {Wang}, W.-H. and {Wong}, K.~C. and 
	{Yabe}, K. and {Yamada}, Y. and {Yamanoi}, H. and {Yasuda}, N. and 
	{Yeh}, S. and {Yonehara}, A. and {Yuma}, S.},
    title = "{The Hyper Suprime-Cam SSP Survey: Overview and survey design}",
  journal = {\pasj},
archivePrefix = "arXiv",
   eprint = {1704.05858},
 primaryClass = "astro-ph.IM",
     year = 2018,
    month = jan,
   volume = 70,
      eid = {S4},
    pages = {S4},
      doi = {10.1093/pasj/psx066},
   adsurl = {http://adsabs.harvard.edu/abs/2018PASJ...70S...4A}
}

@ARTICLE{Aihara2019,
       author = {{Aihara}, Hiroaki and {AlSayyad}, Yusra and {Ando}, Makoto and
         {Armstrong}, Robert and {Bosch}, James and {Egami}, Eiichi and
         {Furusawa}, Hisanori and {Furusawa}, Junko and {Goulding}, Andy and
         {Harikane}, Yuichi and {Hikage}, Chiaki and {Ho}, Paul T.~P. and
         {Hsieh}, Bau-Ching and {Huang}, Song and {Ikeda}, Hiroyuki and
         {Imanishi}, Masatoshi and {Ito}, Kei and {Iwata}, Ikuru and
         {Jaelani}, Anton T. and {Kakuma}, Ryota and {Kawana}, Kojiro and
         {Kikuta}, Satoshi and {Kobayashi}, Umi and {Koike}, Michitaro and
         {Komiyama}, Yutaka and {Li}, Xiangchong and {Liang}, Yongming and
         {Lin}, Yen-Ting and {Luo}, Wentao and {Lupton}, Robert and
         {Lust}, Nate B. and {MacArthur}, Lauren A. and {Matsuoka}, Yoshiki and
         {Mineo}, Sogo and {Miyatake}, Hironao and {Miyazaki}, Satoshi and
         {More}, Surhud and {Murata}, Ryoma and {Namiki}, Shigeru V. and
         {Nishizawa}, Atsushi J. and {Oguri}, Masamune and {Okabe}, Nobuhiro and
         {Okamoto}, Sakurako and {Okura}, Yuki and {Ono}, Yoshiaki and
         {Onodera}, Masato and {Onoue}, Masafusa and {Osato}, Ken and
         {Ouchi}, Masami and {Shibuya}, Takatoshi and {Strauss}, Michael A. and
         {Sugiyama}, Naoshi and {Suto}, Yasushi and {Takada}, Masahiro and
         {Takagi}, Yuhei and {Takata}, Tadafumi and {Takita}, Satoshi and
         {Tanaka}, Masayuki and {Terai}, Tsuyoshi and {Toba}, Yoshiki and
         {Uchiyama}, Hisakazu and {Utsumi}, Yousuke and {Wang}, Shiang-Yu and
         {Wang}, Wenting and {Yamada}, Yoshihiko},
        title = "{Second data release of the Hyper Suprime-Cam Subaru Strategic Program}",
      journal = {\pasj},
         year = "2019",
        month = "Oct",
        pages = {106},
          doi = {10.1093/pasj/psz103},
archivePrefix = {arXiv},
       eprint = {1905.12221},
 primaryClass = {astro-ph.IM},
       adsurl = {https://ui.adsabs.harvard.edu/abs/2019PASJ..tmp..106A}
}

@ARTICLE{Collister2004,
       author = {{Collister}, Adrian A. and {Lahav}, Ofer},
        title = "{ANNz: Estimating Photometric Redshifts Using Artificial Neural Networks}",
      journal = {\pasp},
         year = "2004",
        month = "Apr",
       volume = {116},
       number = {818},
        pages = {345-351},
          doi = {10.1086/383254},
archivePrefix = {arXiv},
       eprint = {astro-ph/0311058},
 primaryClass = {astro-ph},
       adsurl = {https://ui.adsabs.harvard.edu/abs/2004PASP..116..345C}
}

@ARTICLE{DIsantoPolsterer2018,
       author = {{D'Isanto}, A. and {Polsterer}, K.~L.},
        title = "{Photometric redshift estimation via deep learning. Generalized and pre-classification-less, image based, fully probabilistic redshifts}",
      journal = {\aap},
         year = "2018",
        month = "Jan",
       volume = {609},
          eid = {A111},
        pages = {A111},
          doi = {10.1051/0004-6361/201731326},
archivePrefix = {arXiv},
       eprint = {1706.02467},
 primaryClass = {astro-ph.IM},
       adsurl = {https://ui.adsabs.harvard.edu/abs/2018A&A...609A.111D}
}

@ARTICLE{Ilbert2006,
       author = {{Ilbert}, O. and {Arnouts}, S. and {McCracken}, H.~J. and
         {Bolzonella}, M. and {Bertin}, E. and {Le F{\`e}vre}, O. and
         {Mellier}, Y. and {Zamorani}, G. and {Pell{\`o}}, R. and {Iovino}, A. and
         {Tresse}, L. and {Le Brun}, V. and {Bottini}, D. and {Garilli}, B. and
         {Maccagni}, D. and {Picat}, J.~P. and {Scaramella}, R. and
         {Scodeggio}, M. and {Vettolani}, G. and {Zanichelli}, A. and
         {Adami}, C. and {Bardelli}, S. and {Cappi}, A. and {Charlot}, S. and
         {Ciliegi}, P. and {Contini}, T. and {Cucciati}, O. and {Foucaud}, S. and
         {Franzetti}, P. and {Gavignaud}, I. and {Guzzo}, L. and {Marano}, B. and
         {Marinoni}, C. and {Mazure}, A. and {Meneux}, B. and {Merighi}, R. and
         {Paltani}, S. and {Pollo}, A. and {Pozzetti}, L. and {Radovich}, M. and
         {Zucca}, E. and {Bondi}, M. and {Bongiorno}, A. and {Busarello}, G. and
         {de La Torre}, S. and {Gregorini}, L. and {Lamareille}, F. and
         {Mathez}, G. and {Merluzzi}, P. and {Ripepi}, V. and {Rizzo}, D. and
         {Vergani}, D.},
        title = "{Accurate photometric redshifts for the CFHT legacy survey calibrated using the VIMOS VLT deep survey}",
      journal = {\aap},
         year = "2006",
        month = "Oct",
       volume = {457},
       number = {3},
        pages = {841-856},
          doi = {10.1051/0004-6361:20065138},
archivePrefix = {arXiv},
       eprint = {astro-ph/0603217},
 primaryClass = {astro-ph},
       adsurl = {https://ui.adsabs.harvard.edu/abs/2006A&A...457..841I}
}

@ARTICLE{Jarvis2013,
   author = {{Jarvis}, M.~J. and {Bonfield}, D.~G. and {Bruce}, V.~A. and 
	{Geach}, J.~E. and {McAlpine}, K. and {McLure}, R.~J. and {Gonz{\'a}lez-Solares}, E. and 
	{Irwin}, M. and {Lewis}, J. and {Yoldas}, A.~K. and {Andreon}, S. and 
	{Cross}, N.~J.~G. and {Emerson}, J.~P. and {Dalton}, G. and 
	{Dunlop}, J.~S. and {Hodgkin}, S.~T. and {Le}, F.~O. and {Karouzos}, M. and 
	{Meisenheimer}, K. and {Oliver}, S. and {Rawlings}, S. and {Simpson}, C. and 
	{Smail}, I. and {Smith}, D.~J.~B. and {Sullivan}, M. and {Sutherland}, W. and 
	{White}, S.~V. and {Zwart}, J.~T.~L.},
    title = "{The VISTA Deep Extragalactic Observations (VIDEO) survey}",
  journal = {\mnras},
archivePrefix = "arXiv",
   eprint = {1206.4263},
     year = 2013,
    month = jan,
   volume = 428,
    pages = {1281-1295},
      doi = {10.1093/mnras/sts118},
   adsurl = {http://adsabs.harvard.edu/abs/2013MNRAS.428.1281J}
}

@ARTICLE{McCracken2012,
   author = {{McCracken}, H.~J. and {Milvang-Jensen}, B. and {Dunlop}, J. and 
	{Franx}, M. and {Fynbo}, J.~P.~U. and {Le F{\`e}vre}, O. and 
	{Holt}, J. and {Caputi}, K.~I. and {Goranova}, Y. and {Buitrago}, F. and 
	{Emerson}, J.~P. and {Freudling}, W. and {Hudelot}, P. and {L{\'o}pez-Sanjuan}, C. and 
	{Magnard}, F. and {Mellier}, Y. and {M{\o}ller}, P. and {Nilsson}, K.~K. and 
	{Sutherland}, W. and {Tasca}, L. and {Zabl}, J.},
    title = "{UltraVISTA: a new ultra-deep near-infrared survey in COSMOS}",
  journal = {\aap},
archivePrefix = "arXiv",
   eprint = {1204.6586},
 primaryClass = "astro-ph.CO",
     year = 2012,
    month = aug,
   volume = 544,
      eid = {A156},
    pages = {A156},
      doi = {10.1051/0004-6361/201219507},
   adsurl = {http://adsabs.harvard.edu/abs/2012A%26A...544A.156M}
}

@ARTICLE{Sawicki2019,
       author = {{Sawicki}, Marcin and {Arnouts}, Stephane and {Huang}, Jiasheng and
         {Coupon}, Jean and {Golob}, Anneya and {Gwyn}, Stephen and
         {Foucaud}, Sebastien and {Moutard}, Thibaud and {Iwata}, Ikuru and
         {Liu}, Chengze and {Chen}, Lingjian and {Desprez}, Guillaume and
         {Harikane}, Yuichi and {Ono}, Yoshiaki and {Strauss}, Michael A. and
         {Tanaka}, Masayuki and {Thibert}, Nathalie and {Balogh}, Michael and
         {Bundy}, Kevin and {Chapman}, Scott and {Gunn}, James E. and
         {Hsieh}, Bau-Ching and {Ilbert}, Olivier and {Jing}, Yipeng and
         {LeF{\`e}vre}, Olivier and {Li}, Cheng and {Matsuda}, Yuichi and
         {Miyazaki}, Satoshi and {Nagao}, Tohru and {Nishizawa}, Atsushi J. and
         {Ouchi}, Masami and {Shimasaku}, Kazuhiro and {Silverman}, John and
         {de la Torre}, Sylvain and {Tresse}, Laurence and {Wang}, Wei-Hao and
         {Willott}, Chris J. and {Yamada}, Toru and {Yang}, Xiaohu and
         {Yee}, Howard K.~C.},
        title = "{The CFHT large area U-band deep survey (CLAUDS)}",
      journal = {\mnras},
         year = "2019",
        month = "Nov",
       volume = {489},
       number = {4},
        pages = {5202-5217},
          doi = {10.1093/mnras/stz2522},
archivePrefix = {arXiv},
       eprint = {1909.05898},
 primaryClass = {astro-ph.GA},
       adsurl = {https://ui.adsabs.harvard.edu/abs/2019MNRAS.489.5202S}
}

@INPROCEEDINGS{2002ASPC..281..228B,
   author = {{Bertin}, E. and {Mellier}, Y. and {Radovich}, M. and {Missonnier}, G. and 
	{Didelon}, P. and {Morin}, B.},
    title = "{The TERAPIX Pipeline}",
booktitle = {Astronomical Data Analysis Software and Systems XI},
     year = 2002,
   series = {Astronomical Society of the Pacific Conference Series},
   volume = 281,
   editor = {{Bohlender}, D.~A. and {Durand}, D. and {Handley}, T.~H.},
    pages = {228},
   adsurl = {http://adsabs.harvard.edu/abs/2002ASPC..281..228B}
}

@ARTICLE{Ye2025,
       author = {{Ye}, Renhao and {Shen}, Shiyin and {de Souza}, Rafael S. and {Xu}, Quanfeng and {Chen}, Mi and {Chen}, Zhu and {Ishida}, Emille E.~O. and {Krone-Martins}, Alberto and {Durgesh}, Rupesh},
        title = "{From Galaxy Zoo DECaLS to BASS/MzLS: detailed galaxy morphology classification with unsupervised domain adaption}",
      journal = {\mnras},
         year = 2025,
        month = feb,
       volume = {537},
       number = {2},
        pages = {640-649},
          doi = {10.1093/mnras/staf025},
archivePrefix = {arXiv},
       eprint = {2412.15533},
 primaryClass = {astro-ph.GA},
       adsurl = {https://ui.adsabs.harvard.edu/abs/2025MNRAS.537..640Y}
}

@ARTICLE{Belfiore2025,
       author = {{Belfiore}, Francesco and {Ginolfi}, Michele and {Blanc}, Guillermo and {Boquien}, Mederic and {Chevance}, Melanie and {Congiu}, Enrico and {Glover}, Simon C.~O. and {Groves}, Brent and {Klessen}, Ralf S. and {Eduardo M{\'e}ndez-Delgado}, J. and {Williams}, Thomas G.},
        title = "{Machine learning the gap between real and simulated nebulae: A domain-adaptation approach to classify ionised nebulae in nearby galaxies}",
      journal = {\aap},
         year = 2025,
        month = feb,
       volume = {694},
          eid = {A212},
        pages = {A212},
          doi = {10.1051/0004-6361/202451934},
archivePrefix = {arXiv},
       eprint = {2410.16370},
 primaryClass = {astro-ph.GA},
       adsurl = {https://ui.adsabs.harvard.edu/abs/2025A&A...694A.212B}
}

@ARTICLE{Swierc2024,
       author = {{Swierc}, Paxson and {Tamargo-Arizmendi}, Marcos and {{\'C}iprijanovi{\'c}}, Aleksandra and {Nord}, Brian D.},
        title = "{Domain-Adaptive Neural Posterior Estimation for Strong Gravitational Lens Analysis}",
      journal = {arXiv e-prints},
         year = 2024,
        month = oct,
          eid = {arXiv:2410.16347},
        pages = {arXiv:2410.16347},
          doi = {10.48550/arXiv.2410.16347},
archivePrefix = {arXiv},
       eprint = {2410.16347},
 primaryClass = {astro-ph.IM},
       adsurl = {https://ui.adsabs.harvard.edu/abs/2024arXiv241016347S}
}

@ARTICLE{Gilda2024,
       author = {{Gilda}, Sankalp and {de Mathelin}, Antoine and {Bellstedt}, Sabine and {Richard}, Guillaume},
        title = "{Unsupervised Domain Adaptation for Constraining Star Formation Histories}",
      journal = {Astronomy},
         year = 2024,
        month = jul,
       volume = {3},
       number = {3},
        pages = {189-207},
          doi = {10.3390/astronomy3030012},
archivePrefix = {arXiv},
       eprint = {2112.14072},
 primaryClass = {astro-ph.GA},
       adsurl = {https://ui.adsabs.harvard.edu/abs/2024Astro...3..189G}
}

@ARTICLE{Xu2023,
       author = {{Xu}, Quanfeng and {Shen}, Shiyin and {de Souza}, Rafael S. and {Chen}, Mi and {Ye}, Renhao and {She}, Yumei and {Chen}, Zhu and {Ishida}, Emille E.~O. and {Krone-Martins}, Alberto and {Durgesh}, Rupesh},
        title = "{From images to features: unbiased morphology classification via variational auto-encoders and domain adaptation}",
      journal = {\mnras},
         year = 2023,
        month = dec,
       volume = {526},
       number = {4},
        pages = {6391-6400},
          doi = {10.1093/mnras/stad3181},
archivePrefix = {arXiv},
       eprint = {2303.08627},
 primaryClass = {astro-ph.GA},
       adsurl = {https://ui.adsabs.harvard.edu/abs/2023MNRAS.526.6391X}
}

@ARTICLE{Ciprijanovic2021,
       author = {{{\'C}iprijanovi{\'c}}, A. and {Kafkes}, D. and {Downey}, K. and {Jenkins}, S. and {Perdue}, G.~N. and {Madireddy}, S. and {Johnston}, T. and {Snyder}, G.~F. and {Nord}, B.},
        title = "{DeepMerge - II. Building robust deep learning algorithms for merging galaxy identification across domains}",
      journal = {\mnras},
         year = 2021,
        month = sep,
       volume = {506},
       number = {1},
        pages = {677-691},
          doi = {10.1093/mnras/stab1677},
archivePrefix = {arXiv},
       eprint = {2103.01373},
 primaryClass = {astro-ph.IM},
       adsurl = {https://ui.adsabs.harvard.edu/abs/2021MNRAS.506..677C}
}

@ARTICLE{Ciprijanovic2022,
       author = {{{\'C}iprijanovi{\'c}}, Aleksandra and {Kafkes}, Diana and {Snyder}, Gregory and {S{\'a}nchez}, F. Javier and {Perdue}, Gabriel Nathan and {Pedro}, Kevin and {Nord}, Brian and {Madireddy}, Sandeep and {Wild}, Stefan M.},
        title = "{DeepAdversaries: examining the robustness of deep learning models for galaxy morphology classification}",
      journal = {Machine Learning: Science and Technology},
         year = 2022,
        month = sep,
       volume = {3},
       number = {3},
          eid = {035007},
        pages = {035007},
          doi = {10.1088/2632-2153/ac7f1a},
archivePrefix = {arXiv},
       eprint = {2112.14299},
 primaryClass = {cs.LG},
       adsurl = {https://ui.adsabs.harvard.edu/abs/2022MLS&T...3c5007C}
}

@ARTICLE{Ciprijanovic2023,
       author = {{{\'C}iprijanovi{\'c}}, A. and {Lewis}, A. and {Pedro}, K. and {Madireddy}, S. and {Nord}, B. and {Perdue}, G.~N. and {Wild}, S.~M.},
        title = "{DeepAstroUDA: semi-supervised universal domain adaptation for cross-survey galaxy morphology classification and anomaly detection}",
      journal = {Machine Learning: Science and Technology},
         year = 2023,
        month = jun,
       volume = {4},
       number = {2},
          eid = {025013},
        pages = {025013},
          doi = {10.1088/2632-2153/acca5f},
archivePrefix = {arXiv},
       eprint = {2302.02005},
 primaryClass = {astro-ph.GA},
       adsurl = {https://ui.adsabs.harvard.edu/abs/2023MLS&T...4b5013C}
}

@ARTICLE{Huertas-Company2024,
       author = {{Huertas-Company}, M. and {Iyer}, K.~G. and {Angeloudi}, E. and {Bagley}, M.~B. and {Finkelstein}, S.~L. and {Kartaltepe}, J. and {McGrath}, E.~J. and {Sarmiento}, R. and {Vega-Ferrero}, J. and {Arrabal Haro}, P. and {Behroozi}, P. and {Buitrago}, F. and {Cheng}, Y. and {Costantin}, L. and {Dekel}, A. and {Dickinson}, M. and {Elbaz}, D. and {Grogin}, N.~A. and {Hathi}, N.~P. and {Holwerda}, B.~W. and {Koekemoer}, A.~M. and {Lucas}, R.~A. and {Papovich}, C. and {P{\'e}rez-Gonz{\'a}lez}, P.~G. and {Pirzkal}, N. and {Seill{\'e}}, L.-M. and {de la Vega}, A. and {Wuyts}, S. and {Yang}, G. and {Yung}, L.~Y.~A.},
        title = "{Galaxy morphology from z {\ensuremath{\sim}} 6 through the lens of JWST}",
      journal = {\aap},
         year = 2024,
        month = may,
       volume = {685},
          eid = {A48},
        pages = {A48},
          doi = {10.1051/0004-6361/202346800},
archivePrefix = {arXiv},
       eprint = {2305.02478},
 primaryClass = {astro-ph.GA},
       adsurl = {https://ui.adsabs.harvard.edu/abs/2024A&A...685A..48H}
}

@ARTICLE{Alexander2023,
       author = {{Alexander}, Stephon and {Gleyzer}, Sergei and {Parul}, Hanna and {Reddy}, Pranath and {Tidball}, Marcos and {Toomey}, Michael W.},
        title = "{Domain Adaptation for Simulation-based Dark Matter Searches with Strong Gravitational Lensing}",
      journal = {\apj},
         year = 2023,
        month = sep,
       volume = {954},
       number = {1},
          eid = {28},
        pages = {28},
          doi = {10.3847/1538-4357/acdfc7},
archivePrefix = {arXiv},
       eprint = {2112.12121},
 primaryClass = {astro-ph.CO},
       adsurl = {https://ui.adsabs.harvard.edu/abs/2023ApJ...954...28A}
}

@ARTICLE{Farrens2022,
       author = {{Farrens}, S. and {Lacan}, A. and {Guinot}, A. and {Vitorelli}, A.~Z.},
        title = "{Deep transfer learning for blended source identification in galaxy survey data}",
      journal = {\aap},
         year = 2022,
        month = jan,
       volume = {657},
          eid = {A98},
        pages = {A98},
          doi = {10.1051/0004-6361/202141166},
archivePrefix = {arXiv},
       eprint = {2110.08180},
 primaryClass = {astro-ph.IM},
       adsurl = {https://ui.adsabs.harvard.edu/abs/2022A&A...657A..98F}
}

@ARTICLE{Roncoli2023arXiv,
       author = {{Roncoli}, Andrea and {{\'C}iprijanovi{\'c}}, Aleksandra and {Voetberg}, Maggie and {Villaescusa-Navarro}, Francisco and {Nord}, Brian},
        title = "{Domain Adaptive Graph Neural Networks for Constraining Cosmological Parameters Across Multiple Data Sets}",
      journal = {arXiv e-prints},
         year = 2023,
        month = nov,
          eid = {arXiv:2311.01588},
        pages = {arXiv:2311.01588},
          doi = {10.48550/arXiv.2311.01588},
archivePrefix = {arXiv},
       eprint = {2311.01588},
 primaryClass = {astro-ph.CO},
       adsurl = {https://ui.adsabs.harvard.edu/abs/2023arXiv231101588R}
}

@article{tSNE,
  author  = {Laurens van der Maaten and Geoffrey Hinton},
  title   = {Visualizing Data using t-SNE},
  journal = {Journal of Machine Learning Research},
  year    = {2008},
  volume  = {9},
  number  = {86},
  pages   = {2579--2605},
  url     = {http://jmlr.org/papers/v9/vandermaaten08a.html}
}

@article{UMAP, 
doi = {10.21105/joss.00861}, url = {https://doi.org/10.21105/joss.00861}, year = {2018}, publisher = {The Open Journal}, volume = {3}, number = {29}, pages = {861}, author = {McInnes, Leland and Healy, John and Saul, Nathaniel and Großberger, Lukas}, title = {UMAP: Uniform Manifold Approximation and Projection}, journal = {Journal of Open Source Software} }

\begin{appendix}
\onecolumn
\section{Images and PDFs}
\label{sec:images_pdf}

\begin{figure*}[h]
\centering
\includegraphics[width=9.cm]{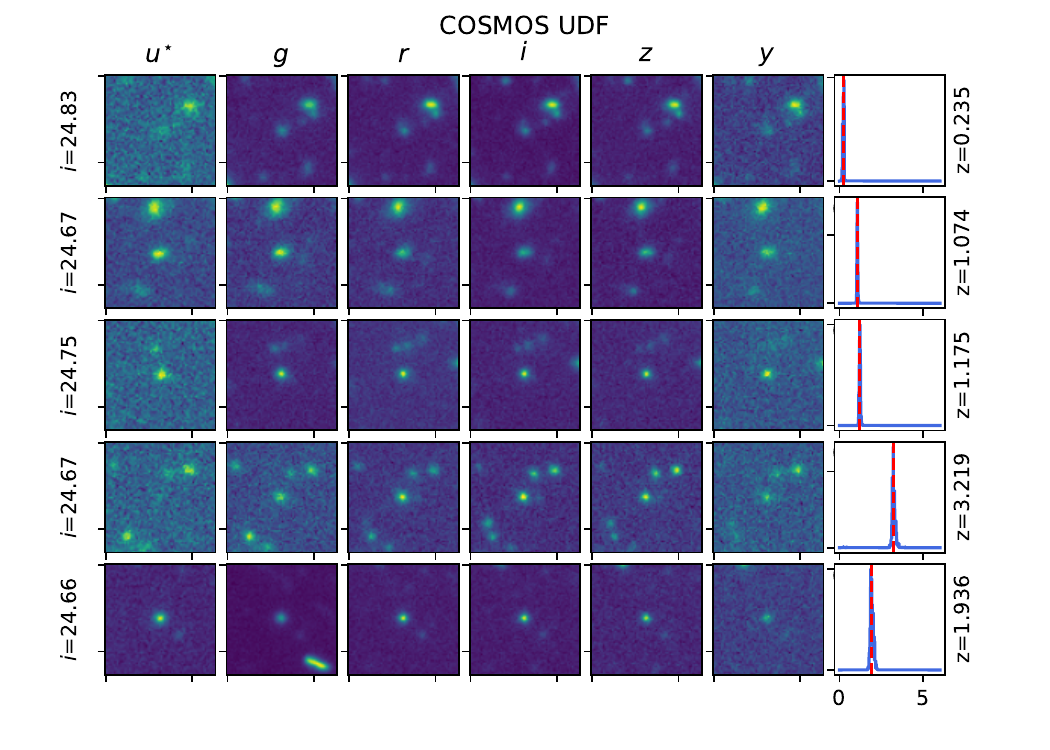}
\includegraphics[width=9.cm]{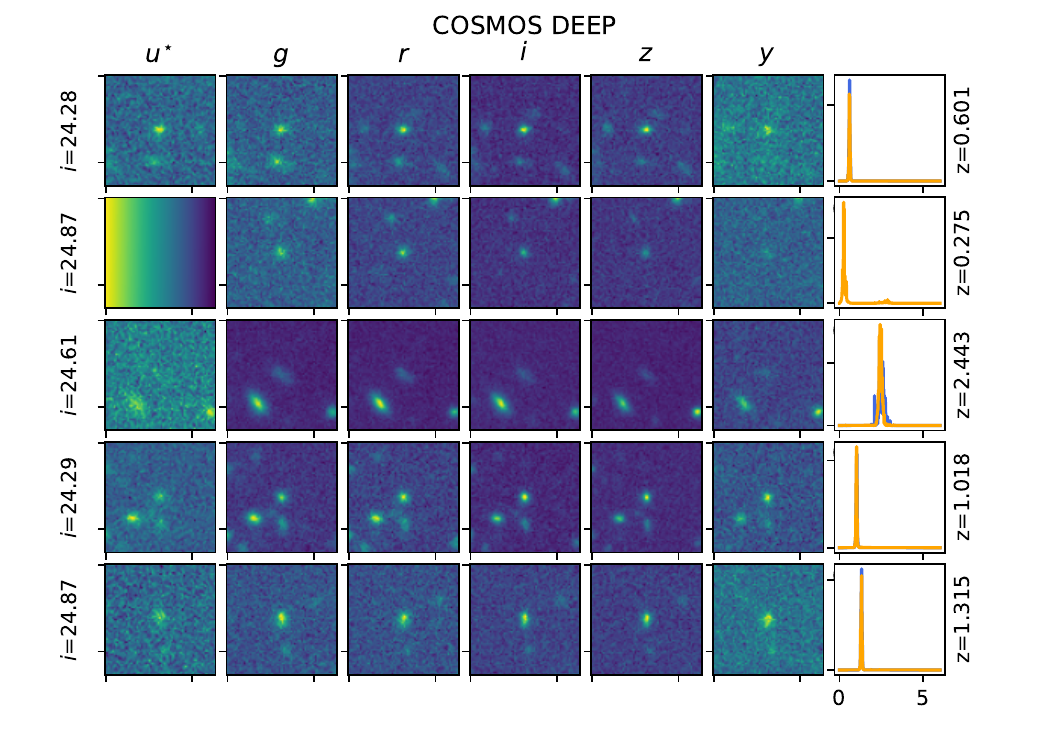}
\includegraphics[width=9.cm]{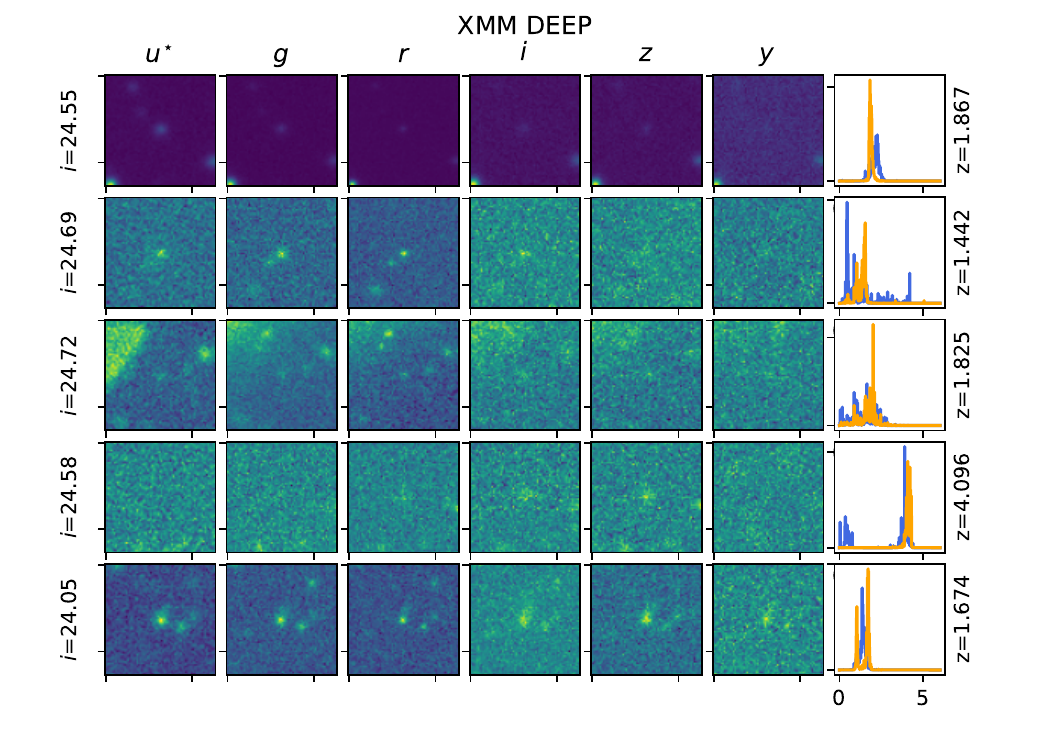}
\includegraphics[width=9.cm]{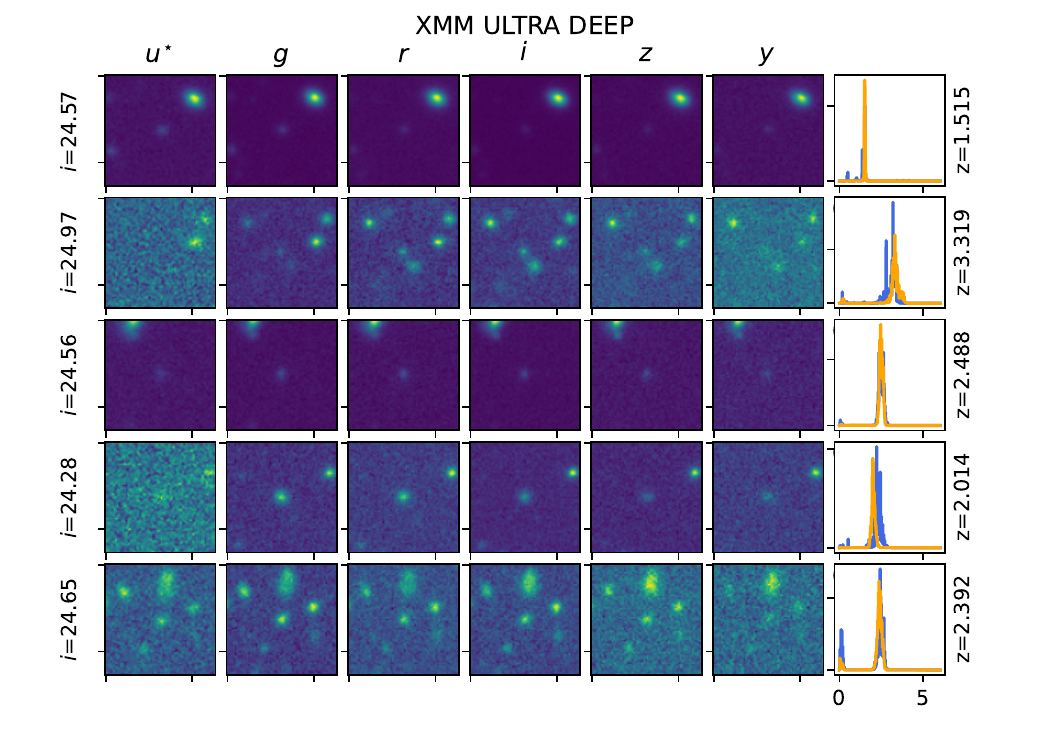}
\includegraphics[width=9.cm]{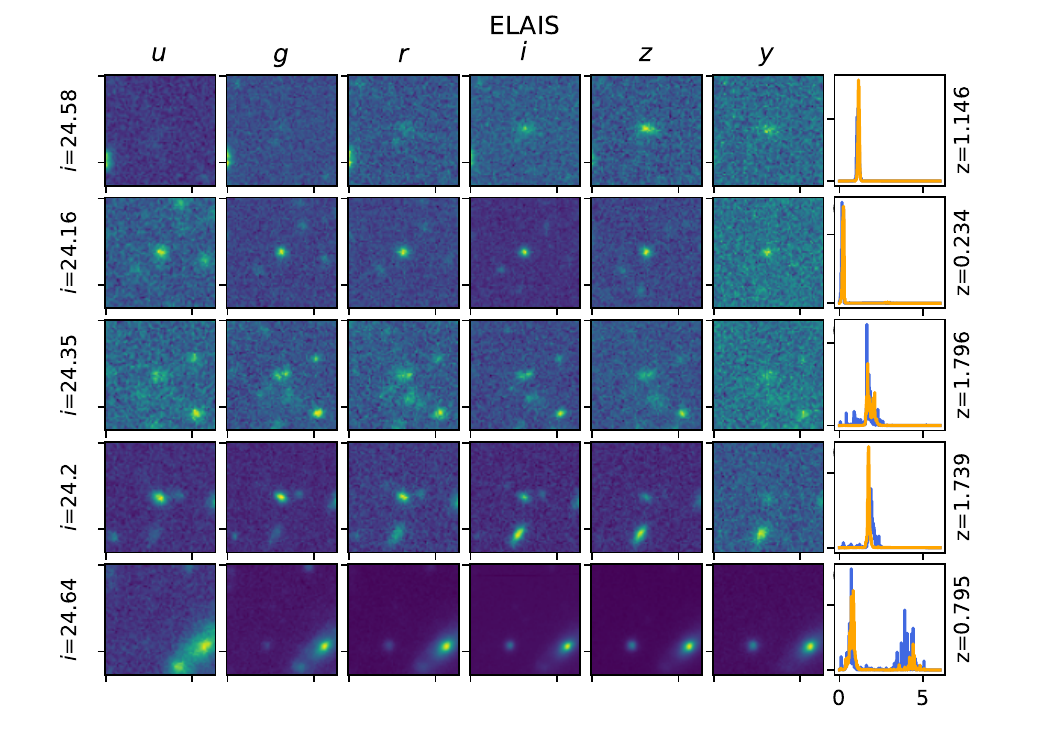}
\includegraphics[width=9.cm]{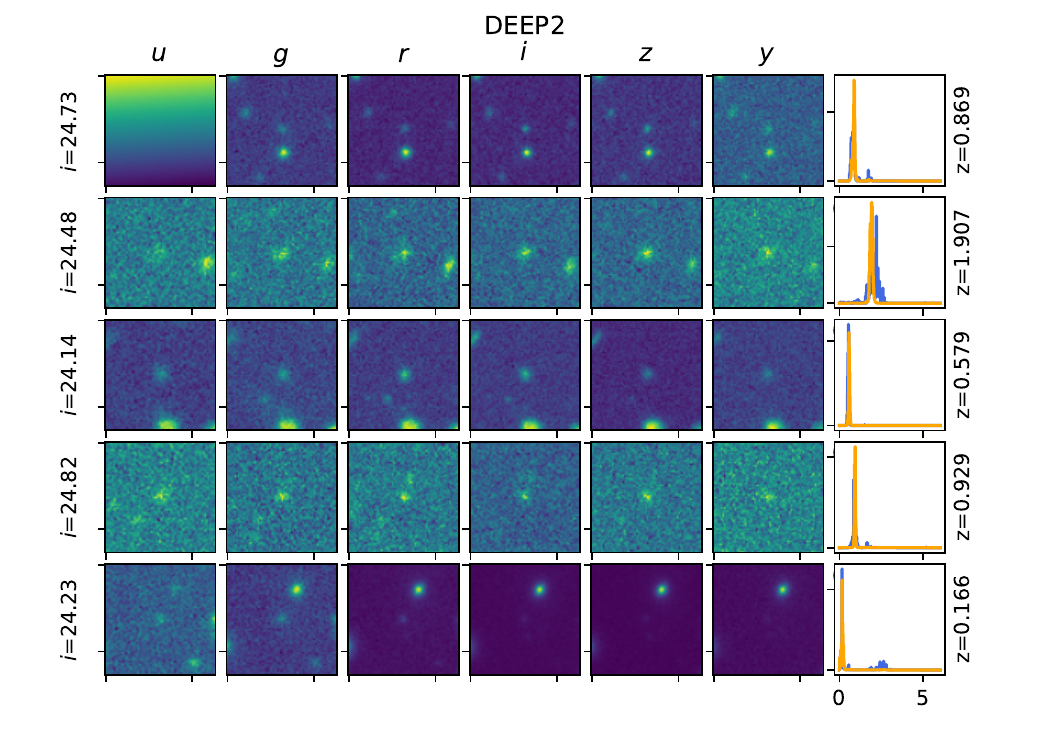}
\caption{A random sample of $u^{\star}grizy$ or $ugrizy$ images of galaxies at $24<i<25$ in the 6 HSC sub-regions. For the COSMOS UDF, the right most panels show the PDFs from the baseline cross-validation training and the $z$ values on the right, marked by vertical red dashed lines, are the redshift labels (spectroscopic or C2020). For the regions other than COSMOS UDF, the right most panels show the PDFs (PDFw $<1.5$) from the v0 model in blue, and for the v2 model in orange. The $z$ values on the right are the v2 redshift estimates.}
\label{fig:images_pdf}
\end{figure*}

\end{appendix}

\end{document}